\documentstyle[12pt]{article}
\newcounter{lyter}[equation]
\newcommand{\ictimes}{\mbox{$\times \hspace{-1em}\supset$}}
\title{General Superfield Quantization Method. I.\\
 Lagrangian Formalism of $\theta$-Superfield Theory of Fields}

\author{A.A. Reshetnyak\thanks{E-mail: reshet@tspu.edu.ru}}

\date{\it Department of Mathematics, Seversk State Technological Institute,
\protect \\
Seversk, {\rm 636036}, Russia}
\begin{document}

\maketitle
\begin{abstract}
The rules for constructing Lagrangian formulation for
$\theta$-superfield theory of fields ($\theta$-STF) are introduced and
considered
on the whole in the framework of proposed here new general superfield
quantization method for general gauge theories.

Algebraic, group-theoretic and analytic description aspects for
supervariables over (Grassmann) algebras containing anticommuting
generating element $\theta$ and interpreted further in particular as an
"odd" time are examined.

Superfunction $S_{L}(\theta) \equiv S_{L}\left({\cal A}(\theta),
\frac{d{{\cal A}}(\theta)}
{d\theta\phantom{xx}}, \theta \right)$ and its global symmetries are defined
 on the extended space (supermanifold)
$T_{odd}{\cal M}_{cl} \times \{\theta\}$ parameterized by local
coordinates: superfields ${\cal A}^{\imath}(\theta), \frac{d{{\cal
A}}^{\imath}(\theta)}{d\theta \phantom{xxx}}, \theta$. Extremality
properties of the superfunctional $Z[{\cal A}]=\int d\theta
S_{L}(\theta)$ are analyzed together with properties of
corresponding Euler-Lagrange equations. System of definitions for
Lagrangian formulation of $\theta$-STF including  extension of
gauge invariance concept is suggested.

The direct and inverse problems of zero locus reduction for extended
(anti)symplectic manifolds over ${\cal M}_{min}$ = $\{({\cal A}^{\imath}(
\theta), C^{\alpha}(\theta))\}$, with (odd) even brackets,
corresponding to initial (special) general type $\theta$-superfield models
are employed to construct iteratively the new interconnected models both
embedded into manifolds above with reduced brackets and enlarging them with
continued ones.

Component (on $\theta$) formulation for $\theta$-STF variables and operations
is produced providing the connection with standard gauge field theory.
Realization
of  $\theta$-STF constructions  is demonstrated on  models of
scalar, spinor, vector superfields which are used to formulate the
$\theta$-superfield model with abelian two-parametric gauge supergroup
generalizing the $\theta$-superfield quantum electrodynamics.
\end{abstract}

\noindent
{\large PACS codes: 03.50.-z, 11.10.Ef, 11.15.-q, 12.20.-m \protect }
\renewcommand{\thesubsection}{\Roman{subsection}}
\renewcommand{\thesubsubsection}{\thesubsection.\arabic{subsubsection}}
\subsection{Introduction}
\renewcommand{\theequation}{\arabic{subsection}.\arabic{equation}}
\renewcommand{\thelyter}{\alph{lyter}}

Investigations in the field of generalization of the Lagrangian
and Hamiltonian quantization methods for gauge theories based on
using special types of supertransformations the such as BRST
symmetry [1] and BRST-antiBRST (extended BRST) symmetry [2] have
been developed in the last 15--20 years sufficiently intensively.

The rules of canonical (BFV [3] and Sp(2) [4]) and Lagrangian (BV [5] and
Sp(2) [6]) quantization methods for gauge theories realizing
above-mentioned symmetry types have become, in the first place, basic for
correct investigations of the quantum properties of concrete modern models
of gauge field theory and, in the second, have been found in fact to be
fundamental as in explosure of algebraic and geometric-differential
aspects inherent in the methods [3--6] as their further advance. As to
latter two sentences, then confining ourselves to Lagrangian methods it
should be noted a use of algebraic and differential structures
(operations) and quantities of the BV method in string theories [7].

Possibilities of deformation theory, deformation quantization and
homological perturbation theory were considered in application to
generating equations of BV and BFV methods for their study with help of
cohomological theory techniques for Lie groups and algebras [8] and for
corresponding obtaining of gauge field theory Lagrangians with interaction
(see refs.[9]). So, for instance, a quantum deformation (on $\hbar$
degree) of generalized symplectic structures by means of star
multiplication [10] was used in order to establish relationship between
Fedosov deformation quantization [11] and BFV-BRST quantization of
dynamical systems with II class constraints [12].

Ingredients of BV method considered from viewpoint of supermathematics and
theory of supermanifolds [13] have found more or less clear classified
geometric matter [14] (see paper [15] and references therein as well)
being by more complicated analog of symplectic geometry.

The development of quantization scheme [5] in the field of introduction a more
general class of gauge conditions than commuting with respect to (w.r.t.)
antibracket, i.e. nonabelian hypergauges, was suggested in ref.[16]. In its
turn, Sp(2)-covariant Lagrangian quantization method was intensively
improved that was expressed both in its generalization in the form of
Sp(2)-symmetric method [17], then of triplectic [18], recently modified
triplectic [19] (in connection with Fedosov supermanifolds concept [11]) ones
and in creation
of some variants of the corresponding differential geometry [20].
A presence of $Z_{2}$-graded differential structures and quantities on
superspaces and, more generally, on supermanifolds permitted one to
consider the generalization of (ordinary) differential equations concept
by means of introduction of the so-called "Shander's supertime" [21]
{\boldmath $\Gamma$}$=(t,\theta)$.  {\boldmath $\Gamma$} includes together
with even parameter $t \in {\bf R}$, in the sense of fixed
$Z_{2}$-grading, the odd parameter $\theta$ $({\theta}^{2}=0)$ as well.
The fact that variable $\theta$ can be used as "odd time" for BV method
formulation was noted by ${\rm \ddot{O}.F.Dayi}$ [22].

The realization of the fact that BRST (extended BRST) symmetry can be
realized, in a some sense, in the form of translations along variable $\theta$
($\theta,\bar{\theta}$) [23] had led, in first, to extension of
$D$-dimensional
Minkovski space to superspace parametrized by sets of supernumbers
$z^{M}=(x^{\mu},\theta)$ $\left(
z^{M}=(x^{\mu},\theta,\overline{\theta})\right)$, where $x^{\mu},{}\mu=0,
1, \ldots, D-1$ are coordinates in ${\bf R}^{1,D-1}$ and, in second, to
special superfield construction  of action functional for Yang-Mills type
theories [24]. By distinctive feature of that construction is the point
that gauge superfield multiplet has consisted of gauge classical and ghost
fields.

One from versions of superfield generalization of the Lagrangian BRST
quantization method for arbitrary gauge theories was given in ref.[25] and
in [26] for the case of (Sp(2)) BLT one [6]. The  extrapolation of the
superfield method [25] of BV quantization for nonabelian hypergauges [16]
is proposed in ref.[27].

However, given quantization scheme [25] has a number of problematic
places (including problems of fundamental nature), which, in general,
reduce on component (on $\theta$) formulation level to discrepancy with
quantities and relations of BV method. Besides, a superfield realization
of some(!) ingredients of BV method both on the whole and for concrete
field theory models was considered in ref.[28].

In Hamiltonian formalism of quantization for dynamical systems with
constraints (BFV method) the version of superfield quantization was
suggested as well (among them in operator formulation) with its
generalization to the case of arbitrary phase space (i.e. with local
coordinates not corresponding to Darboux theorem) [29]. This formulation
of quantization (with modifications) and its connection with objects and
relations from BV method have been considered in ref.[30].
Recently, a so-called superfield algorithm for constructing the master actions
in terms of superfields in the framework of BV formulation for a class of
special gauge field theories was proposed in refs.[31,32], being based on
geometrical description of sigma models given in [33] developed later to the
case of models above on manifold with boundary [34].

Quantization variant [25] has important methodological significance
consisting in fact that the multiplet contents of superfields and
superantifields, defined on the superspace ${\bf R}^{1,D-1 \mid 1}$ with
coordinates $(x^{\mu},\theta)$, include into themselves, in a natural way,
componentwise w.r.t. expansion in powers of $\theta$ the  sets
of such (anti)fields which can be identified with all variables of BV
method [5] (fields ${\phi}^{A}$, auxiliary fields ${\lambda}^{A}$,
antifields ${\phi}^{\ast}_{A}$ and sources $J_{A}$ to fields
${\phi}^{A}$). In the second place, an action of all differential-algebraic
structures: superantibracket, odd operators $U,{}V,{}\Delta$ was realized
in the explicit superfield form on superalgebra of functionals with derivation
(locally) defined on a supermanifold with coordinates
${\Phi}^{A}(\theta),{}{\Phi}^{\ast}_{A}(\theta)$.  In the third, the
generating equation is formulated in terms of above-mentioned objects and
generating functional of Green's functions $Z[{\Phi}^{\ast}]$ (in
notations of paper [25]) is constructed, and its properties formally
repeating a number of ones for corresponding functional from BV method are
established.

The work opens by itself a number of papers devoted to development of new
general superfield quantization method for gauge theories in Lagrangian
formalism of their description. The complete and noncontradictory formulation
of all the statements of the method requires accurate and successive
introduction in superfield form of all the quantities being used in
quantum field theory, for example, generating functionals of Green's
functions, including effective action, together with correct study of
their properties such as gauge invariant renormalization, gauge dependence
and so on.

Existence theorems for solutions of generating equations being used in the
method and for similar statements in further development of this approach,
for instance, for nonabelian hypergauges are the key and more complicated
objects for investigation, than in BV scheme.

The correct superfield formulation for classical theory based on variational
principle and having by one's definitely chosen restriction the usual
quantum field theory model with standard classical action functional
${{\cal S}}_{0}(A)$ of classical gauge fields $A^{\imath}$, composing the
zero component w.r.t. expansion on $\theta$ in the superfield
multiplet ${\cal A}^{\imath}(\theta)$, is the necessary(!) condition for
accurate establishment of general rules for general superfield
quantization method (GSQM) in Lagrangian formalism.

Similar realization of the classical objects including additionally to an
action the definition of all the gauge algebra structural functions, among
them the generators of gauge transformations can be carried out by the way to
be different from the construction of the action functionals for standard
superfield SUSY field theory models as the superfunctionals over usual
superspace coordinatized by $z^a$. By the key point on this way it appears the
enlargement of initial ${{\cal S}}_{0}(A)$ to superfunction with values in
Grassmann algebra with one generating element $\theta$ and depending upon
${\cal A}^{\imath}(\theta)$,
their derivatives w.r.t. $\theta$ and  $\theta$. To realize in this
direction the noncontradictory description means now of the more general
field theory models including the standard ones for  $\theta$ = $\partial_{
\theta}{\cal A}^{\imath}(\theta)$ =$0$ it will be widely used the analogy with
Lagrangian classical mechanics and field theory.

Quantities and
relations of above-described classical theory will have the adequate
correspondence with BV quantization objects and operations. Additionally,
given $\theta$-superfield formulation will have ensured the more
significant results of the new gauge models construction on the basis of
$\theta$-superfield zero locus reduction (ZLR) direct [35] and so-called inverse
problems by means of the duality between odd and even Poisson
brackets (see [36] and references therein) being embedded each in other
in definite sequences on the corresponding manifolds.

The purpose of present work is the construction according to what has been
said above of the Lagrangian formulation\footnote{the term "Lagrangian
formulation" of classical theory or
$\theta$-STF does not coincide w.r.t.
sense with notion "Lagrangian formalism" in the set expression: GSQM in the
Lagrangian formalism} for $\theta$-STF together with its some extensions and
field-theoretic examples. The paper is written in the
following way.

In Sec.II elements of algebra on Grassmann
algebra ${\Lambda}_ {1}(\theta)$ with a single generating element $\theta$
are considered together with canonical  realization of superspace ${\cal
M}$ coordinatized by sets
$(z^{a},\theta)$, where $z^{a}$ are the coordinates of usual superspace with
space-time supersymmetry. Superfield (in mentioned sense)
representations, including (ir)reducible ones, of corresponding supergroup
in superspace of superfunctions on ${\cal M}$ are shortly examined.
In addition, some technically main questions of algebra and analysis on
superalgebras of special superfunctions on ${\cal M}$ are analyzed here.

Section III is devoted to study of algebraic properties of the first order
differential operators acting on superalgebra of superfunctions on
$T_{odd}{\cal M}_{cl} \times \{\theta\}$.  Properly Lagrangian formulation
for $\theta$-STF is defined in Sec.IV and is directly connected with
possibility of
representation of special superfunction $S_{L}\left( {\cal A} (\theta),
\frac{d{{\cal A}}(\theta)}{d\theta\phantom{xx}}, \theta \right)$ defined
on $T_{odd}{\cal M}_{cl} \times \{\theta\}$ together with its maximal
global symmetry group.

The detailed systematic research of the Lagrangian formulation for
$\theta$-STF is carried out here, being concentrated on the study for
Euler-Lagrange equations for
superfunctional $Z[{\cal A}]=\int d \theta S_{L}(\theta)$ endowed with
introduction a concept on constraints and ideas concerning gauge theories
and gauge transformations of general and special types. In Sec.V it is
shown as
gauge invariance
permits to use the BV and BFV generating superfunction(al)s and
equations constructions with help of $\theta$-superfield brackets
introduction in order to realize, in general, qualitatively the algorithms
of the new $\theta$-superfield models obtaining.

The component (on $\theta$) formulation for objects and relations
of Lagrangian formalism for $\theta$-STF is suggested in Sec.VI. An
application of
general statements of Secs.II--VI is demonstrated in Sec.VII on a number of
$\theta$-superfield models starting with five basic
simple $\theta$-STF ones, among them interacting, describing a massive complex
spinless
scalar superfield, massive spinor superfield of spin $\frac{1}{2}$, massless
real (of helicity 1 for $D=4$) and massive complex vector superfields.
Given models can be directly generalized to the case when the corresponding
superfields take values in an arbitrary semisimple Lie superalgebra forming
some isotopic vectors.  In this connection note, that
mentioned
models appear, in fact, by the base ones for construction of the interacting
$\theta$-superfield  Yang-Mills type models in realizing of the gauge
principle [40]. The such programm is completely realized for the case of
superfield theory generalizing the model of Quantum
Electrodynamics as well.

Finally, concluding propositions
and analogy for $\theta$-STF in Lagrangian formalism with usual classical
mechanics complete the paper in Sec.VIII.

Necessary questions from theory of ordinary
differential equations with odd differential operator $\frac{d}{d\theta}$
are considered in appendix A.

For satisfaction to requirements of mathematical correctness it
is necessary to note the conditions in framework of which the work is made.
It is supposed that on supermanifold of classical superfields ${\cal
A}^{\imath}(\theta)$ (usually one considers a vector bundle with ${\cal
M}$ as a base) a trivial atlas is given or its consideration is bounded by
a definite neighbourhood in ignoring the topological aspects. As
consequence the local supermanifold coordinates are
defined globally, and therefore the elements of differential geometry on given
supermanifold are not considered in an invariant coordinate free form.

In paper it is used the standard condensed De Witt's notations
[37]. The total left derivative of superfunction $f(\theta)$ w.r.t.
$\theta$ and superfield partial right derivative of differentiable
superfunction ${\cal J}(\theta) \equiv {\cal J}\bigl( {\cal A}(\theta),
{\stackrel{\ \circ}{{\cal A}}}(\theta), \theta \bigr)$ w.r.t. superfield
${\cal A}^{\jmath}(\theta)$ for fixed $\theta$ are denoted by
means of  conventions
\begin{eqnarray}
{\stackrel{\;\circ}{f}}(\theta) \equiv  \frac{d_l f(\theta)}{d\theta
\phantom{xx}} \equiv \frac{d f(\theta)}{d\theta\phantom{xx}}
\equiv \partial_{\theta} f(\theta),\
\frac{{\partial}_{\theta,r}{\cal J}(\theta)}{\partial {\cal
A}^{\jmath}(\theta) \phantom{x}}  \equiv  \frac{{\partial}_{r}{\cal
J}(\theta)}{\partial {\cal A}^{\jmath}(\theta)} \equiv
\frac{{\partial}{\cal J}(\theta)}{\partial {\cal A}^{\jmath}(\theta)}
\equiv {\cal J},_{\jmath}(\theta)\,.
\end{eqnarray}
\subsection{Mathematical grounds}
\setcounter{equation}{0}

Let us consider a supergroup $J$ being by the direct product
of Lie supergroup $\bar{J}$ and one-parameter supergroup $P$
\begin{eqnarray}
{J} = \bar{J} \times P,\
P = \{h \in P\; \vert\; h(\mu) = \exp{(\imath \mu p_{\theta})}\}\; , 
\end{eqnarray}
with $\mu \in {}^{1}\Lambda_{1}(\theta)$, being by subspace of odd
elements  w.r.t. generating element $\theta$ from
2-dimensional Grassmann algebra $\Lambda_{1}(\theta)$ over number
field ${\bf K}{}({\bf R}\; {\rm or}\; {\bf C})$, and quantity
$p_{\theta}$ $(p_{\theta}^{2} = \frac{1}{2}[p_{\theta},p_{\theta}]_{+}=0)$ as
the basis element of Lie superalgebra corresponding to $P$. The latter can
be realized as the translation supergroup acting on Grassmann algebra
over $\Lambda_{1}(\theta)$ by the formula
\begin{eqnarray}
{} & {}  & h(\mu)g(\theta) = g(\theta + \mu), \ h(\mu) \in P,
\  g(\theta) \in {\tilde{\Lambda}}_{1}(\theta)\,, \nonumber \\
{} & {} & {\tilde{\Lambda}}_{1}(\theta) = \{g(\theta) \mid g(\theta)= g_0 +
g_1\theta, g_0,g_1 \in  E_{{\bf K}}\}\,, 
\end{eqnarray}
where $E_{\bf K}$ is the algebra of functions over ${\bf K}$.
 From Eq.(2.2) it follows the translation  generator
$p_{\theta}$ may be realized by means of
$\frac{d}{d\theta}$: $p_{\theta} = -\imath\frac{d}{d\theta}$.
Regarding that $\bar{J}$ is a semidirect product of Lie
supergroup $\overline{M}$ on a some Lie subsupergroup
${\bar{J}}_{\tilde{A}}$ from the supergroup ${\bar{J}}_{A}$ of all
automorphisms of $\overline{M}$: $\bar{J} = \overline{M}
\ictimes {{\bar{J}}_{\tilde{A}}}$ and taking into
account that ${J}_{\tilde{A}} \simeq \bigl( e, {\bar{J}}_{\tilde{A}}
\bigr)$\footnote{$e_{P}, e_{\bar{J}}, e$ appear by  the units in $P$,
$\bar{J}$, $\overline{M}$ respectively, and "$\simeq$" is the
sign of group isomorphism} is the Lie subsupergroup in $\bar{J}$, we obtain
the canonical realization of superspace $\tilde{\cal M}$ as the quotient
space $\bar{J}/{J_{\tilde{A}}}$.

In view of $P$ group commutability it follows the representation for
supergroup $J$ and superspace ${\cal M}$ in the form
\begin{eqnarray}
J = (\overline{M} \ictimes {{\bar{J}}_{\tilde{A}}})
\times P \simeq (\overline{M} \times P) \ictimes
{{\bar{J}}_{\tilde{A}}},\; {\cal M} =
{J}/{J_{\tilde{A}}} = \tilde{{\cal M}}\times \tilde{P}\,, 
\end{eqnarray}
where sign "$\times$" for ${\cal M}$ denotes a
Cartesian product of the superspaces $\tilde{{\cal M}}$ and $\tilde{P}$
\begin{eqnarray}
\tilde{{\cal M}}  =
\bar{J}/{J_{\tilde{A}}} \simeq (\bar{J} \times \{e_{P}\}) \big/
J_{\tilde{A}},\
\tilde{P}  \simeq  (\{e\} \times
P) \ictimes {\bar{J}}_{\tilde{A}} \big/J_{\tilde{A}}\;. 
\end{eqnarray}
Next, consider as $\bar{J}$
the group of space-time supersymmetry, the such that $\tilde{{\cal M}}$ is the
real superspace, with which
one deals in the superfield formulations of supersymmetric field
theory models. So choosing $\bar{J}$ in the form of Poincare type
supergroup acting in
\begin{eqnarray}
\tilde{{\cal M}} = {\bf {R}}^{1,D-1\mid N c}, c = 2^{[D/2]}\;, 
\end{eqnarray}
with $D$, $N$, $[x]$ being by the dimension of Minkowski space, the number of
supersymmetries and the integer part of  $x \in {\bf R}$ respectively, the
global symmetry supergroup can be realized under
validity of representation (2.3).

More general symmetry supergroups being encountered, for instance, in
(super)gravity and (super)string theories can be obtained by localization
of $\bar{J}$ up to supergroup of general coordinate transformations
simultaneously with introduction of Riemann metric on $\tilde{{\cal M}}$.

The elements from ${\cal M}$ are parametrized  in a basis determined by
generators from $\overline{M}$ and $\imath p_{\theta}$ by the coordinates
\begin{eqnarray}
(z^a,\theta) = \left(
x^{\mu},{\theta}^{Aj},\theta\right),\;\mu = 0, 1,\ldots,D-1,\;A =
1,\ldots,2^{[D/2]},\;j = 1,\ldots,N\,,
\end{eqnarray}
where  $\mu, A$ correspond to usual vector ($\mu$) and
spinor $(A)$ Lorentz indices.

The actions of supergroups $\bar{J}$, $P$ on the points from ${\cal M}$
follows from definitions (2.1), (2.2) and
identities $\bar{g} \equiv (\bar{g} ,e_{P})$, $\bar{g}\in \bar{J}$,
$h(\mu) \equiv \bigl( e_ {\bar{J}}, h(\mu)\bigr)$
\begin{eqnarray}
\forall \bar{g} \in \bar{J}: \bar{g}(z^a,\theta) =
(\bar{g}z^a,\theta)\,,
\forall h(\mu) \in P: h(\mu)(z^{a},\theta) =
 (z^a,\theta + \mu)\,.
\end{eqnarray}
Presence of $Z_{2}$-grading w.r.t. $\theta$ in
${\cal M}$ makes the following representation by one-valued
\begin{eqnarray}
{\cal M} = {}^{0}{\cal M} \oplus {}^{1}{\cal M} \equiv \tilde{{\cal M}}
\oplus \tilde{P},\; \dim{\cal M} = (\dim{\tilde{{\cal M}}},
\dim{\tilde{P}}) = (D|N c, 1)\,. 
\end{eqnarray}
The action of boson projectors $P_{a}(\theta), a=0, 1$ is defined on
$\Lambda_{1}(\theta)$ with standard properties
\begin{eqnarray}
P_{a}(\theta)P_{b}(\theta)  =   {\delta}_{ab}P_{a}(\theta),\, a, b= 0, 1\;,\;
\sum\limits_{a} P_{a}(\theta) =
1\,,
\end{eqnarray}
dividing $\Lambda_{1}(\theta)$ into direct sum of their proper subsuperspaces
${}^{a}\Lambda_{1}(\theta)$ = $P_{a}(\theta)\Lambda_{1}(\theta)$.
Their action  is continued in a natural way to the
action
on $\tilde{\Lambda}_{1}(\theta)$, so that for any Grassmann function
$g(\theta) \in \tilde{\Lambda}_{1}(\theta)$  (in what follows
called the superfunction) the equalities hold
\begin{eqnarray}
P_{0}(\theta)g(\theta) = g_{0}\;,\;P_{1}(\theta)g(\theta)
= g_{1}\theta\,, 
\end{eqnarray}
remaining valid if instead of
$E_{\bf K}$ one considers the algebra of functions over
$\tilde{\cal M}$ (${}^{a}{\cal M}$ = $P_{a}(\theta){\cal M}$).

From the various realization for $\Lambda_1(\theta)$ elements we will use
the representability for any $a(\theta)$ $\in$ $\Lambda_1(\theta)$ as the
series in powers of $\theta$ with trivial differentiability w.r.t.
this element  [13], so that projectors have the form of the
1st order differential operators
\begin{eqnarray}
P_0(\theta) = 1 - \theta
\partial_{\theta} ,\;P_1(\theta) = \theta \partial_{\theta}\,.
\end{eqnarray}
This realization of $\Lambda_{1}(\theta)$
is transferred without modifications on
$\tilde{\Lambda}_{1}(\theta)$ and $\Lambda_{D\vert Nc+1}(z^{a},\theta;$
${\bf K})$ being by Grassmann algebra over ${\bf K}$ with $D$ even
$x^{\mu}$ and $(Nc+1 )$ odd $\theta^{Aj}, \theta$ generating elements
[13].

From the main problem of supergroup $J$ finite-dimensional irreducible
representations
(irreps) study we only note that due to the
triviality of group $P$ occurence into $J$ given question, in fact, is
reduced to the study of supergroup $\bar{J}$  finite-dimensional
irreps.
So group $J$ superfield irreps  are realized (among them) on
the superfields of "Lorentz" ($\bar{J}$) type [38]
\begin{eqnarray}
{\cal A}^{\imath}(\theta),\; \imath & = & \bigl({\mu}_1, \ldots ,{\mu}_k,
(A_{1}j_{1}), \ldots ,(A_{m}j_{m}), z^{a}\bigr),\;{\mu}_{p} = 0,
1,\ldots,D-1\;,\nonumber\\
A_{r} & = & 1,\ldots,2^{[D/2]}, \;j_{s} =
1,\ldots,N,\;p = \overline{1,k},\; r,s = \overline{1,m}\, ,
\end{eqnarray}
to be regarded as superfunctions on $\Lambda_{D\vert Nc+1}(z^{a},\theta;{
\bf K})$ with
values in the corresponding representation space. Superfields ${\cal
A}^{\imath}(\theta)$ are homogeneous w.r.t. Grassmann parity
operator $\varepsilon$ acting on  $\tilde{\Lambda}_{D\vert Nc+1}(z^{a},
\theta;{\bf K})$, being by the superalgebra of superfunctions defined on
$\Lambda_{D\vert Nc+1}(z^{a},\theta;{\bf K})$,
\begin{eqnarray}
\varepsilon : \tilde{\Lambda}_{D\vert Nc+1}(z^{a},\theta;{\bf K})
\rightarrow {\bf Z}_2\,,
\end{eqnarray}
being considered as the additive homomorphism
of superalgebras. Grassmann parity (grading) $\varepsilon$ can be represented
in the form of direct sum of Grassmann gradings $ \varepsilon_{\bar{J}}$,
$\varepsilon_{P}$
\begin{eqnarray}
\varepsilon =
\varepsilon_{\bar{J}} + \varepsilon_{P},\ \ \varepsilon_{\bar{J}} :
\tilde{\Lambda}_{D\vert Nc}(z^{a};{\bf K}) \rightarrow {\bf Z}_2,\ \
\varepsilon_{P} : \tilde{\Lambda}_{1}(\theta;{\bf K}) \rightarrow {\bf
Z}_2\,, 
\end{eqnarray}
trivially continued up to mappings on
$\tilde{\Lambda}_{D\vert Nc+1}(z^{a},\theta; {\bf K})$. Thus,
$\varepsilon_{\bar{J}}$, $\varepsilon_{P}$ are the Grassmann parities
w.r.t. generating elements $z^{a}$ and $\theta$
respectively. Elements from $\tilde{\Lambda}_{D\vert Nc}(z^{a};{\bf K})$
are the superfunctions, which the $J$ (ir)reducible superfield representation
 is realized on,  being by restriction  of the supergroup $J$
representation $T$  onto $\bar{J}$: $T_{\mid \bar{J}}$.

In accordance with (2.13), (2.14) $\vec{\varepsilon}$ =
($\varepsilon_{P}$, $\varepsilon_{\bar{J}}$, $\varepsilon)$ are
defined on the generating elements $z^{a}, \theta$ in the
following way
\begin{eqnarray}
\vec{\varepsilon}(x^{\mu})=\vec{\varepsilon}({\theta}^{Ai}) + (0,1,1) =
\vec{\varepsilon}(\theta)+(1,0,1)= (0,0,0)=\vec{0}\,.
\end{eqnarray}
Contents of component fields in ${\cal A}^{\imath}(\theta)$ are given by the
expansion in powers of $\theta$ [25] together with the values
of their $\vec{\varepsilon}$ parity
\begin{eqnarray}
{\cal A}^{\imath}(\theta) = A^{\imath} + {\lambda}^{\imath}\theta,\ \;
\vec{\varepsilon}(A^{\imath}) = \vec{\varepsilon}({\cal A}^{\imath}(\theta))=
\vec{\varepsilon}({\lambda}^{\imath}) + (1,0,1) =
((\varepsilon_{P})_{\imath}, (\varepsilon_{\bar{J}})_{\imath},
 {\varepsilon}_{\imath})\,. 
\end{eqnarray}
Thus, the homogeneous w.r.t. $\varepsilon$ superfield ${\cal A}^{
\imath}(\theta)$ has $\varepsilon_{\bar{J}}$, $\varepsilon_{P}$ parities
as for one's $P_{0}(\theta) $-component field $A^{\imath}$. In addition to
gradings above  define the $\vec{\varepsilon}$ values for differentials
$(dz^{a}, d\theta)$ and for
differential operators  $(\partial_{z^{a}}, \partial_{\theta})$ to be the
same as in (2.15).

The parities spectrum  shows that for
$A^{\imath}$, $\lambda^{\imath}$ the connection
between spin and statistic is standard w.r.t.
$\varepsilon_{\bar{J}}$ for $\varepsilon_{P} = 0$, but w.r.t. $\varepsilon$
for
$\lambda^{\imath}$ is wrong being corresponding as the rule to unphysical
degrees of freedom. The latter reflects the nontrivial fact
of the  generating element $\theta$ presence and
$\varepsilon_{P}\not\equiv 0$.

Whereas classical superfields ${\cal A}^{\imath}(\theta)$ are transformed
w.r.t. a some, in general, group $J$ reducible superfield finite-dimensional
representation $T$, the group $P$ irrep is one-dimensional and operators
$T(h(\mu))$ act on ${\cal A}^{\imath}(\theta)$ as translations along
$\theta$.

The transformation laws
\begin{eqnarray}
{\cal A}^{\imath}(\theta) & \mapsto & {{\cal A}'}^{\imath}({\theta}') =
\bigl( T(e,\tilde{g}){\cal A} {\bigr)}^{\imath}(\theta),\;\tilde{g} \in
{\bar{J}}_{\tilde{A}}\,,\\
{\cal A}^{\imath}(\theta) &
\mapsto & {{\cal A}'}^{\imath}({\theta}) = \bigl( T(e,\tilde{g}){\cal A}
{\bigr)}^{\imath}\bigl( T(h^{-1}(\mu))\theta\bigr) = ( T(e,\tilde{g}){\cal
A} {\bigr)}^{\imath}(\theta - \mu)\,,
\end{eqnarray}
realize the finite-dimensional and infinite-dimensional superfield
representations respectively with generator of translations along $\theta$:
$T(p_{\theta})$ = $- \imath \frac{d}{d\theta}$.

Note, firstly, that we do not consider here the other possibilities for
supergroup $\bar{J}$
nontrivial extension being analogous to the way of $N=1$ supersymmetry
group construction and, secondly,
the following permutability rule for any $\varepsilon$-homogeneous
elements  holds
\begin{eqnarray}
a(\theta)b(\theta) =
(-1)^{{\varepsilon}(a(\theta))\varepsilon(b(\theta))}
b(\theta)a(\theta),\
a(\theta), b(\theta)\in \tilde{\Lambda}_{D\vert Nc+1}(z^{a},\theta;{\bf K})
\,.
\end{eqnarray}

Starting from supermanifold ${\cal M}_{cl}$ coordinatized by ${\cal
A}^{\imath}(\theta)$ (formally $\imath = 1,\ldots, n,\,
n=(n_{+}, n_{-})$; $n_{+}(n_{-})$ is the number of boson (fermion) w.r.t.
$\varepsilon$ degrees of freedom entering in
condensed index\footnote{besides of formula (2.12),
index $\imath$ can certainly contain the discrete indices, characterizing
the belonging of ${\cal A}^{\imath} (\theta)$ to representation space of
some other groups, for example, of the Yang-Mills type} $\imath$),
 being more precisely by a special
tensor bundle over ${\cal M}$, let us formally
construct the following
supermanifolds $T_{odd}{\cal
M}_{cl}$, $T_{odd}{\cal M}_{cl} \times \{\theta\}$ parametrized by
$\bigl( {\cal A}^{\imath}(\theta), \partial_{\theta}{\cal
A}^{\imath}(\theta)\bigr)$,  $\bigl( {\cal A}^{\imath}(\theta),
\partial_{\theta}{\cal A}^{\imath}(\theta), \theta \bigr)$
respectively.
\sloppy
\begin{sloppypar}
Define the superalgebra of
superfunctions ${\bf K}[[T_{odd}{\cal M}_{cl} \times \{\theta\}]]$ given
on $T_{odd}{\cal M}_{cl} \times \{\theta\}$ with elements being by formal
power series w.r.t.  ${\cal
A}^{\imath}(\theta), \partial_{\theta}{\cal A}^{\imath}(\theta),
\theta$. This set con\-ta\-ins the su\-per\-al\-geb\-ra
${\bf K}[T_{odd}{\cal M}_{cl} \times \{\theta\}]$
of fi\-ni\-te po\-ly\-no\-mi\-als in po\-wers of ${\cal
A}^{\imath}(\theta), \partial_{\theta}{\cal A}^{\imath}(\theta)$.
For ar\-bit\-ra\-ry  ${\cal F}(\theta)$ $\in$
${\bf K}[[T_{odd}{\cal M}_{cl} \times \{\theta\}]]$ the
trans\-for\-ma\-ti\-on laws hold in ac\-ting of the
re\-pre\-sen\-ta\-tion $T$ ope\-ra\-tors, en\-su\-ing from Eqs.(2.17), (2.18)
res\-pec\-ti\-ve\-ly
\end{sloppypar}
\begin{eqnarray}
{\cal F}\left( {\cal A}'({\theta}'), \partial_{\theta'}{\cal A}'(
{\theta}'), {\theta}'\right) & = & {\cal F}\left( \bigl(
T(e,\tilde{g}){\cal A}\bigr)(\theta), \bigl( T(e,\tilde{g}) \partial_{\theta}{
\cal A}\bigr)(\theta), \theta\right)\,,\\
{\cal F}\left( {\cal A}'(\theta), \partial_{\theta}{\cal A}'(\theta),
{\theta}\right) & = & {\cal F}\left( \bigl( T(e,\tilde{g}){\cal
A}\bigr)(\theta - \mu), \bigl( T(e,\tilde{g}) \partial_{\theta}{\cal
A}\bigr)(\theta), \theta - \mu\right).
\end{eqnarray}
To obtain
(2.20), (2.21) we have used the formulae for the transformations of
$\partial_{\theta}{\cal A}^{\imath}(\theta)$
\begin{eqnarray}
{} & {} & \partial_{\theta'}{\cal A}'^{\imath}({\theta}')
= (\partial_{\theta'}{\theta})
\partial_{\theta}{\cal A}'^{\imath}({\theta}')
= {\left(T(e,\tilde{g})\partial_{\theta}{\cal A}\right)^{\imath}(\theta)}
,\; (\partial_{\theta'}{\theta}) = (\partial_{\theta}({\theta}+\mu))^{-1}\,,
\\
{} & {} & \partial_{\theta}{\cal A}'^{\imath}(\theta) = {\left(
T(e,\tilde{g}) \partial_{\theta}{\cal A}\right)^{\imath}(\theta-\mu)} =
{\left( T(e,\tilde{g}) \partial_{\theta}{\cal
A}\right)^{\imath}(\theta)},\;\tilde{g} \in {\bar{J}}_{\tilde{A}}
\,.
\end{eqnarray}
By definition, (local superfunction) ${\cal F}(\theta)$ is expanded in
(finite sum) formal power series
\begin{eqnarray}
{\cal F}\left({\cal A}(\theta),
{\stackrel{\ \circ}{\cal A}}(\theta), \theta \right) =
\sum_{l=0}\frac{1}{l!} {\cal F}_{(\jmath)_l}\bigl({\cal A} (\theta),
\theta \bigr)\vec{\stackrel{\ \circ}{\cal A}\,}{}^{(\jmath)_l}(\theta) =
\sum_{k,l=0}\frac{1}{k!l!} {\cal F}_{(\imath)_{k} (\jmath)_{l}}(\theta )
\vec{\stackrel{\ \circ}{\cal
A}\,}{}^{(\jmath)_l}(\theta)\vec{{\cal A}}^{(\imath)_k}(\theta)\,,
\end{eqnarray}
where we  have introduced the notations
\renewcommand{\theequation}{\arabic{subsection}.\arabic{equation}\alph{lyter}}
\begin{eqnarray}
\setcounter{lyter}{1}
{} & {} & {\cal F}_{(\jmath)_l}\bigl({\cal A}(\theta),
\theta {\bigr)} \equiv {\cal F}_{ \jmath_1 \ldots \jmath_l}\bigl({\cal A}
(\theta), \theta {\bigr)}\;,\;{\cal F}_{(\imath)_{k} (\jmath)_{l}}(\theta )
\equiv {\cal F}_{\imath_1 \ldots \imath_k \;\jmath_1 \ldots
 \jmath_l}(\theta)\,,  \\
\setcounter{equation}{25}
\setcounter{lyter}{2}
{} & {} &
\vec{\stackrel{\ \circ}{\cal A}\,}{}^{(\jmath)_l}(\theta) \equiv
\prod\limits_{p=1}^{l}
{\stackrel{\ \circ}{\cal A}}{}^{\jmath_p}(\theta),\
\vec{{\cal A}}^{(\imath)_k}(\theta) \equiv \prod\limits_{p=1}^{k} {\cal
A}^{\imath_p}(\theta)\,.  
\end{eqnarray}
Expansion coefficients in (2.24) appear themselves by superfunctions on
${\bf
K}[[{\cal M}_{cl} \times \{\theta\}]]$ or $\tilde{\Lambda}_{D
\vert Nc+1}(z^{a},\theta;{\bf K})$, and obey to the generalized symmetry
properties
\renewcommand{\theequation}{\arabic{subsection}.\arabic{equation}}
\begin{eqnarray}
 {\cal F}_{(\imath)_{k} (\jmath)_{l}}(\theta ) & = &
(-1)^{(\varepsilon_{\jmath_r}+1)(\varepsilon_{
\jmath_{r-1}}+1)}{\cal F}_{(\imath)_k\;\jmath_1 \ldots \jmath_r
\jmath_{r-1}\ldots\ \jmath_{l}}(\theta )
\nonumber \\
{} & = &
(-1)^{\varepsilon_{\imath_s}\varepsilon_{ \imath_{s-1}}}{\cal F}_{\imath_1
\ldots \imath_s \imath_{s-1}\ldots \imath_{k} (\jmath)_{l}}(\theta),\
s=\overline{2,k},\;r=\overline{2,l}\,. 
\end{eqnarray}

The superalgebra of the $k$-times differentiated superfunctions
$C^{k}\bigl( T_{odd}{\cal M}_{cl} \times \{\theta\}\bigr)$
$\equiv$ $C^k$, $k\le {\infty}$ can be obtained from ${\bf
K}[[T_{odd}{\cal M}_{cl} \times \{\theta\}]]$ by means of
compatible introduction on the latter the operation of
differentiation, the norm structure and the corresponding
convergence of  series in (2.24). Then for arbitrary ${\cal
F}(\theta) \in C^{k}$ we will suppose to be valid the expansion in
functional Taylor's series in powers of $\delta{\cal
A}^{\imath}(\theta)$ = $\bigl( {\cal A}^{\imath}(\theta)$ $-$
${\cal A}^{\imath}_{0} (\theta)\bigr)$ and $\delta{\stackrel{\
\circ}{\cal A}}{}^{\jmath}(\theta)$ $=$ $\bigl( {\stackrel{\
\circ}{\cal A}}{}^{\jmath}(\theta)$ $-$ ${\stackrel{\ \circ}{\cal
A}}{}^{\jmath}_{0}(\theta)\bigr)$ in a some neighbourhood ${\cal
A}^{\imath}_{0}(\theta)$, ${\stackrel{\ \circ}{\cal
A}}{}^{\jmath}_{0}(\theta)$
\begin{eqnarray}
{} & {} & {\cal F}\left( {\cal
A}(\theta), {\stackrel{\ \circ}{\cal A}}(\theta), \theta \right) =
 \sum\limits_{k,l=0}\frac{1}{k!l!}
{\cal  F}_{(\imath)_k\;(\jmath)_l} \left( {\cal A}_0(\theta), {\stackrel{\
\circ}{\cal A}}_0(\theta), \theta \right)\delta
\vec{\stackrel{\ \circ}{\cal A}\,}{}^{(\jmath)_l}(\theta)
\delta\vec{{\cal A}}^{(\imath)_k}(\theta)\,,
\nonumber \\
{} & {} &
{\cal  F}_{(\imath)_k\;(\jmath)_l} \left( {\cal A}(\theta), {\stackrel{\
\circ}{\cal A}}(\theta), \theta \right) =
\left( \prod\limits_{t=0}^{k-1}
\frac{{\partial}_r \phantom{xxxxx}}{\partial {\cal
A}^{\imath_{k-t}}(\theta)} \prod\limits_{u=0}^{l-1}
\frac{{\partial}_r
\phantom{xxxxx}}{\partial{\stackrel{\ \circ}{\cal
A}}{}^{\jmath_{l-u}}(\theta)}{\cal F}\left( {\cal A}(\theta),
{\stackrel{\ \circ}{\cal A}}(\theta), \theta \right) \right). 
\end{eqnarray}
The nonzero action of the partial right superfield derivatives w.r.t.
${\cal A}^{\imath}(\theta)$, $\partial_{\theta}{\cal A}^{\imath}(\theta)$
introduced according to (1.1) and acting nontrivially  on
${\cal F}(\theta)$ only for coinciding $\theta$, is defined as follows
\begin{eqnarray}
\frac{{\partial}_r {\cal
A}^{\jmath}(\theta)}{\partial {\cal A}^{\imath}(\theta)} =
{\delta}^{\jmath}{}_{ \imath}\;,\; \frac{{\partial}_r
\partial_{\theta}{\cal A}^{\jmath}
(\theta)}{\partial \partial_{\theta}{\cal A}^{\imath}(\theta)} =
\delta^{\jmath}{}_{\imath}\,.
\end{eqnarray}
At last, it is useful  to
combine the expansions (2.24),  (2.27) for corresponding arguments
${\cal A}^{\imath}(\theta)$,
$\partial_{\theta}{\cal A}^{\imath}(\theta)$.

The action of $\{P_{a}(\theta)\}$ is naturally continued onto
${\cal M}_{cl}$ ($P_{a}(\theta){\cal M}_{cl}$ = ${}^{a}{\cal M}_{cl}$) and
$C^{k}$.
Besides, it is convenient to
introduce  a more detailed system of projectors
$\{\tilde{P}_{a}(\theta), U(\theta)\}, a = 0, 1$ with $C^{k}$
decomposition\footnote{the only
${}^{0,0}C^{k}$ appears by nontrivial subsuperalgebra, whereas
${}^{1,0}C^{k}$, ${}^{0,1}C^{k}$ are  nilpotent ideals in $C^{k}$}
\begin{eqnarray}
C^{k} & =  &
C^{k}\bigl(P_0(T_{odd}{\cal M}_{cl})\bigr) \oplus C^{k}\bigl( P_1(T_{odd}{\cal
M}_{cl})\bigr) \oplus C^{k}\bigl( P_0(T_{odd}{\cal M}_{cl}) \times
\{\theta\}\bigr) \nonumber \\
{} &  \equiv & {}^{0,0}C^{k} \oplus {}^{1,0}C^{k}
\oplus {}^{0,1}C^{k},\ \; \tilde{P}_a(\theta) C^k =
{}^{0,a}C^{k},\  U(\theta) C^k = {}^{1,0}C^{k}\,.
\end{eqnarray}
By definition $\{\tilde{P}_{a}(\theta), U(\theta)\}$ are characterized
by the relations
\begin{eqnarray}
\tilde{P}_a(\theta)\tilde{P}_b(\theta) =
\delta_{ab}\tilde{P}_b(\theta),\;\tilde{P}_a(\theta) U(\theta) =
0,\;U^2(\theta) = U(\theta),\;\Bigl(\sum_a \tilde{P}_a(\theta)\Bigr)
+U(\theta)=1\,. 
\end{eqnarray}
The analytic notation  of ${\cal
F}(\theta)$ by means of relation (2.27) results in representation of the
projectors under their action on $C^{k}$ in the form of the 1st order
differential operators
\begin{eqnarray}
\tilde{P}_0(\theta)={P}_0(\theta)=1-\theta\frac{d}{d\theta} ,\;
U(\theta)=P_1(\theta){\cal A}^{\imath}(\theta)\frac{\partial_l
\phantom{xxx}}{\partial{\cal A}^{\imath}(\theta)}\footnotemark ,\;
\tilde{P}_1(\theta)=\theta\frac{\partial}{\partial\theta}\,,
\end{eqnarray}
\footnotetext{$P_{1}(\theta){\cal
A}^{\imath}(\theta)$ is regarded as indivisible object in the sense, that
$\frac{\partial}{ \partial \theta} P_{1} (\theta){\cal A}^{\imath}(\theta)
= 0$, but $\frac{\partial}{\partial \theta}\bigl( {\theta}
{\stackrel{\,\circ}{{\cal A}^{\imath}}}(\theta)\bigr) =
{\stackrel{\,\circ}{{\cal A}^{\imath}}}(\theta)$}permitting to conclude that
$\tilde{P}_{1}(\theta),
U(\theta), P_{1}(\theta)$ appear by the derivations on $C^{k}$ whereas the
${P}_{0}(\theta), (\tilde{P}_{0}(\theta))$
action on the product of  ${\cal F}(\theta), {\cal
J}(\theta)$ from $C^{k}$ is equal to the product of its action on these
elements. In turn, the connection between derivatives $\frac{d}{d\theta}$,
$\frac{\partial}{\partial \theta}$ under their action on  $C^{k}$ is
defined by the formulae
\begin{eqnarray}
\displaystyle\frac{d}{d\theta} =
\frac{\partial}{\partial\theta} + {\stackrel{\;\circ}{{\cal
A}^{\imath}}}(\theta) P_0(\theta)\frac{\partial_l
\phantom{xxx}}{\partial{\cal A}^{\imath}(\theta)} \equiv
\frac{\partial}{\partial\theta} +
P_0(\theta){\stackrel{\circ}{U}}(\theta),\ {\stackrel{\circ}{U}}(\theta) =
\left[\displaystyle\frac{d}{d\theta}, U(\theta)\right]_s . 
\end{eqnarray}

Define a class $C_{F}$ of analytic over  ${\bf K}$
superfunctionals on  ${\cal M}_{cl}$ by means of the formula
\renewcommand{\theequation}{\arabic{subsection}.\arabic{equation}}
\begin{eqnarray}
F[{\cal A}] = \int d\theta {\cal F}\bigl({\cal A}
(\theta), \partial_{\theta}{\cal A}(\theta), \theta\bigr) \equiv
\partial_{\theta}{\cal F}(\theta),\; F[{\cal A}] \in C_{F}, {\cal
F}(\theta) \in C^{k}\,,
\end{eqnarray}
the such that, only $({\cal F}^{1,0}, {\cal F}^{0,1})$ = $(U,
\tilde{P}_1){\cal F}(\theta)$ parts of
${\cal F}(\theta)$ give the nontrivial
contribution into $F[{\cal A}]$. As far as the operator
$\partial_{\theta}$ does not lead out ${\cal F}(\theta)$ from $C^{k}$,
then $F[{\cal A}]$ belongs to $C^{k}$ as well. Besides, $F[A]$ appears now
by the scalar under action of representation $T$ operators, if ${\cal
F}(\theta)$ is transformed according to the rules (2.20) or (2.21).

With help of $\theta$-superfield analog of variational calculus basic lemma
one can establish for superfunctionals from $C_F$ the relation between
left (right) superfield variational derivative of $F[{\cal A}]$ w.r.t. ${\cal
A}^{\imath}(\theta)$  and left (right) partial superfield derivatives
of its ($F[{\cal A}]$) density ${\cal F}(\theta)$ w.r.t.
${\cal A}^{\imath}(\theta)$, $\partial_{\theta}{\cal
A}^{\imath}(\theta)$
\begin{eqnarray}
\frac{\delta_{l(r)} F[{\cal A}]}{\delta{\cal A}^{\imath}(\theta)\phantom{x}}
= \left[\frac{\partial_{l(r)} \phantom{xx}}{\partial{\cal A}^{\imath}(\theta)}
 -(-1)^{\varepsilon_{\imath}}\partial_{\theta}^{l(r)}\frac{
 \partial_{l(r)}
\phantom{xxxxxx}}{\partial
\left(\partial_{\theta}^{l(r)}{\cal
A}^{\imath}(\theta)\right)}\right]
{\cal F}(\theta) \equiv {\cal  L}_{\imath}^{l(r)}(\theta){\cal F}(\theta)\,.
\end{eqnarray}
Note, if to start from priority of  the density ${\cal F}(\theta)$ in
comparison with $F[{\cal A}]$ then the preservation of the $J$-covariant
total derivative w.r.t. $\theta$ by fixed, if othewise it is not said,
do not gives the nonvanishing contribution in $F[{\cal A}]$ and to
left-hand side of (2.34) providing the consistensy and additional special
properties for ${\cal F}(\theta)$.
Superfield derivatives have the following values of Grassmann parities
according to (2.14)-(2.16)
\renewcommand{\theequation}{\arabic{subsection}.\arabic{equation}}
\begin{eqnarray}
\vec{\varepsilon}\frac{\delta_l \phantom{xxx}}{\delta{\cal
A}^{\imath}(\theta)} =
\vec{\varepsilon}\frac{\partial_l \phantom{xxxx}
 }{\partial (\partial_{\theta}{\cal A}^{\imath}(\theta))}=
\vec{\varepsilon}\frac{\partial_l \phantom{xxx}}{\partial{\cal A}^{\imath}(
\theta)} + (1,0,1) =
((\varepsilon_{P})_{\imath}+1, (\varepsilon_{\bar{J}})_{\imath},
\varepsilon_{\imath}+1).
\end{eqnarray}
The $k$th superfield variational derivative of superfunctional $F[{\cal A}]$
w.r.t. ${\cal A}^{{\imath}_{1}} (\theta_{1}),\ldots,
{\cal A}^{{\imath}_{k}}(\theta_{k})$ is ex\-pres\-sed thro\-ugh par\-tial
su\-per\-field de\-ri\-va\-ti\-ves w.r.t. ${\cal A}^{{\imath}_{1}}
(\theta),\ldots, {\cal A}^{{\imath}_{k}}(\theta)$, $\partial_{\theta}{
\cal A}^{{\imath} _{1}}(\theta), \ldots , \partial_{\theta}{\cal
A}^{{\imath}_{k}}(\theta)$ of its ($F[{\cal A}]$) density ${\cal
F}(\theta)$ by the formula
\begin{eqnarray}
{} & {} & \prod\limits_{l=0}^{k-1}
\frac{\delta_r F[{\cal A}]\phantom{xxx}}{\delta{\cal
A}^{\imath_{k-l}}(\theta_{k-l})}
= \Bigl( \prod\limits_{l=0}^{k-2}{\cal
L}_{\imath_{k-l}}^{r}(\theta_k)\delta(\theta_k - \theta_{k-l-1})\Bigr)
{\cal L}_{\imath_1}^{r}(\theta_k){\cal F}\bigl({\cal A}(\theta_k),
\partial_{\theta_k}{{\cal A}}(\theta_k), \theta_k \bigr)\,, \\
{} & {} & \delta({\theta}' -\theta)= {\theta}'
-\theta\;,\;\int d{\theta}'\delta({\theta}' -\theta)y(\theta')=
y(\theta)\;,\;\vec{\theta}_k \equiv \theta_1, \ldots,\theta_k\,.
\end{eqnarray}
The superfield variational derivative w.r.t. ${\cal
A}^{\imath}(\theta)$ of  ${\cal F}\bigl({\cal
A}({\theta}^{\prime}), \partial_{\theta'}{\cal A}({\theta}^{\prime}),
{\theta}^{\prime}\bigr)$ for compulsory not coinciding
$\theta$, ${\theta}^{\prime}$ is determined by (2.36) as well.
For its calculation from arbitrary ${\cal F}\bigl({\cal
A}({\theta}^{\prime}), \partial_{\theta'}{\cal A}({\theta}^{\prime})$,
${\theta}^{\prime}; {\vec{\theta}}_{k}\bigr)$ $\in$ $C^{k}\bigl( T_{odd}{\cal
M}_{cl}$ $\times$ $\{{\theta}^{\prime}\} \times \{{\vec{\theta}}_{k}\}\bigr)$
it is sufficient to know the variational derivatives of
${\cal A}^{\imath}({\theta}^{\prime})$, $\partial_{\theta'}{\cal
A}^{\imath}({\theta}^{\prime})$ whose values follow from (2.34),
(2.36) according to (2.28), (2.37)
\begin{eqnarray}
\frac{\delta_l {\cal
A}^{\jmath}(\theta')}{\delta{\cal A}^{\imath}(\theta)} =
(-1)^{\varepsilon_{\jmath}} \delta^{\jmath}{}_{\imath}\delta(\theta -
\theta')\ ,\; \frac{\delta_l (\partial_{\theta'}{\cal
A}^{\jmath}(\theta'))}{
\delta{\cal A}^{\imath}(\theta)\phantom{xxxx}}
= \delta^{\jmath}{}_{\imath}\,.
\end{eqnarray}
\subsection{Operatorial superalgebra ${\cal A}_{cl}$}
\setcounter{equation}{0}

A connection between partial superfield derivative
of superfunction ${\cal F}(\theta)$ w.r.t. ${\cal
A}^{\imath}(\theta)$ and partial derivatives for fixed $\theta$  w.r.t.
component fields $P_{0}(\theta){\cal
A}^{\imath}(\theta)$ and $P_{1}(\theta){\cal A}^{\imath}(\theta)$ may be
established in many ways. The resultant operatorial formulae being true on
$C^k$ read as follows
\begin{eqnarray}
\frac{\partial_{r(l)} \phantom{xx}}{\partial{\cal
A}^{\imath}(\theta)} = \frac{\partial_{r(l)} \phantom{xxxxxx}}{\partial
P_{0}(\theta){\cal A}^{\imath}(\theta)} + \frac{\partial_{r(l)}
\phantom{xxxxxx}}{\partial P_{1}(\theta){\cal A}^{\imath}(\theta)}\,.
\end{eqnarray}
The formulae (3.1) correctness can be readily determined in acting of
$\frac{\partial_{l} \phantom{xxx}}{\partial{\cal A}^{\imath}(\theta)}$
on  ${\cal F}(\theta) \in C^{k}$
taking account of the properties
for projectors $P_{a}(\theta)$ and the following summary of formulae
\begin{eqnarray}
\frac{\partial_l P_{a}{\cal
A}^{\imath}(\theta)}{\partial P_{b}{\cal A}^{\jmath}(\theta)} =
P_{a}(\theta){\delta}_{ab}\delta^{\imath}{}_{\jmath}\ ,\;
\frac{\partial_l {\cal A}^{\imath}(\theta)\phantom{x}
}{\partial P_{b} {\cal A}^{\jmath}(\theta)} =
P_{b}(\theta)\delta^{\imath}{}_{\jmath}\ ,\; \frac{\partial_l
P_{a}{\cal A}^{\imath} (\theta)}{\partial {\cal
A}^{\jmath}(\theta)\phantom{xx}} =
P_{a}(\theta){\delta}^{\imath}{}_{\jmath}\,.  
\end{eqnarray}
In fact, it is sufficient to prove the formula (3.1) for
superfunction ${\cal J}(\theta) \in C^{k}$ of the  form
\begin{eqnarray}
{} & {} & {\cal J}\left({\cal A}({\theta}), \partial_{\theta}{
\cal A}({\theta}),{\theta}\right)  = {\cal J}_1\left({\cal
A}({\theta}), \partial_{\theta}{\cal A}({\theta}), 0\right) + {\cal
J}_2\left({\cal A}({\theta}), \partial_{\theta}{\cal
A}({\theta}),{\theta}\right) \nonumber \\
{} & {} & \hspace{0.5em} =
\textstyle\frac{1}{n!}f_{(\imath)_n} \left(\partial_{\theta}{\cal
A}({\theta})\right) \vec{{\cal A}}^{(\imath)_n}(\theta) +
 \textstyle\frac{1}{l!} g_{(\imath)_l} \left(\partial_{\theta}{\cal
A}(\theta), \theta \right) \vec{{\cal A}}^{(
\imath)_l}(\theta),\; {\tilde{P}}_1 (\theta)g_{(\imath)_l}(\theta) =
g_{(\imath)_l}(\theta)\,,
\end{eqnarray}
with coefficients
$f_{(\imath)_{n}}(\theta)$,  $g_{(\imath)_{l}}(\theta)$ satisfying to
properties (2.26). We have successively for right derivatives
\renewcommand{\theequation}{\arabic{subsection}.\arabic{equation}\alph{lyter}}
\begin{eqnarray}
\setcounter{lyter}{1}
\Bigl({\cal J}_1{},_{\imath}, {\cal J}_2{},_{\imath}\Bigr)(\theta) &
\hspace{-0.3em}
= & \hspace{-0.3em}\Bigl(\textstyle\frac{1}{(n-1)!}f_{(\imath)_{n-1}\imath}
\left(
\partial_{\theta}{\cal A}({\theta})\right) \vec{{\cal
A}}^{(\imath)_{n-1}}, \textstyle\frac{1}{(l-1)!}
g_{(\imath)_{l-1}\imath}\left(\partial_{\theta}{\cal
A}({\theta}), \theta\right) \vec{{\cal
A}}^{(\imath)_{l-1}}\Bigr)(\theta),\\ 
\setcounter{equation}{4}
\setcounter{lyter}{2}
\frac{\partial_r {\cal
J}_1(\theta)\phantom{xxx}}{\partial P_{0}(\theta) {\cal
A}^{\imath}(\theta)} & \hspace{-0.3em}= &\hspace{-0.3em}
\textstyle\frac{1}{(n-1)!}f_{(\imath)_{n-1}\imath}
\left(\partial_{\theta}{\cal A}({\theta})\right) \vec{{\cal
A}}^{(\imath)_{n-1}}(\theta)P_0(\theta)\,, \\ 
\setcounter{equation}{4}
\setcounter{lyter}{3}
\frac{\partial_r {\cal
J}_1(\theta)\phantom{xxx}}{\partial P_{1} (\theta){\cal
A}^{\imath}(\theta)} & \hspace{-0.3em}= &\hspace{-0.3em}
\textstyle\frac{1}{(n-1)!}f_{(\imath)_{n-1}\imath}
\left(\partial_{\theta}{\cal A}({\theta})\right) \left(P_0{\vec{\cal
A}}\right)^{(\imath)_{n-1}}(\theta)P_1 (\theta)\,, \\ 
\setcounter{equation}{4}
\setcounter{lyter}{4}
\frac{\partial_r {\cal
J}_2(\theta)\phantom{xxx}}{\partial P_{0}(\theta){ \cal
A}^{\imath}(\theta)} &\hspace{-0.3em} = &\hspace{-0.3em}
\textstyle\frac{1}{(l-1)!}\left(
P_1(\theta)g_{(\imath)_{l-1}\imath} \left( \partial_{\theta}{\cal
A}(\theta), \theta\right) \right)\left(P_0 \vec{{\cal A}}\right)^{
(\imath)_{l-1}}(\theta)P_0(\theta)  \nonumber \\
{} & \hspace{-0.3em}\equiv &\hspace{-0.3em}
 \textstyle\frac{1}{(l-1)!}g_{(\imath)_{l-1}\imath} \left(
 \partial_{\theta}{\cal A}({\theta}), \theta\right)
 \vec{{\cal A}}^{(\imath)_{l-1}}(\theta)\,,
\end{eqnarray}
with derivative of ${\cal J}_2(\theta)$ w.r.t. $P_1(\theta){\cal A}^{
\imath}(\theta)$ identical vanishing, that proves the formula (3.1).

With use of the connection for derivatives given by (3.1) it is easily to
obtain the component representation for superfield partial derivatives of
${\cal F}(\theta)\in C^k$ w.r.t. ${\cal A}^{\imath}(\theta)$
\begin{eqnarray}
\frac{\partial_{r(l)} {\cal F}(\theta)}{\partial {\cal A}^{\imath}(\theta)
\phantom{x}} =
\frac{\partial_{r(l)} P_0(\theta){\cal F}(\theta)}{\partial P_0(\theta){
\cal A}^{\imath}(\theta)\phantom{x}} +
\frac{\partial_{r(l)} P_1(\theta){\cal F}(\theta)}{\partial P_0(\theta){
\cal A}^{\imath}(\theta)\phantom{x}} +
\frac{\partial_{r(l)} P_1(\theta){\cal F}(\theta)}{\partial P_1(\theta){
\cal A}^{\imath}(\theta)\phantom{x}}\,. 
\end{eqnarray}

Consider on $\Lambda_{D\vert Nc+1}(z^{a},\theta;{\bf K})$ the special
involution acting as isomorphism
\renewcommand{\theequation}{\arabic{subsection}.\arabic{equation}}
\begin{eqnarray}
{} & {} & (z^a,\theta)^{\ast} = ({z^a}^{\ast},\theta^{\ast}) =
(z^a,-\theta) = (z^a,\bar{\theta})\,, \\
{} & {} &\bigl(g(z,\theta)\cdot f(z,\theta)\bigr)^{\ast} = \bigl(g(z,\theta)
\bigr)^{\ast} \bigl(f(z,\theta) \bigr)^{\ast} = g(z,\bar{\theta})\cdot
f(z,\bar{\theta}),\
\bigl(\bigl( g(z,\theta)\bigr)^{\ast}\bigr)^{\ast} =
g(z,\theta)\,,
\end{eqnarray}
being easily continued
onto $\tilde{\Lambda}_{D\vert
Nc+1}(z^{a}, \theta;{\bf K})$,  $C^{k}$  by the expressions
\renewcommand{\theequation}{\arabic{subsection}.\arabic{equation}}
\begin{eqnarray}
\bigl({\cal A}^{\imath}(\theta)\bigr)^{\ast} = {\cal
A}^{\imath}(\bar{\theta}) \equiv \overline{{\cal A}^{\imath}(\theta)},\
\left(\partial_{\theta}{\cal A}^{\imath}(\theta)\right)^{\ast} =
\partial_{\theta}{\cal A}^{\imath}(\theta)\,,
\end{eqnarray}
where $\overline{{\cal A}^{\imath}(\theta)}$ is the superfield being
conjugate to ${\cal A}^{\imath}(\theta)$ w.r.t. $\ast$
with components
\begin{eqnarray}
\overline{{\cal
A}^{\imath}(\theta)} =  P_0(\theta){\cal A}^{\imath}(\theta) -
P_1(\theta){\cal A}^{\imath}(\theta),\; P_1(\theta)\overline{{\cal
A}^{\imath}(\theta)} = - P_1(\theta) {\cal A}^{\imath}(\theta),\;
\overline{P_a(\theta)} = P_a(\theta)\,. 
\end{eqnarray}
The restriction of involution onto ${}^{0,0}C^{k}$
appears by identity mapping so that this superalgebra
is coordinatized by following invariant w.r.t. $*$ superfields
\begin{eqnarray}
P_0(\theta){\cal
A}^{\imath}(\theta) = \textstyle\frac{1}{2}\left({\cal A}^{\imath}(\theta) +
\overline{{\cal A}^{\imath}(\theta)}\right) ,\ \partial_{\theta}{\cal
A}^{\imath}(\theta)\,.
\end{eqnarray}
Consider the superspace {\boldmath ${\cal A}_{cl}$} of the 1st order
differential operators acting on $C^k$ of the form
\begin{eqnarray}
\mbox{\boldmath ${\cal A}_{cl}$} =
\bigl\{\alpha^aU_a(\theta)+\alpha^{\pm}U_{\pm}(\theta)+\beta^{a}{
\stackrel{\circ}{U}}_a(\theta) +
\beta^{\pm}{\stackrel{\circ}{U}}_{\pm}(\theta),\;a=0,1,\;
\alpha^a,\alpha^{\pm},\beta^a, \beta^{\pm}\in {\bf K} \bigr\}\,,
\end{eqnarray}
whose elements are given by means of the formulae
\begin{eqnarray}
U_a(\theta) & = &
P_1(\theta){\cal A}^{\imath}(\theta)
\displaystyle\frac{\partial_{l} \phantom{xxxxxxx}}{ \partial
P_a(\theta){\cal A}^{\imath}(\theta)},\
 U_{+}(\theta)  =  \sum\limits_a U_a(\theta)
= P_1(\theta){\cal A}^{\imath}(\theta)
\displaystyle\frac{\partial_{l} \phantom{xxx}}{\partial {\cal
A}^{\imath}(\theta)}\,, \\ 
U_{-}(\theta) & = & \bigl(U_0 - U_1\bigr)(\theta) =
P_1(\theta){\cal A}^{\imath}(\theta)
\displaystyle\frac{\partial_{l} \phantom{xxx}}{\partial \overline{{\cal
A}^{\imath}(\theta)}} = -\bigl(U_{+}(\theta)\bigr)^{\ast}\,,
\\ 
{\stackrel{\circ}{U}}_{j}(\theta) & = & \left[
\displaystyle\frac{d}{d\theta},U_j(\theta)\right]_{s} =
{\stackrel{\ \circ}{\cal A}}{}^{\imath}(\theta)
\displaystyle\frac{\partial_{l} \phantom{xxx}}{\partial K^{\imath}{}_j
(\theta)},\
{\stackrel{\circ}{U}}_{-}(\theta) =
\left({\stackrel{\circ}{U}}_{+}(\theta)\right)^{\ast}\,,  \\
{} & {} &\hspace{-2em}\frac{\partial_{l} \phantom{xxx}}{\partial
\overline{{\cal A}^{\imath}(\theta)}} =
\left(\frac{\partial_{l}
\phantom{xxx}}{\partial {\cal A}^{\imath}(\theta)}\right)^{\ast},\
K^{\imath}{}_j(\theta)=\{P_a(\theta){\cal
A}^{\imath}(\theta), {\cal A}^{\imath}(\theta), \overline{{\cal A}^{
\imath}(\theta)},\, j=a,+,-\}\,. \nonumber
\end{eqnarray}
The only ${\stackrel{\circ}{U}}_{+}(\theta)$ and
${\stackrel{\circ}{U}}_{-}( \theta)$ from all these operators  are
compatible with supergroup $J$ superfield representation. In
particular, ${\stackrel{\circ}{U}}_{+}(\theta)$ does not lead out
the any ${\cal F}(\theta) \in C^{k}$ from $C^{k}$. The only
operators $U_{1}(\theta)$, ${\stackrel{\circ}{U}}_{0} (\theta)$
appear to be invariant w.r.t. involution continued by means of
relations (3.8), (3.14) onto {\boldmath ${\cal A}_{cl}$}.

Elements from {\boldmath ${\cal A}_{cl}$} satisfy to the following algebraic
relationships in omitting of fixed $\theta$ in arguments and intensive use
of the formulae (3.2)
\renewcommand{\theequation}{\arabic{subsection}.\arabic{equation}\alph{lyter}}
\begin{eqnarray}
\setcounter{lyter}{1}
1) \hspace{2em} & {} & U_a^2 = \delta_{a1}U_a\ ,\ U_{\pm}^2 = \pm
U_{\pm},\
U_{+}U_{-} = U_{-}\;,\;U_{-}U_{+} = -U_{+}\,; \\ 
\setcounter{equation}{15}
\setcounter{lyter}{2}
2) \hspace{2em} &{} & [U_a , U_b]_{-} = \varepsilon_{a b}U_0,\
[U_{+},U_{-}]_{-} = 2U_0,\ \varepsilon_{a b} = -\varepsilon_{b
 a},\;\varepsilon_{10} = 1,\ a,b=0,1\,,   \nonumber \\
\phantom{2) \hspace{2em}} & {} & [U_{+},U_{-}]_{+} =
-2U_1,\ [U_{+},U_{a}]_{-} = (-1)^a U_0,\ [U_{-},U_{a}]_{-} = -U_0\,;
   \\ 
\setcounter{equation}{15}
\setcounter{lyter}{3}
3) \hspace{2em} &  {} & \bigl[{\stackrel{\circ}{U}}_{i},{\stackrel{\circ}{U
}}_{j}\bigr]_{+} =0,\
 \bigl[{\stackrel{\circ}{U}}_{j},
{\stackrel{\phantom{\circ}}{U}_{i}}\bigr]_{-} =
j{{\stackrel{\circ}{U}}_{i}},\  i,j \in
\{0,1,+,-\phantom{1}\hspace{-0.5em}\}\,. 
\end{eqnarray}
Relationships
(3.15a,b) demonstrate the subset
{\boldmath $\vec{{\cal A}}_{cl}$} in {\boldmath ${\cal
A}_{cl}$}
\renewcommand{\theequation}{\arabic{subsection}.\arabic{equation}}
\begin{eqnarray}
\mbox{\boldmath$\vec{{\cal A}}_{cl}$} =
\bigl\{\alpha^i U_i(\theta),\;i \in \{0,1,+,-\},\;\alpha^i \in {\bf K}
\bigr\} \subset \mbox{\boldmath${\cal A}_{cl}$} 
\end{eqnarray}
appears, in first,
by superalgebra w.r.t. usual multiplication, in
second, by module over $C^{k}$, in third, by
resolvable Lie subsuperalgebra w.r.t. commutator $[\; ,\; ]_{-}$
with radical spanned on $U_{0}(\theta)$.  The sets $\{U_{a}(\theta)\}$
and $\{U_{a}(\theta), {\stackrel{\circ}{U}}_a(\theta)\}$, $a=0, 1$
are the bases in
{\boldmath $\vec{{\cal A}}_{cl}$} and {\boldmath ${\cal A}_{cl}$}
respectively.

It should be noted that ${\stackrel{\circ}{U}}_{0}(\theta)$ coincides with
the operator $U$ from ref.[25] introduced there from the other grounds.
Besides, some analogs of operators from {\boldmath${\cal A}_{cl}$},
namely ${\stackrel{\circ}{U}}_{i}(\theta)$,  are intensively made use in
Sp(2)-covariant Lagrangian and (modified) triplectic quantization methods
[6,17-20] and
their superfield extension [26].
\subsection{Foundations of  $\theta$-STF}
\setcounter{equation}{0}

Let us consider a scalar $\Lambda_{1} (\theta;{\bf R})$-valued superfunction
$S_{L}\bigl({\cal A} (\theta),
\partial_{\theta}{\cal A}(\theta), \theta \bigr) \equiv
S_{L}(\theta)\in C^{k}$, in what follows called the classical action
$\bigl(\vec{\varepsilon}S_{L}(\theta) =
\vec{0}\bigr)$.

$S_{L}(\theta)$ is invariant under action of the restricted representation
$T_{\mid \bar{J}}$  operators
\begin{eqnarray}
S_{L}\left({\cal
A}'(\theta), \partial_{\theta}{\cal A}'(\theta), \theta \right) =
S_{L}\left(T(e,\tilde{g}){\cal A}(\theta), T(e,\tilde{g})\partial_{\theta}{
\cal A}(\theta), \theta \right) = S_{L}\bigl({\cal A}(\theta),
\partial_{\theta}{\cal A}(\theta), \theta \bigr)\,, 
\end{eqnarray}
whereas w.r.t. action of the $T_{\mid P}$ operators $S_{L}(\theta)$ is
transformed according to the rules (translation along $\theta$)
(2.18), (2.21)
\begin{eqnarray}
{} & {} & S_{L}\left({\cal A}'(\theta),
\partial_{\theta}{\cal A}'(\theta), {\theta}\right) =
S_{L}\bigl({\cal A}(\theta - \mu), \partial_{\theta}{\cal A}(
\theta), \theta -\mu\bigr) \equiv S_{L}(\theta')\,, \\ 
{} & {} & \delta S_L(\theta) = S_L({\theta}') -S_L(\theta) = - \mu
\partial_{\theta}S_L(\theta) = -\mu \left(\frac{\partial}{\partial \theta}
+ P_0(\theta){\stackrel{\circ}{U}}_+(\theta)\right)S_L(\theta)\,. 
\end{eqnarray}
Formula (4.1) provides the fact that $\bar{J}$ is the maximal in $J$
global symmetry group  for $S_{L}(\theta)$. It should be noted that one can
write equivalently in Eq.(4.3)  instead
of $P_0(\theta){\stackrel{\circ}{U}}_+(\theta)$
 the operator $P_{0}(\theta){\stackrel{\circ}{U}}_0(\theta)$, or
${\stackrel{\circ}{U}}_1(\theta)$, but the latter ones are given in
nonsuperfield form.

From the assumption on existence of critical point for the fermion
superfunctional, being together with $S_{L}(\theta)$ by the central, but
not equivalent, $\theta$-STF objects
\begin{eqnarray}
Z[{\cal A}]=\int
d\theta S_{L}\bigl({\cal A}(\theta), \partial_{\theta}{\cal
A}(\theta), \theta \bigr)\ ,\ \vec{\varepsilon} Z[{\cal A}] = (1, 0, 1)
\,,
\end{eqnarray}
it follows the validity of Euler-Lagrange type equations
(see Eqs.(2.34))
\begin{eqnarray}
\frac{\delta_l Z[{\cal A}]}{\delta{\cal
A}^{\imath}(\theta)} = \left(\frac{\partial_l \phantom{xxx}}{
\partial{\cal A}^{\imath}(\theta)}
 -(-1)^{\varepsilon_{\imath}}\partial_{\theta}\frac{\partial_l
\phantom{xxxx}}{\partial(\partial_{\theta}{\cal
A}^{\imath}(\theta))}\right)S_{L}(\theta) = 0\,.
\end{eqnarray}

Formally, the relations above from the  differential equations theory
viewpoint are the 2nd order w.r.t. derivatives of superfields ${\cal
A}^{\imath}(\theta)$ on $\theta$(!)
system of $n$ ordinary differential equations (ODE), in spite of the identity
fulfilment
\begin{eqnarray}
\partial_{\theta}^2{\cal A}^{\imath}(\theta) \equiv
{\stackrel{\;\circ\circ}{{\cal A}^{\imath}}}(\theta) \equiv 0\,.
\end{eqnarray}
Abstracting from the fact that system (4.5), in general,  is a
complicated system of partial differential equations defined by
differential operators w.r.t. $(z^{a},\theta)$ let us
single out from them the only operators with
$\partial_{\theta}$ considering others as the zero order on $\theta$
operators.

The analysis of Eqs.(4.5) is based on general statements about
the 1st and 2nd orders on $\theta$ system of $n$
ODE  and on assumptions concerning the $S_{L}(\theta)$ structure.
Therefore the results of the ODE investigation fulfilled in Appendix
permit one to
lead immediately a classified study
of the Euler-Lagrange equations simultaneously with additional conditions
specification for $S_{L}(\theta)$ or almost equivalently for
$Z[{\cal A}]$.

System (4.5) has the form of the Eqs.(A.10)  not given
in NF w.r.t. unknowns ${{\cal A}}^{\imath}(\theta)$, being coordinates in
superdomain $V \subset {\cal M}_{cl}$
\begin{eqnarray}
\hspace{-1.5em}(-1)^{\varepsilon_{\imath}} {\stackrel{\ \circ\circ}{\cal
A}}{}^{\jmath}(\theta) \frac{\partial^{2}_{l}
S_{L}(\theta)\phantom{xxxx}}{ {\partial{\stackrel{\ \circ}{\cal
A}}{}^{\imath}(\theta)}{\partial{\stackrel{\ \circ}{\cal A}}{}^{\jmath}(
\theta)}} = \frac{\partial_l S_{L}(\theta)}{\partial {\cal A}^{\imath}
(\theta)} - (-1)^{\varepsilon_{\imath}} \left(
\displaystyle\frac{\partial}{\partial\theta} \frac{\partial_l
S_{L}(\theta)}{{\partial{\stackrel{\ \circ}{\cal A}}{}^{\imath}(\theta)}} +
{\stackrel{\ \circ}{\cal A}}{}^{\jmath}(\theta)
\frac{\partial_l\phantom{xxx}}{ \partial {\cal A}^{\jmath}(\theta)}
\frac{\partial_l S_{L}(\theta)}{{\partial{\stackrel{\ \circ}{\cal
A}}{}^{\imath}(\theta)}}\right). 
\end{eqnarray}
System (4.7) is equivalent to one of the 2nd order on $\theta$ $2n$ ODE
\renewcommand{\theequation}{\arabic{subsection}.\arabic{equation}\alph{lyter}}
\begin{eqnarray}
\setcounter{lyter}{1}
{} & {} & \hspace{-1.5em}
{\stackrel{\ \circ\circ}{\cal A}}{}^{\jmath}(\theta) \displaystyle\frac{
\partial^{2}_{l} S_{L}(\theta) \phantom{xxxx}}
{{\partial(\partial_{\theta}{\cal
A}^{\imath}(\theta))}{\partial(\partial_{\theta}{\cal A}^{\jmath}
(\theta))}} = 0\,, \\ 
\setcounter{equation}{8}
\setcounter{lyter}{2}
{} & {} & \hspace{-1.5em}
{\Theta}_{\imath}\bigl( {{\cal A}}(\theta), \partial_{\theta}{
\cal A}(\theta), \theta \bigr) \equiv \displaystyle\frac{
\partial_l S_{L}(\theta) }{\partial{\cal A}^{\imath}(\theta)}
 -(-1)^{\varepsilon_{\imath}}\left(\frac{\partial_l}{\partial\theta}\frac{
 \partial_l
S_{L}(\theta)}{{\partial(\partial_{\theta}{\cal A}^{\imath}(\theta))}} +
{\stackrel{\circ}{U}}_{+}(\theta)\displaystyle\frac{
\partial_l S_{L}(\theta)}
{{\partial(\partial_{\theta}{\cal A}^{\imath}(\theta))}} \right) = 0.
\end{eqnarray}
\underline{\bf Definition:} Equations ${\Theta}_{\imath} (\theta) \equiv
{\Theta}_{\imath} \bigl( {{\cal
A}}(\theta), \partial_{\theta}{{\cal A}}(\theta), \theta \bigr)  = 0$ in
fulfilling of the conditions
\renewcommand{\theequation}{\arabic{subsection}.\arabic{equation}}
\begin{eqnarray}
{\rm deg}_{\partial_{\theta}{{\cal A}}(\theta)}
{\Theta}_{\imath}(\theta) \neq 0,\ \; {\rm deg}_{\partial_{\theta}{{\cal
A}}(\theta)} {\Theta}_{\imath}(\theta) = 0  
\end{eqnarray}
we will call the {\it differential constraints in Lagrangian formalism}
(\underline{DCLF}) and
{\it holonomic  constraints in Lagrangian formalism} (\underline{HCLF})
respectively for
 Eqs.(4.5).  The system (4.8)  on the whole  let us call the {\it
 Lagrangian system} (\underline{LS}) as well.

By virtue of remarks (A.6), (A.14)  the solvable DCLF  are equivalent to
the 1st order on $\theta$ $2n$ ODE system
\begin{eqnarray}
{\Theta}_{\imath}\bigl( {{\cal A}}(\theta), \partial_{\theta}{\cal
A}(\theta), \theta \bigr) = 0,\ \;
\partial_{\theta}\frac{\partial_l
S_{L}(\theta)}{\partial{\cal A}^{\imath}(\theta)} = 0\,.  
\end{eqnarray}
Thus, the solvable LS is equivalent
to the 2nd order on $\theta$ $3n$ ODE system
(4.8a), (4.10). Naturally, the 2nd subsystem in (4.10) would not be necessary,
if to consider the Eqs.(4.5) as the type (A.1) system.

DCLF restricts an admissible arbitrariness in the choice of $2n$ initial
conditions determining the Cauchy problem for LS for $\theta = 0$
\begin{eqnarray}
\left(\bar{{\cal A}}^{\imath}, \overline{\partial_{\theta}{\cal A}}{}^{
\imath}\right) \in T_{odd}V, \;
0 \in \Gamma_{(0,1)}
\subset {}^1\Lambda_1(\theta)\,. 
\end{eqnarray}

In its turn, subsystem (4.8a) are not found  in NF w.r.t.
$\partial_{\theta}^2{\cal A}^{\imath}(\theta)$. The possibility to pass to NF is
controlled
by rank value for the supermatrix
\begin{eqnarray}
K(\theta) = \left\|K_{\imath \jmath}\bigl( {{\cal A}}(\theta),
\partial_{\theta}{{\cal A}}(\theta), \theta) \right\| \equiv
\left\|\frac{\partial_{r}\phantom{xx}}{{\partial(\partial_{\theta}{\cal
A}^{\jmath}(\theta))}} \frac{\partial_l
 S_{L}(\theta)}{{\partial(\partial_{\theta}{\cal A}^{\imath}(\theta))}}
 \right\|\,. 
\end{eqnarray}
If ${\rm rank}K(\theta) < n$, then there
are some constraints, in general, independent from DCLF must be imposed on
subsystem (4.8a),  complicating the analysis of LS.

The problem of independence for  ${\Theta}_{\imath}(\theta)$, appearing by
the most important one, requires for its effective resolution in accordance
with (A.18), (A.19) to specify the initial assumptions on $Z[{\cal A}]$
$(S_{L}(\theta))$

\noindent
{\bf 1.} There exists $\left({{\cal A}}^{\imath}_{0}(\theta),
\partial_{\theta}{\cal A}^{\imath}_{0}(\theta)\right)
\in T_{odd}{\cal M}_{cl}$ that
\begin{eqnarray}
{\Theta}_{\imath} \bigl( {{\cal A}}(\theta),
\partial_{\theta}{{\cal A}}(\theta), \theta \bigr)_{\Bigl| \left({{\cal
A}}(\theta), \partial_{\theta}{{\cal A}}(\theta)\right) = \left( {{\cal
A}_0}(\theta), \partial_{\theta}{\cal A}_0(\theta)\right)} = 0\,;
\end{eqnarray}
{\bf 2.} There exists a smooth supersurface $\Sigma \subset {{\cal
M}_{cl}}$ at least in a some neighbourhood of
$\left({{\cal A}}^{\imath}_{0}(\theta),
\partial_{\theta}{\cal A}^{\imath}_{0}(\theta)\right)$
the such that,
\renewcommand{\theequation}{\arabic{subsection}.\arabic{equation}\alph{lyter}}
\begin{eqnarray}
\setcounter{lyter}{1}
{} & {} & \left( {{\cal
A}}^{\imath}_{0}(\theta), \partial_{\theta}{\cal A}^{\imath}_{0}(
\theta)\right) \in T_{odd}\Sigma,\ \
\Theta_{\imath}(\theta)_{\mid T_{odd}\Sigma} = 0\,,  \\ 
\setcounter{equation}{14}
\setcounter{lyter}{2}
{} & {} & \dim{\Sigma} = m = (m_+,m_-),\
\dim{T_{odd}\Sigma} = (m_+ + m_-, m_- + m_+)\,. 
\end{eqnarray}
Index $\imath$ can be divided into 2 groups
\begin{eqnarray}
\setcounter{lyter}{1}
\imath = (A , \alpha){},\ A = 1,\ldots,n-m,\ \alpha = n-m+1,\ldots,n\;
\end{eqnarray}
in such a way, that the following condition almost everywhere on $\Sigma$
holds
\begin{eqnarray}
\setcounter{equation}{15}
\setcounter{lyter}{2}
{\rm rank} \left\| \frac{\delta_l \phantom {xxxx}}{\delta{\cal
A}^{\jmath}(\theta_1)} \frac{\delta_l Z[{\cal A}]}{\delta{\cal
A}^{\imath}(\theta)}\right\|_{\textstyle\mid\Sigma} = {\rm rank} \left\|
\frac{\delta_l \phantom {xxxx}}{\delta{\cal A}^{\jmath}(\theta_1)}
\frac{\delta_l Z[{\cal A}]}{\delta{\cal
A}^{A}(\theta)}\right\|_{\textstyle\mid\Sigma} = n-m\,;  
\end{eqnarray}
{\bf 3.} There exists a separation of index $\imath$ to be consistent with one
from (4.15a)
\renewcommand{\theequation}{\arabic{subsection}.\arabic{equation}}
\begin{eqnarray}
\imath = (\underline{A} , \underline{\alpha}){},\ \underline{A} = 1,\ldots, n
- \stackrel{\;\circ}{m} {},\ \underline{\alpha} =
n-\stackrel{\;\circ}{m}+1,\ldots,n{},\ \stackrel{\;\circ}{m}=
({\stackrel{\;\circ}{m}}_{+},{\stackrel{\;\circ}{m}}_{-})\;, 
\end{eqnarray}
that the relation is valid in superdomain $V \supset \Sigma$
\begin{eqnarray}
{\rm rank}\left\|\frac{\partial_{r}\phantom{xxx}}{{\partial(\partial_{\theta}
{\cal A}^{\jmath}(\theta))}}
\frac{\partial_l S_{L}(\theta)}{{\partial(\partial_{\theta}{\cal A}^{
\imath}(\theta))}}\right\|_{\mid T_{odd} V} =
{\rm rank}\left\|\frac{\partial_{r}\phantom{xxx}}{{\partial(\partial_{\theta}{
\cal A}^{\jmath}(\theta))}}
\frac{\partial_l
S_{L}(\theta)}{{\partial(\partial_{\theta}{\cal A}^{\underline{A}}
(\theta))}}\right\|_{\mid T_{odd} V} = n - \stackrel{\circ}{m}
.
\end{eqnarray}
So, the conditions (4.13)--(4.15) mean that for the superfields
\begin{eqnarray}
\tilde{{\cal A}}^{\imath}(\theta) = {{\cal
A}}^{\imath}(\theta) - {{\cal A}}^{\imath}_{0}(\theta){}, \ \tilde{{\cal
A}}^{\imath}_{0}(\theta) = 0 \in \Sigma\,, 
\end{eqnarray}
the following representation as in (A.22) is valid providing, at least,
quadratic dependence upon $
\tilde{\cal A}{}^{\imath}(\theta),
\partial_{\theta}\tilde{\cal A}{}^{\imath}(\theta)$ for $S_{L}(\theta)$
\begin{eqnarray}
{} & {} & {\Theta}_{\imath} \bigl(
{\cal A}(\theta), \partial_{\theta}{\cal A}(\theta), \theta \bigr) =
{\Theta}_{\imath\ {\rm lin}} \bigl(
\tilde{\cal A}(\theta),
\partial_{\theta}\tilde{\cal A}(\theta),
\theta \bigr) + {\Theta}_{\imath\ {\rm nl}} \bigl( \tilde{\cal A}(\theta),
\partial_{\theta}\tilde{\cal A}(\theta),
\theta \bigr)\;,\nonumber \\
{} & {} & \Bigl(\min{\rm deg}_{\tilde{\cal
A}(\theta)\partial_{\theta}\tilde{\cal
A}(\theta)}, {\rm deg}_{\tilde{{\cal
A}}(\theta)\partial_{\theta}\tilde{\cal
A}(\theta)}\Bigr) {\Theta}_{\imath\ {\rm lin}}(\theta) = (1, 1),\ \min{\rm
deg}_{\tilde{{\cal A}}(\theta)\partial_{\theta}\tilde{\cal
A}(\theta)} {\Theta}_{\imath\ {\rm nl}}(\theta) \geq 2
\,.
\end{eqnarray}
Whereas the hypothesis 2 gives the possibility to represent
DCLF in the form of 2 subsystems being especially important for the
field (infinite-dimensional) case, when the requirements of locality
and covariance w.r.t. index
$\imath$ appears by obstacles to the condition (4.15) fulfilment.

With help of general relation (2.36) in accordance with (A.19), (A.20)
we mean under rank (4.15b) calculation the rule
\begin{eqnarray}
\hspace{-1.5em}{\rm rank} \left\| \frac{\delta_l
\phantom{xxxx}}{\delta{\cal A}^{\jmath}(\theta_1)} \frac{\delta_l Z[{\cal
A}]}{\delta{\cal A}^{\imath} (\theta)}\right\|_{\mid \Sigma} \hspace{-0.5em}
= {\rm rank}\left\|{\cal L}^l_{\jmath}(\theta_1)\Bigl[\bigl({\cal
L}^l_{\imath}S_{L}\bigr)(\theta_1)\right\|_{\mid T_{odd}
\Sigma}
\delta(\theta_1 - \theta) \Bigr](-1)^{\varepsilon_{\imath}} = n-m.
\end{eqnarray}
In particular, in view of the same definition (A.20) and its
consequence (A.21), the following formula holds for $F[{\cal A}]$ of the
form (2.33) subject to condition below
\begin{eqnarray}
{\rm rank}
\left\|\frac{ \delta_l \phantom{xxx}}{\delta{\cal
A}^{\imath}(\theta)} \frac{\delta_l F[{\cal A}]}{\delta{\cal
A}^{\jmath}(\theta_1)}\right\| = {\rm rank}\left\| \frac{\partial_l ({\cal
L}^l_{\jmath}{\cal F})(\theta)}{\partial {{\cal A}}^{\imath}(\theta)
\phantom{xxx}}\right\|\delta(\theta -
\theta_1)(-1)^{\varepsilon_{\jmath} + \varepsilon({\cal F})}
\hspace{-0.2em} ,\
{\rm deg}_{\partial_{\theta}{\cal A}}({\cal
L}^l_{\jmath}{\cal F})(\theta) = 0.  
\end{eqnarray}

In the framework of assumptions 1--3  the
following fundamental theorem about structure of
DCLF ${\Theta}_{\imath}(\theta)$ is valid

\noindent
\underline{\bf Theorem} (on reduction of the 1st order on $\theta$
$n$ ODE system to equivalent equations in GNF) \\
A nondegenerate parametrization for
superfields ${{\cal A}}^{\imath}(\theta)$ exists
\renewcommand{\theequation}{\arabic{subsection}.\arabic{equation}}
\begin{eqnarray}
{} & {} & \hspace{-4em}{{\cal A}}^{\imath}(\theta) = \bigl(
\delta^{\bar{\imath}}(\theta), \beta^{\underline{\imath}} (\theta),
\xi^{\alpha}(\theta)\bigr) \equiv \bigl(\varphi^{A}(\theta), \xi^{
\alpha}(\theta)\bigr),\ \imath = (\bar{\imath}, {\underline{\imath}}, \alpha)
\equiv (A, \alpha),\ \bar{\imath} = 1,\ldots, n - \underline{m}, \nonumber \\
{} & {} & \hspace{-4em} {\underline{\imath}} = n- \underline{m}+1, \ldots,
n-m,\;\underline{m}=(\underline{m}_+ , \underline{m}_-),\
 A=1,\ldots,n-m,\;\alpha = n-m+1,\ldots,n, 
\end{eqnarray}
so that   the 1st order on $\theta$ $n$ ODE system w.r.t. uknowns $
{\cal A}^{\imath}(\theta)$ (4.8b) is equivalent to the
following independent ODE in generalized normal form (GNF)
\begin{eqnarray}
\partial_{\theta}{\delta}^{\bar{\imath}}(\theta) =  \phi^{\bar{\imath}}
\bigl( \delta(\theta),\partial_{\theta}{\xi}(\theta), \xi(\theta),
\theta\bigr),\ \;
{\beta}^{\underline{\imath}}(\theta) = \kappa^{\underline{\imath}}
 \bigl( \delta(\theta), \xi(\theta), \theta\bigr) 
\end{eqnarray}
with  ${\phi}^{\bar{\imath}}(\theta), {\kappa}^{\underline{\imath}}(\theta)
\in C^{k}$ and with arbitrary superfields
${\xi}^{\alpha}(\theta): [{\xi}(\theta)] = m,\  0 < m < n$. Their
$\bigl({\xi} ^{\alpha}(\theta)\bigr)$ number coincides with one of the
differential identities among Eqs.(4.8b)
\begin{eqnarray}
{} & {} & \int
 d\theta \displaystyle\frac{\delta_r Z[{\cal A} ]}{\delta
{\cal A}^{\imath}(\theta)}{}{\hat{{\cal R}}}^{\imath}_{\alpha} \left(\theta;
{\theta}' \right) = 0\,, \nonumber \\
{} & {} & \vec{\varepsilon}{\hat{{\cal
R}}}^{\imath}_{\alpha}(\theta; {\theta}') = (1 + (\varepsilon_{P})_{\alpha}
+ (\varepsilon_{P})_{\imath}, (\varepsilon_{\bar{J}})_{\imath}+
(\varepsilon_{\bar{J}})_{\alpha} , \varepsilon_{\imath} +
\varepsilon_{\alpha} + 1)
\end{eqnarray}
with a) local and b) functionally independent
operators ${\hat{{\cal R}}}^{\imath}_{\alpha} \left( {\cal A}(\theta),
\partial_{\theta}{\cal A}(\theta), \theta; {\theta}' \right) \equiv
{\hat{{\cal R}}}^{\imath}_{\alpha}(\theta; {\theta}')$:
\renewcommand{\theequation}{\arabic{subsection}.\arabic{equation}\alph{lyter}}
\begin{eqnarray}
\setcounter{lyter}{1}
{\rm a)}  & {} & {\hat{{\cal R}}}^{\imath}_{
\alpha}(\theta; {\theta}') =
\sum\limits_{k=0}^{1} \left(\left(\partial_{\theta} \right)^k
\delta(\theta - \theta')\right) {\hat{ {\cal R}}}_{k}{}^{\imath}_{\alpha}
\bigl( {\cal A}(\theta), \partial_{\theta}{\cal A}(\theta), \theta\bigr),
\\ 
\setcounter{equation}{25}
\setcounter{lyter}{2}
\phantom{\rm a)} & {} &\vec{\varepsilon}{\hat{{\cal R}}}_{k}{}^{
\imath}_{ \alpha}(\theta) =
(\delta_{1k} + (\varepsilon_{P})_{\imath} +
(\varepsilon_{P})_{\alpha}, (\varepsilon_{\bar{J}})_{\imath} +
(\varepsilon_{\bar{J}})_{\alpha} ,
\varepsilon_{\imath}+ \varepsilon_{\alpha} + \delta_{1k})\,,
\end{eqnarray}
b) functional equation
\renewcommand{\theequation}{\arabic{subsection}.\arabic{equation}}
\begin{eqnarray}
\int d\theta' {\hat{{\cal R}}}^{\imath}_{\alpha}(\theta;
{\theta}') u^{\alpha} \bigl( {\cal A}(\theta'),
\partial_{\theta'}{\cal A}(\theta'), \theta'
\bigr) = 0 
\end{eqnarray}
has the unique vanishing solution.

The theorem above appearing by the special case for Theorem from Appendix A
and therefore the important consequences follow from it.

\noindent
\underline{\bf Corollary 1} \\
In fulfilling of the condition (4.21), written for $Z[{\cal A}]$,
 indicating on HCLF  dependence
\begin{eqnarray}
{\rm
rank}\left\|\frac{\partial_l {\Theta}_{\imath}\bigl({\cal A}(\theta), \theta
\bigr)}{ \partial {\cal
A}^{\jmath}(\theta)\phantom{xxxxx}}\right\|_{\textstyle \mid \Sigma} = n-m <
 [{\Theta}_{\imath}(\theta)]\,, 
\end{eqnarray}
a nondegenerate
parametrization for superfields ${{\cal A}}^{\imath}(\theta)$ exists
\begin{eqnarray}
{\cal A}^{\imath}(\theta) = \bigl(\varphi^{A}(\theta),
\xi^{\alpha}(\theta)\bigr),\;A=1,\ldots,n-m,\; \alpha = n - m + 1,\ldots,n\,,
\end{eqnarray}
so that HCLF are equivalent to the system of algebraically
independent constraints, in the sense of differentiation w.r.t.  $\theta$
\begin{eqnarray}
\tilde{\Theta}_{A}\bigl(\varphi(\theta), \xi(\theta),
\theta\bigr) = 0\,. 
\end{eqnarray}
The number of superfields
$[{\xi}(\theta)]$ coincides with one of algebraic (on $\theta$)
identities among ${\Theta}_{\imath}(\theta)$
\begin{eqnarray}
{\Theta}_{\imath}\bigl({\cal A}(\theta), \theta \bigr) {{\cal
R}}_{0}{}^{\imath}_{\alpha}\bigl( {\cal A}(\theta), \theta\bigr) = 0\,,
\end{eqnarray}
where operators ${{\cal
R}}_{0}{}^{\imath}_{\alpha} \bigl( {\cal A}(\theta), \theta \bigr) $ can be
chosen in the form being consistent with (4.25)
\begin{eqnarray}
{\hat{{\cal R}}}^{\imath}_{\alpha} \bigl( {\cal A}(\theta),
\theta; \theta'\bigr) = - \delta(\theta - \theta') {{\cal
R}}_{0}{}^{\imath}_{\alpha}\bigl( {\cal
A}(\theta), \theta\bigr)\,. 
\end{eqnarray}
Their linear independence means that equation
\begin{eqnarray}
{{\cal R}}_{0}{}^{\imath}_{\alpha}\bigl( {\cal A}(\theta), \theta\bigr)
u^{\alpha}\bigl( {\cal A}(\theta), \theta\bigr) = 0   
\end{eqnarray}
has the unique trivial solution.

One from the realization for Corollary 1 is the

\noindent
\underline{\bf Corollary 2} \\
If the model of $\theta$-STF is represented by almost natural system defined
in the form
\begin{eqnarray}
{} & {} & S_L \bigl( {\cal A}(\theta), \partial_{\theta}{\cal A}(\theta),
\theta\bigr) = T \bigl( {\cal A}(\theta),
\partial_{\theta}{\cal A}(\theta)\bigr) - S\bigl( {\cal A}(\theta), \theta
\bigr) \equiv T(\theta) - S(\theta)\,, \\  
{} & {} & T(\theta) = T_1\bigl(\partial_{\theta}{\cal A}(\theta)\bigr) +
\partial_{\theta}{\cal A}^{\jmath}(\theta)g_{\jmath
\imath}(\theta){\cal A}^{\imath} (\theta), \
 g_{\jmath
\imath}(\theta) = P_0(\theta)g_{\jmath\imath}(\theta)
= (-1)^{\varepsilon_{\jmath}\varepsilon_{\imath}}g_{\imath\jmath}(\theta)
\,, 
\end{eqnarray}
where $(\min{\rm deg}_{{\cal A}(\theta)}S(\theta) =
\min{\rm deg}_{\partial_{\theta}{\cal A}(\theta)}T_1(\theta) = 2)$, then DCLF
\begin{eqnarray}
{\Theta}_{\imath} \bigl( {{\cal A}}(\theta), \partial_{\theta}{{\cal A}}(
\theta), \theta \bigr) \equiv \tilde{\Theta}_{\imath}
\bigl( {{\cal A}}(\theta), \theta \bigr) = - S,_{\imath} \bigl( {{\cal
A}}(\theta), \theta \bigr)(-1)^{ \varepsilon_{\imath}} = 0   
\end{eqnarray}
appear by the HCLF explicitly depending upon $\theta$.
Condition (4.27) in question has the form
\begin{eqnarray}
{\rm rank}\left\|S,_{\imath
\jmath} \bigl( {{\cal A}}(\theta),\theta)\right\|_{ \textstyle \mid \Sigma}
= n-m < n\,,
\end{eqnarray}
and expressions (4.30)--(4.32) remain valid
in this case.

If for DCLF (4.8b) $\bigl({\rm HCLF}\ {\Theta}_{\imath}\bigl({\cal
A}(\theta), \theta \bigr)\bigr)$ the following conditions are fulfilled
almost everywhere in any neighbourhood $U:$ ${\cal M}_{cl} \supset U
\supset \Sigma$ respectively
\begin{eqnarray}
{\rm rank} \left\| \frac{\delta_l
\phantom{xxx}}{\delta{\cal A}^{\imath}(\theta)}\frac{\delta_l Z[{\cal
A}]}{\delta{\cal A}^{\jmath}(\theta_1)} \right\|_{\textstyle \mid U} = n,\
\ \left({\rm rank}\left\|\Theta_{\imath},_{\jmath} \bigl( {\cal
A}(\theta),\theta\bigr)\right\|_{\textstyle\mid U} = n\right), 
\end{eqnarray}
then ${\Theta}_{\imath}(\theta)$ appear by functionally (linearly)
independent and have been already found in GNF.

The performed investigation of LS makes to be justified an introduction of
the following terminology:

\noindent
{\bf 1)} The model of $\theta$-superfield theory of fields (mechanics)
given by superfunction $S_{L}(\theta) \in C^{k}$ (or, almost
equivalently, by superfunctional $Z[{\cal A}] \in C_{F}$) satisfying to
the postulates 1--3 ((4.13)--(4.17)) for $m >
0$ is called the {\it gauge
theory of general type} (\underline{GThGT}) for superfields ${\cal A}^{
\imath}(\theta)$,
and in fulfilling of the 1st condition in (4.37) the
{\it nondegenerate theory of general type} (\underline{ThGT});

\noindent
{\bf 2)} If, in addition, the  Corollary 2 conditions on HCLF (4.27) for
$m>0$ are fulfilled, then the  model of $\theta$-superfield theory of fields
(mechanics) is called
the {\it gauge theory of special type} (\underline{GThST}), and in realizing
of the 2nd condition in (4.37) the {\it nondegenerate
theory of special type} (\underline{ThST});

\noindent
{\bf 3)} Formulation of GThGT and GThST defined by means of $S_{L}(\theta)
\in C^{k}$ ($Z[{\cal A}] \in C_{F}$) let us call the Lagrangian formalism of
description for GThGT and GThST, or equivalently the {\it Lagrangian
formalism  (formulation)  of $\theta$-STF}.

Identities (4.24) for GThGT ((4.30) for GThST) with operators ${\hat{{\cal
R}}}^{\imath}_ {\alpha}(\theta; {\theta}')$ $\bigl({{\cal
R}}_{0}{}^{\imath}_{\alpha} \bigl( {\cal A}(\theta), \theta \bigr) \bigr)$,
whose set is complete and functionally (linearly) independent, i.e. is the
basis in linear space $Q(Z)= {\rm Ker} \left\{
\displaystyle\frac{{\delta}_{l}Z[{\cal A}]} {\delta {{\cal
A}}^{\imath}(\theta)}\right\}$, $\Bigl( Q(S_{L}) = {\rm Ker}\bigl\{
{\Theta}_{\imath}(\theta) \bigr\} \Bigr)$, make to be possible the following
interpretation for quantities $\hat{{\cal R}}^{\imath}_{\alpha}(\theta;
{\theta}'), \bigl({{\cal R}}_{0}{}^{\imath}_{\alpha} (\theta) \bigr)$:

\noindent
{\bf 1)} Any quantities ${\hat{{\cal R}}}^{\imath}\bigl({\cal A}(\theta),
\partial_{\theta}{\cal A}(\theta), \theta \bigr) \equiv {\hat{{\cal
R}}}^{\imath}(\theta), {\cal R}_{0}^{\imath} \bigl( {\cal A}(\theta),
\theta \bigr) \equiv {\cal R}_{0}^{\imath} (\theta)$ satisfying to
identities
\begin{eqnarray}
\int d\theta{}\frac{\delta_r Z[{\cal A}]}{\delta
{\cal A}^{\imath}(\theta)}{}{\hat{{\cal R}}}^{\imath}(\theta) =  0,\
{\Theta}_{\imath}\bigl({\cal A}(\theta),
\theta \bigr) {\cal R}_{0}^{\imath}(\theta) = 0,\
\vec{\varepsilon}{\hat{{\cal R}}}^{\imath}(\theta) =
\vec{\varepsilon}{{\cal R}}^{\imath}_0(\theta)
= ((\varepsilon_{P})_{\imath},{}
(\varepsilon_{\bar{J}})_{\imath},{} \varepsilon_{\imath}) 
\end{eqnarray}
are called the {\it generator of general type gauge
transformations} (\underline{GGTGT}) and {\it generator of special type
gauge transformations} (\underline{GGTST}) respectively;

\noindent
{\bf 2)} The quantities
${\hat{\tau}}^{\imath}\bigl({\cal A}(\theta),
\partial_{\theta}{\cal A}(\theta), \theta \bigr) \equiv
{\hat{\tau}}^{\imath}(\theta)$, ${\tau}_{0}^{\imath} \bigl( {\cal
A}(\theta), \theta \bigr) \equiv {\tau}_{ 0}^{\imath}(\theta)$
being by particular cases for
${\hat{{\cal R}}}^{\imath}(\theta)$, ${{\cal R}}^{\imath}_0(\theta)$
\begin{eqnarray}
{\hat{\tau}}^{\imath}(\theta)  =  \int
d\theta'{}\frac{\delta_r Z[ {\cal A}]}{\delta{\cal A}^{\jmath}(\theta')}{}
\hat{E}^{\imath \jmath}\bigl({\cal A}(\theta), \partial_{\theta}{ \cal
A}(\theta), \theta; {\theta}' \bigr),\
{\tau}_{0}^{\imath}(\theta) =
{\Theta}_{\jmath} \bigl({\cal A} (\theta), \theta \bigr) E_{0}^{\imath
 \jmath}\bigl({\cal A}(\theta), \theta\bigr)\,,
\end{eqnarray}
are called the {\it trivial GGTGT},
{\it GGTST} respectively,  where the superfunctions
$E_{0}^{\imath \jmath}(\theta) \in
C^{k}\bigl( {{\cal M}_{cl}}$ $\times$ $\{\theta\}\bigr)$ and $\hat{E}^{\imath
\jmath}(\theta,{\theta}') \in C^{k}\bigl( T_{odd}{{ \cal M}_{cl}}$ $\times$
$\{\theta, {\theta}'\}\bigr)$ obey to the properties
\renewcommand{\theequation}{\arabic{subsection}.\arabic{equation}\alph{lyter}}
\begin{eqnarray}
\setcounter{lyter}{1}
\hat{E}^{\imath \jmath}(\theta,
{\theta}') & = & -(-1)^{(\varepsilon_{\jmath} + 1) (\varepsilon_{\imath} +
1)}\hat{E}^{\jmath \imath}({\theta}', \theta),\ E_{0}^{\imath \jmath}(\theta)
= -(-1)^{\varepsilon_{\jmath}\varepsilon_{ \imath}} E_{0}^{\jmath
\imath}(\theta)\,, \\ 
\setcounter{equation}{40}
\setcounter{lyter}{2}
\vec{\varepsilon}(E_{0}^{\imath \jmath}(\theta)) & = &
\vec{\varepsilon}(\hat{E}^{\imath \jmath}(\theta,\theta')) + (1,0,1) =
((\varepsilon_{P})_{\imath}+  (\varepsilon_{P})_{\jmath},
(\varepsilon_{\bar{J}})_{\imath} +
(\varepsilon_{\bar{J}})_{\jmath},
\varepsilon_{\imath} + \varepsilon_{\jmath})\,.
\end{eqnarray}
Trivial GGTST ${\tau}_{0}^{\imath}(\theta)$ defined by means of
$E_{0}^{\imath \jmath}(\theta)$ can be always represented in the form of
trivial GGTGT
${\hat{\tau}}^{\imath}\bigl({\cal A}(\theta),\theta \bigr)$ with
corresponding $\hat{E}^{\imath \jmath}\bigl({\cal A}(\theta),\theta;
{\theta}' \bigr) \in C^{k}\bigl( {{\cal M}_{cl}} \times \{\theta,
{\theta}'\}\bigr)$\footnote{to GThST quantities  of the form
$y_{0}{}^{(\imath)_{n}}_{\alpha}\bigl({\cal A}(\theta),\theta \bigr)$ one
corresponds the GThGT ones ${\hat{y}}^{(\imath)_{n}}_{\alpha}\bigl({\cal A}
(\theta),\theta ; \vec{\theta}_{n}\bigr) \approx \prod_{k=1}^{n} \delta
({\theta}_{k} - \theta)y_{0}{}^{ (\imath)_{n}}_{\alpha}\bigl( {\cal
A}(\theta),\theta \bigr)$ in the so-called ultralocal representation on
$\theta$ with accuracy up to special sign factor $(-1)^{R},{} R \in \mbox{
\boldmath$N$}$ in the last expression}
\renewcommand{\theequation}{\arabic{subsection}.\arabic{equation}}
\begin{eqnarray}
{\hat{E}}^{\imath\jmath} \bigl( {\cal A}(\theta), \theta;
\theta'\bigr)  = - \delta(\theta - \theta') {E}_{0}^{\imath \jmath}\bigl(
{\cal A}(\theta), \theta\bigr),\
{\hat{\tau}}^{\imath}\bigl({\cal A}(\theta),\theta \bigr) =
{\tau}_{0}^{\imath}\bigl({\cal A} (\theta),\theta \bigr)\,. 
\end{eqnarray}
Taking account of completeness for quantities ${\hat{{\cal
R}}}^{\imath}_{\alpha}(\theta; {\theta}')$ and ${{\cal
R}}_{0}{}^{\imath}_{\alpha}(\theta)$ and definitions (4.38),(4.39) the any
GGTGT, GGTST one can represent by the corresponding formulae
\begin{eqnarray}
{} & {} & \hspace{-1em}{\hat{{\cal R}}}^{\imath}\left({\cal A}(\theta),
\partial_{\theta}{\cal A}
(\theta), \theta \right)  =  \int d\theta' {\hat{{\cal
R}}}^{\imath}_{\alpha} \left( {\cal A}(\theta),
\partial_{\theta}{\cal A}(\theta), \theta; {\theta}' \right)
\hat{\xi}^{ \alpha}\bigl( {\cal A}(\theta'),
\partial_{\theta'}{\cal A}(\theta'), {\theta}' \bigr) +
{\hat{\tau}}^{\imath}(\theta),  \\ 
{} & {} & \hspace{-1em}
{{\cal R}}_{0}^{\imath}\bigl( {\cal A}(\theta), \theta\bigr) =
{{\cal R}}_{0}{}^{\imath}_{\alpha} \bigl({\cal A}(\theta), \theta\bigr)
\xi_{0}^{\alpha} \bigl( {\cal A}(\theta), \theta\bigr) +
{\tau}_{0}^{\imath} \bigl( {\cal A}(\theta), \theta \bigr) \,,\\
{} & {} & \hspace{-1em}(\hat{\xi}^{\alpha},\xi_{0}^{\alpha})(\theta) \in
(C^{k}, C^{k}\bigl({{\cal M}_{cl}} \times \{\theta\}),\
\vec{\varepsilon}{\xi}_{0}^{\alpha}(\theta) =
\vec{\varepsilon}\hat{\xi}^{\alpha}(\theta) =
((\varepsilon_P)_{\alpha},(\varepsilon_{\bar{J}})_{\alpha},
\varepsilon_{\alpha})\,, \nonumber
\end{eqnarray}
which turn the linear spaces $Q(Z)$, $Q(S_{L})$ into affine $C^{k}\bigl(
T_{odd}{{\cal M}_{cl}} \times \{\theta \}\bigr)$-,
$C^{k} \bigl({{\cal M}_{cl}} \times \{\theta\}\bigr)$-modules
respectively.

At last, GGTGT and GGTST are defined (as the basis elements of $Q$) up to
affine transformations of modules $Q(Z)$ and $Q(S_{L})$ respectively
(so-called equivalence transformations)
\renewcommand{\theequation}{\arabic{subsection}.\arabic{equation}\alph{lyter}}
\begin{eqnarray}
\setcounter{lyter}{1}
{} & {} & \hspace{-2.5em}\hat{{\cal R}}'^{\imath}_{\alpha}
\left(\theta;{\theta}'\right) \hspace{-0.15em} = \hspace{-0.15em}
\int \hspace{-0.25em}d\theta_1 \hspace{-0.15em}
\Bigl[{\hat{\cal R}}^{\imath}_{\beta}(\theta;
{\theta}_1) \hat{\xi}^{\beta}_{\alpha}\bigl({\cal A}(\theta_1),
{\stackrel{\ \circ}{\cal A}}(\theta_1), {\theta}_1 ;\theta'\bigr)
\hspace{-0.2em}
+\hspace{-0.2em}
\frac{\delta_r Z[{\cal A}]}{\delta {\cal
A}^{\jmath}(\theta_1)} \hat{E}^{\imath \jmath}_{\alpha}\bigl(
{\cal A}(\theta), {\stackrel{\ \circ}{\cal A}}(\theta), \theta, \theta_1;
{\theta}'\bigr)\hspace{-0.1em}\Bigr], \\
\setcounter{equation}{44}
\setcounter{lyter}{2}
{} & {} & \hspace{-2.5em}{{\cal R}'}_{0}{}^{\imath}_{\alpha}\bigl( {\cal
A}(\theta),\theta\bigr) = {{\cal R}}_{0}{}^{\imath}_{\beta} \bigl( {\cal
A}(\theta), \theta\bigr) \xi_{0}{}^{\beta}_{\alpha} \bigl( {\cal A}(\theta),
\theta\bigr) +
{\Theta}_{\jmath}\bigl( {\cal A} (\theta), \theta
\bigr) E_{0}{}^{\imath \jmath}_{{}\alpha}\bigl({\cal A}(\theta),
\theta\bigr)\,, 
\end{eqnarray}
\vspace{-4ex}
\renewcommand{\theequation}{\arabic{subsection}.\arabic{equation}}
\begin{eqnarray}
{} & {} & \vec{\varepsilon}
({\hat{E}}^{\imath \jmath}_{\alpha}(\theta, \theta_1;\theta'),
{E}_{0}{}^{\imath\jmath}_{{}\alpha}(\theta)) =(
(\varepsilon_{P})_{
\imath}+ (\varepsilon_{P})_{\jmath} + (\varepsilon_P)_{\alpha},
(\varepsilon_{\bar{J}})_{\imath} + (\varepsilon_{\bar{J}})_{\jmath} +
(\varepsilon_{\bar{J}})_{\alpha}, \varepsilon_{\imath} +
\varepsilon_{\jmath} + \varepsilon_{\alpha})\,,\nonumber\\
{} & {} &
\vec{\varepsilon}{\xi}_{0}{}^{\beta}_{\alpha}(\theta) =
\vec{\varepsilon}\hat{\xi}^{\beta}_{\alpha}(\theta;\theta') + (1,0,1) =
((\varepsilon_P)_{\alpha}+ (\varepsilon_P)_{\beta},
(\varepsilon_{\bar{J}})_{\alpha} +
(\varepsilon_{\bar{J}})_{\beta},
\varepsilon_{\alpha} + \varepsilon_{\beta}), 
\end{eqnarray}
where superfunctions ${\hat{\xi}}^{\beta}_{\alpha}(\theta;
{\theta}')$ belong to $C^{k}\bigl( T_{odd}{{\cal M}_{cl}} \times \{\theta,
{\theta}'\}\bigr)$, ${\hat{E}}^{\imath \jmath}_{\alpha} ({\vec{\theta}}_{2};
{\theta}') \in C^{k}\bigl(T_{odd}{{\cal M}_{cl}} \times \{{\vec{\theta}}_{2},
{\theta}'\}\bigr)$, ${\xi}_{0}{}^{\beta}_{\alpha}(\theta)$, $E_{0}{}^{\imath
\jmath}_{{}\alpha}(\theta) \in C^{k}\bigl( {{\cal M}_{cl}} \times
\{\theta\}\bigr)$ and possess by the properties
\begin{eqnarray}
{} & {} & {E}_{0}{}^{\imath\jmath}_{{}\alpha}({\cal A}(\theta),\theta) =
-(-1)^{\varepsilon_{\jmath}\varepsilon_{\imath}} {E}_{0}{}^{\jmath\imath}_{
{}\alpha}({\cal A}(\theta), \theta),\; {\rm
 rank}\left\|\hat{\xi}^{\beta}_{\alpha}(\theta;\theta')\right\|  = {\rm
rank}\left\|{\xi}_{0}{}^{\beta}_{\alpha}(\theta)\right\| = [\alpha] = m\;,
\nonumber \\
{} & {} & {\hat{E}}^{\imath \jmath}_{\alpha}\left({\cal
A}(\theta), \partial_{\theta}{\cal A}(\theta), \theta, \theta_1;
{\theta}'\right) = -(-1)^{(\varepsilon_{\jmath} + 1)(\varepsilon_{ \imath} +
1)} {\hat{E}}^{\jmath \imath}_{\alpha}\left({\cal A}(\theta),
\partial_{\theta}{\cal A}(\theta), \theta_1, \theta; {\theta}'\right). 
\end{eqnarray}
The general type quantities  ${\hat{E}}^{\imath
\jmath}_{\alpha}\bigl( {\cal A}(\theta_1), {\vec{\theta}}_{2}; {\theta}'\bigr),
{\hat{\xi}}^{\beta}_{\alpha}\bigl( {\cal A}(\theta),\theta; \theta'\bigr)$
correspond to the special type ones $E_{0}{}^{\imath
\jmath}_{{}\alpha}\bigl( {\cal A}(\theta),\theta\bigr),
{\xi}_{0}{}^{\beta}_{\alpha}\bigl( {\cal A}(\theta),\theta \bigr)$ (with
accuracy up to sign factor $(-1)^{K}, K \in \mbox{\boldmath$N$}$)
\begin{eqnarray}
{\hat{\xi}}^{\beta}_{\alpha}\bigl(\theta; {\theta}'\bigr) =
\delta(\theta - \theta')
{\xi}_{0}{}^{\beta}_{\alpha}\bigl(\theta \bigr),\;
{\hat{E}}^{\imath
\jmath}_{\alpha}\bigl( {\cal A}(\theta_1),{\vec{\theta}}_{2}; {\theta}'\bigr)
&\hspace{-0.5em} = & \hspace{-0.5em}- \delta(\theta_1 -
\theta')\delta(\theta_2 - \theta')E_{0}{}^{\imath
\jmath}_{{}\alpha}\bigl(\theta_1\bigr)(-1)^{\varepsilon_{\jmath}}\,, 
\end{eqnarray}
providing together with relations (4.5), (4.31), (4.41)
the conversion of $Q(S_{L})$ into
$C^{k}\bigl( {{\cal M}_{cl}} \times \{\theta,
{\theta}'\}\bigr)$-submodule of  $C^{k}\bigl( T_{odd}{{\cal M}_{cl}}
\times \{\theta, {\theta}'\}\bigr)$-module $Q(Z)$. In its turn
the locality w.r.t. $\theta$ for transformed
$\hat{{\cal R}}'^{\imath}_{\alpha}(\theta; {\theta}')$ (4.44) will be
guaranteed by locality of
${\hat{\xi}}^{\beta}_{\alpha}(\theta; {\theta}')$,
${\hat{E}}^{\imath \jmath}_{\alpha} ({\vec{\theta}}_{2}; {\theta}')$.

Finally, relations (4.24) are easily interpreted for GThGT as $Z[{\cal A}]$
invariance   w.r.t. infinitesimal transformations of superfields ${\cal
A}^{\imath}(\theta)$
\renewcommand{\theequation}{\arabic{subsection}.\arabic{equation}}
\begin{eqnarray}
{\cal A}^{\imath}(\theta) \mapsto {{\cal A}'}^{\imath}(\theta)  =  {\cal
A}^{\imath}(\theta) + \delta_{g}{\cal A}^{\imath}(\theta),\;\delta_{g}{\cal
A}^{\imath}(\theta) = \int d\theta' {\hat{{\cal R}}}^{\imath}_{\alpha}
(\theta; {\theta}'){\xi}^{\alpha}(\theta')\,, 
\end{eqnarray}
with arbitrary superfields ${\xi}^{\alpha}({\theta}')$ $\in$
$\tilde{\Lambda}_{D\vert Nc+1}(z^{a},\theta; {\bf K})$
whose parities defined as in (4.43).
Really, the formula holds
\begin{eqnarray}
Z[{\cal A}'] & = & Z[{\cal A}] + \int d\theta \frac{\delta_r Z[{\cal
A}]}{\delta{\cal A}^{\imath}(\theta)} \int d\theta'
{\hat{\cal R}}^{\imath}_{\alpha}(\theta; {\theta}'){\xi}^{\alpha}(\theta')
+ F[{\cal A};\xi]  \nonumber \\
{} & = & Z[{\cal A}] + F[{\cal A};\xi] ,\; \min{\rm
deg}_{\xi(\theta)}F = 2\,. 
\end{eqnarray}
The relation (4.30) can not, in
general, be interpreted for GThST as the invariance of $S_{L}\bigl({\cal A}
(\theta),$ $\partial_{\theta}{\cal A}(\theta), \theta \bigr)$
w.r.t. the transformations with arbitrary ${\xi}_{0}^{\alpha}(\theta)$
\begin{eqnarray}
{\cal A}^{\imath}(\theta)
\mapsto {{\cal A}'}^{\imath}(\theta) = {\cal A}^{\imath}(\theta) +
\delta{\cal A}^{\imath}(\theta),\;\delta{\cal A}^{\imath}(\theta) =
{\cal R}_{0}{}^{\imath}_{\alpha} \bigl({\cal A}(\theta), \theta
\bigr){\xi}_{0}^{\alpha}(\theta)\,.
\end{eqnarray}
However, for
superfunction $S\bigl({\cal A}(\theta), \theta \bigr)$ defined as in
Corollary 2  the real invariance takes place
\begin{eqnarray}
S\bigl({\cal
A}'(\theta), \theta \bigr) & = & S\bigl({\cal A}(\theta), \theta \bigr) +
S,_{\imath}\bigl({\cal A}(\theta), \theta \bigr)
{\cal R}_{0}{}^{\imath}_{\alpha} \bigl({\cal A}(\theta), \theta\bigr)
{\xi}_{0}^{\alpha}(\theta) + {\cal F}\bigl({\cal A}(\theta),
 \xi_{0}(\theta), \theta \bigr)  \nonumber\\
{} & = & S\bigl({\cal A}(\theta),
\theta \bigr) + {\cal F}\bigl({\cal A}(\theta), \xi_{0}(\theta), \theta
\bigr)
 ,\; \min{\rm deg}_{\xi_0(\theta)}{\cal F} = 2 \,. 
\end{eqnarray}
Therefore in view of definitions above it is natural to call the
transformations (4.48), (4.50) as
the {\it general type gauge transformations}  (\underline{GTGT}) for
$Z[{\cal A}]$ and
the {\it special type gauge transformations}  (\underline{GTST}) for
$S\bigl({\cal A}(\theta), \theta \bigr)$ respectively.
The identities (4.24), (4.30) interpretation  in
terms of GTGT, GTST permit to call them by Noether's identities for
GThGT,  GThST respectively.

Further to be more simplifiedly one can regard that GThGT (GThST) is defined
by relations (4.4), (4.15), (4.24) ((4.27), (4.30), (4.17)) and for the
existence of  only trivial solution  for Eqs.(4.26) (Eqs.(4.32)) is called
the {\it irreducible GThGT (GThST)}, otherwise the
{\it reducible GThGT (GThST)}.
\subsection{Generating Equations. Zero Locus
Reduction Problems}
\setcounter{equation}{0}

Quantities $Z[{\cal A}]$, $S({\cal A}(\theta),\theta)$, GGTGT $
{\hat{\cal R}}^{\imath}_{\alpha}(\theta; {\theta}')$, GGTST
${\cal R}_{0}{}^{\imath}_{\alpha}(\theta)$ together with identities (4.24),
(4.30) as the first structural relations of the corresponding general and
special types {\it gauge algebras} (\underline{GA}) being by
differential-algebraic systems on $Q(Z)$, $Q(S_L)$ are effectively described
by means of 2 special generating (master) equations for not uniquely defined
superfunctional $Z_{(1)}[\Gamma_{min}]\equiv Z_{(1)}$ and superfunction
$S_{(1)}(\Gamma_{min}(\theta),\theta)$ $\equiv$ $S_{(1)}(\theta)$.
The latters
appear by corresponding deformations in powers of ghost superfields
$(C^{\alpha}(\theta)$, $
\partial_{\theta}C^{\alpha}(\theta))$ and only $C^{\alpha}(\theta)$
respectively into the supermanifolds with following local coordinates
specified here for the case of irreducible GThGT, GThST
\begin{eqnarray}
{} & {} & T_{odd}(T^{\ast}_{odd}{\cal M}_s)\times \{\theta\} =
\bigl\{((\Gamma^p_s, \partial_{\theta}\Gamma^p_s)(\theta), \theta)\bigr\},\
p=\overline{1,2(n+m)},\; s=min\,, \nonumber \\
{} & {} & T^{\ast}_{odd}{\cal M}_s\times \{\theta\} =
\bigl\{(\Gamma^p_s(\theta) = (\Phi^B_s, \Phi^{\ast}_{B{}s})(\theta),
\theta)\bigr\},\; \Phi^B_s(\theta) = ({\cal A}^{\imath}, C^{\alpha})(\theta),
 B=\overline{1,n+m}.  
\end{eqnarray}
The 1st geometric object above may be considered as so-called odd
tangent bundle over odd cotangent bundle $T^{\ast}_{odd}{\cal
M}_{min}$, in turn over supermanifold ${\cal M}_{min}$
coordinatized by $\Phi^B_{min}(\theta)$ in the so-called minimal
sector [5].

The new superfields $C^{\alpha}(\theta)$, superantifields $({\cal A}^{\ast}_{
\imath}, C^{\ast}_{\alpha})(\theta)$ as the elements from
$\tilde{\Lambda}_{D\vert Nc+1}(z^{a}$, $\theta; {\bf K})$ are transformed
w.r.t. $J$ superfield representations $T_{\xi}$, $T^{\ast}$,
$T^{\ast}_{\xi}$ connected with $T$ and transformations for $\xi^{\alpha}(
\theta)$ (for instance, $T^{\ast}$ is conjugate to $T$ w.r.t. a some
bilinear form). Grassmann parities and expansion in powers of $\theta$ for
$\Phi^B_s(\theta)$, $\Phi^{\ast}_{B{}s}(\theta)$ are given as follows
\begin{eqnarray}
{} & {} & \bigl(\Phi^B_s, \Phi^{\ast}_{B{}s}\bigr)(\theta) = \bigl(
\phi^B_s + \lambda^B_s\theta,\
 \phi^{\ast}_{B{}s} - \theta J_{B{}s}\bigr),\ \;
\vec{\varepsilon} C^{\alpha}(\theta) = ((\varepsilon_P)_{\alpha} + 1,
(\varepsilon_{\bar{J}})_{\alpha}, \varepsilon_{\alpha}+1)\,,
 \nonumber \\
{} & {} &
\vec{\varepsilon}(\Phi^B_s(\theta), J_{B{}s}) =
\vec{\varepsilon}(\Phi^{\ast}_{B{}s}(\theta), \lambda^B_s) + (1,0,1) =
((\varepsilon_P)_{B},
(\varepsilon_{\bar{J}})_{B}, \varepsilon_{B})\,.
\end{eqnarray}
The superfunctional $Z_{(1)} =
\partial_{\theta} S_{(1){}L}\bigl(\Gamma_s(\theta),
\partial_{\theta}\Gamma_s(\theta), \theta\bigr)$
($\vec{\varepsilon}Z_{(1)}=(1,0,1)$) for GThGT,
its generating equation and $P, \bar{J}$-even superfunction $S_{(1)}(\theta)$
together with one's $\theta$-local master equation for GThST have the
representations for $s=min$ with accuracy up to 1st degree in
$C^{\alpha}(\theta)$ (therefore exact for abelian GAs)
\begin{eqnarray}
Z_{(1)}[\Gamma_{s}] &=& Z[{\cal A}] +\int d\vec{\theta}_2{\cal
A}^{\ast}_{\imath}(\theta_1){\hat{\cal
R}}^{\imath}_{\alpha}(\theta_1;
{\theta}_2)C^{\alpha}(\theta_2)(-1)^{\varepsilon_{\imath}} +
O(C^2(\theta))
, \nonumber \\
{} & {} & \bigl\{Z_{(1)}[\Gamma_{s}], Z_{(1)}[\Gamma_{s}]\bigr\} = 0\,; \\
S_{(1)}\bigl(\Gamma_s(\theta),\theta\bigr) & = &
S({\cal A}(\theta),
\theta) +
{\cal A}^{\ast}_{\imath}(\theta){\cal R}_{0}{}^{\imath}_{\alpha}(\theta)C^{
\alpha}(\theta)
+ O(C^2(\theta))\,, \nonumber \\
{} & {} & (S_{(1)}(\theta), S_{(1)}(\theta))_{\theta} = 0\,. 
\end{eqnarray}
The even $\{\ ,\ \}$ and odd $\theta$-local Poisson brackets are defined
on class of superfunctionals $C_{F{}min} \supset C_F$ on ${\cal M}_{min}$
given as in (2.33) via densities  on $\theta$, i.e. superfunctions from
superalgebra $C^k(T_{odd}(T^{\ast}_{odd}{\cal M}_{min})\times \{\theta\})$
$\equiv$ $D^{k}_{min}$, with expansion properties to be analogous
to (2.24), (2.27) w.r.t. all supervariables, and on the superalgebra
$C^k(T^{\ast}_{odd}{\cal M}_{min}\times \{\theta\})$
$\equiv$ $C^{k\ast}_{min}$ respectively by the formulae
\begin{eqnarray}
\bigl\{F[\Gamma_s], G[\Gamma_s]\bigr\} & = &\int d\theta\Bigl[
\frac{\delta F\phantom{x}}{\delta \Phi^B_s(\theta)}
\frac{\delta_l G\phantom{x}}{\delta \Phi^{\ast}_{B{}s}(\theta)} -
\frac{\delta_r F\phantom{x}}{\delta \Phi^{\ast}_{B{}s}(\theta)}
\frac{\delta_l G\phantom{x}}{\delta \Phi^B_s(\theta)}\Bigr]\,,
 \\
({\cal F}(\Gamma_s(\theta),\theta), {\cal G}(\Gamma_s(\theta),\theta))_{
\theta} & = &
\frac{\partial {\cal F}(\theta)}{\partial\Phi^B_s(\theta)}
\frac{\partial_l {\cal G}(\theta)}{\partial\Phi^{\ast}_{B{}s}(\theta)} -
\frac{\partial_r {\cal F}(\theta)}{\partial\Phi^{\ast}_{B{}s}(\theta)}
\frac{\partial_l {\cal G}(\theta)}{\partial\Phi^B_s(\theta)}\,, 
\end{eqnarray}
defining the corresponding ${\bf QP}$-structure [14] on the bundles
$T_{odd}(T^{\ast}_{odd}{\cal M}_{min})$, $T^{\ast}_{odd}{\cal M}_{min}$.

The calculation rules for superfield variational in (5.5) and partial (5.6)
derivatives w.r.t. $\Phi^*_B(\theta)$ are based on the expressions (with
omitting of index "min")
\begin{eqnarray}
\left( \frac{\partial_l \Phi^{\ast}_C(\theta)}{\partial
\Phi^{\ast}_B(\theta) \phantom{x}}, \frac{\partial_l
(\partial_{\theta} \Phi^{\ast}_C(\theta))}{\partial
(\partial_{\theta}\Phi^{\ast}_B(\theta)) \phantom{x}},
\frac{\delta_l \Phi^{\ast}_C(\theta_1)}{\delta
\Phi^{\ast}_B(\theta) \phantom{x}}, \frac{\delta_l
(\partial_{\theta_1} \Phi^{\ast}_C(\theta_1))}{\delta
\Phi^{\ast}_B(\theta) \phantom{xxxx}} \right) = \bigl(1,
1,(-1)^{\varepsilon_C +1}
\delta(\theta - \theta_1), 1\bigr)\delta_C{}^B\,,
\end{eqnarray}
added by the Euler-Lagrange operator w.r.t. $\Phi^*_B(\theta)$ written in
terms of superfunctional $F[\Gamma]$ and its density ${\cal F} \in D^k_{min}$
\begin{eqnarray}
\frac{\delta_{l(r)} F[\Gamma]}{\delta\Phi^{*}_{B}(\theta)\phantom{x}}
= \left[\frac{\partial_{l(r)} \phantom{xx}}{\partial\Phi^{*}_B(\theta)}
 -(-1)^{\varepsilon_{B}+1}\partial_{\theta}^{l(r)}\frac{
 \partial_{l(r)}
\phantom{xxxxxx}}{\partial
\left(\partial_{\theta}^{l(r)}\Phi^{*}_{B}(\theta)\right)}\right]
{\cal F}(\theta) \equiv {\cal  L}^{*{}B}_{l(r)}(\theta){\cal F}(\theta)\,.
\end{eqnarray}
The even bracket (5.5) may be expressed through  the new $\theta$-local
antibracket (5.6) provided that $D[\Gamma]$ are written via one's density
${\cal D}(\theta)\in D^k_{min}$ taking the formula (5.8) into account
\begin{eqnarray}
\{F[\Gamma],G[\Gamma]\} = \int d\theta ({\cal F}(\theta), {\cal G}(\theta))_{
\theta}^{(\Gamma, \partial_{\theta}\Gamma)} = \int d\theta\Bigl(\Bigl[
{\cal L}^r_B{\cal F}{\cal L}_l^{*{}B}{\cal G}
-{\cal L}_r^{*{}B}{\cal F}{\cal L}^l_B{\cal G}\Bigr](\theta)\Bigr)\,.
\end{eqnarray}
Relationship (5.9) appears by natural generalization for the connection of
odd and even Poisson brackets from ref.[30] for the case of the densities
dependence  upon superfields $\partial_{\theta}\Gamma^p(\theta)$ and contains,
in fact, 4 antibrackets.

As the consequence the generating equation (5.3) written for the
superfunctional
$Z_{(0)}[\Gamma]$ $=$ $-\int d\theta S_{(1)}(\theta)$ is embedded into the same
equation but for $Z_{(1)}[\Gamma]$  providing, at least for abelian GAs of
GTGT and GTST, the embedding of the latter GA into former one
\begin{eqnarray}
\bigl\{Z_{(0)}[\Gamma_{min}], Z_{(0)}[\Gamma_{min}]\bigr\} =
\int d\theta (S_{(1)}(\theta), S_{(1)}(\theta))_{\theta} = 0,\,
Z_{(1)}[\Gamma_{min}]=Z_{(0)}[\Gamma_{min}] + \ldots.
\end{eqnarray}

Numbering even and odd Poisson brackets as $k=0$, $k=1$
respectively it is easy to check the standard properties validity  of
generalized antisymmetry, Leibnitz rule, Jacobi identity
\renewcommand{\theequation}{\arabic{subsection}.\arabic{equation}\alph{lyter}}
\begin{eqnarray}
\setcounter{lyter}{1}
{} & {} & \bigl\{D,E\bigr\}_k = -(-1)^{(\varepsilon(D) + k)(\varepsilon(E)+k)}
\bigl\{E,D\bigr\}_k\,,\\
\setcounter{equation}{11}
\setcounter{lyter}{2}
{} & {} & \bigl\{DE, K\bigr\}_k = D\bigl\{E, K\bigr\}_k + \bigl\{D, K
\bigr\}_k E(-1)^{\varepsilon(E)(\varepsilon(K) + k)}\,, \\ 
\setcounter{equation}{11}
\setcounter{lyter}{3}
{} & {} &
\bigl\{\bigl\{D,E\bigr\}_k, K\bigr\}_k(-1)^{(\varepsilon(D) + k)(
\varepsilon(K) + k)} + {\rm cycl. perm.}(D, E, K) = 0\,, 
\end{eqnarray}
with superfunctionals for $k=0$ and $\theta$-local superfunctions for $k=1$
instead of $D, E, K$.

The master equations (5.3), (5.4) allow to solve the $\theta$-superfield
problem of ZLR  [35] (through duality between
superbrackets [36]) simultaneously with new procedure to obtain the
$\theta$-superfield (and therefore almost standard for $\theta=0$)
models on the
reduced (anti)symplectic spaces starting from initial GThGT, GThST.
Indeed, from validity of (5.3), (5.4) it follows the definition for new
superfunctional antibracket and even $\theta$-local bracket on the
corresponding (at least locally) supermanifolds  given by zero
locus ${\cal Z}_{Q^b}$ for $(\varepsilon_P,\varepsilon)$ odd interrelated
nilpotent vector fields ${\bf Q}^b$, $b=0,1$
\renewcommand{\theequation}{\arabic{subsection}.\arabic{equation}}
\begin{eqnarray}
{} & {} & \hspace{-0.5em}
{\bf Q}^0(\theta) = (S_{(1)}(\theta),\ \;)_{\theta},\ \
{\bf Q}^1 = \bigl\{Z_{(1)}[\Gamma],\ \;\bigr\}\,, \\ 
{} & {} & \hspace{-1.5em}
\bigl\{f(\theta), g(\theta)\bigr\}_{\theta} = ({\cal F}(\theta),
(S_{(1)}(\theta), {\cal G}(\theta))_{\theta})_{\theta}{}_{\mid
{\cal Z}_{Q^0}},\;
{\cal F}, {\cal G} \in C^{k{}*}_{min},
(f;g) = ({\cal F}; {\cal G})_{\mid{\cal Z}_{Q^0}},\\ 
{} & {} & \hspace{-1.5em}
\bigl(\tilde{f}[\Gamma],\tilde{g}[\Gamma]\bigr) =
\bigl\{F[\Gamma],\bigl\{Z_{(1)}[\Gamma], G[\Gamma]\bigr\}\bigr\}_{
\mid{\cal Z}_{Q^1}},\
(\tilde{f};\tilde{g}) = ( F;  G)_{\mid{\cal Z}_{Q^1}}\,.
\end{eqnarray}
Given this, the explicit form for nonlinear $\theta$-local even bracket
together with specification for ${\cal Z}_{Q^0}$ structure in $T^*_{odd}{\cal
M}_{min}$ are defined by the equations  with accuracy up to
$O({\cal A}^{\ast}C)$ and terms, at least, linear in $C^*, C$
\begin{eqnarray}
{\bf Q}^0(\theta)\Gamma^p_{min}(\theta) = 0 \Rightarrow
\left\{
\begin{array}{l}
{\cal R}_{0}{}^{\imath}_{\alpha}(\theta)C^{\alpha}(\theta) =0\,,\\
S,_{\imath}(\theta) + {\cal A}^{\ast}_{\jmath}(\theta)
{\cal R}_{0}{}^{\jmath}_{\alpha},_{\imath}(\theta)C^{\alpha}(\theta)(
-1)^{\varepsilon_{\imath}(\varepsilon_{\alpha}+1)} =0\,,\\
{\cal A}^{\ast}_{\imath}(\theta)
{\cal R}_{0}{}^{\imath}_{\alpha}(\theta)=0\,.
\end{array}\right.
\end{eqnarray}
The irreducibility of GGTST (4.32) permits to solve (5.15) in the form
\begin{eqnarray}
C^{\alpha}(\theta) = 0,\ \;  S,_{\imath}(\theta) = 0,\ \;
{\cal A}^{\ast}_{\imath}(\theta)
{\cal R}_{0}{}^{\imath}_{\alpha}(\theta)_{\mid S,_{\imath}(\theta) = 0}=0\,,
\end{eqnarray}
and therefore to get the representation for $\{\ ,\ \}_{\theta}$
\begin{eqnarray}
\bigl\{f(\theta), g(\theta)\bigr\}_{\theta}& \hspace{-0.5em}= &\hspace{-0.5em}
\left[-\frac{\partial_r {\cal F}(\theta)}{\partial{\cal A}^{\ast}_{\imath}(
\theta)}S,_{\imath\jmath}(\theta)
\frac{\partial_l {\cal G}(\theta)}{\partial{\cal A}^{\ast}_{\jmath}(\theta)}
(-1)^{\varepsilon_{\imath}} + \left(\hspace{-0.1em}\Bigl(
\frac{\partial {\cal F}(\theta)}{\partial{\cal A}^{\imath}(\theta)}
{\cal R}_{0}{}^{\imath}_{\alpha}(\theta)
\frac{\partial_l {\cal G}(\theta)}{\partial C^{\ast}_{\alpha}(\theta)}
\right.\right.\nonumber \\
{} &\hspace{-0.5em} - & \hspace{-0.5em}\left.\left.
\frac{\partial_r {\cal F}(\theta)}{\partial{\cal A}^{\ast}_{\imath}(\theta)}
{\cal A}^{\ast}_{\jmath}(\theta){\cal R}_{0}{}^{\jmath}_{\alpha},_{\imath}(
\theta)\frac{\partial_l {\cal G}(\theta)}{\partial
C^{\ast}_{\alpha}(\theta)}(-1)^{\varepsilon_{\imath}\varepsilon_{
\alpha}}\Bigr)  - (-1)^{\varepsilon({\cal F})\varepsilon({\cal
G})}({\cal F} \leftrightarrow {\cal G})\hspace{-0.1em}\right)
\hspace{-0.1em}\right]_{\mid
{\cal Z}_{Q^0}}\hspace{-0.9em}.
\end{eqnarray}
In turn, the structure of ${\cal Z}_{Q^1}$ embedded in $T_{odd}(T^*_{odd}{\cal
M}_{min})$ are given by the equations with the same accuracy as for
${\cal Z}_{Q^0}$ in (5.15) extended
for $\partial_{\theta}({\cal A}^{\ast}, C, C^*)(\theta)$ as well
\begin{eqnarray}
{\bf Q}^1\Gamma^p_{min}(\theta) = 0 \Rightarrow
\left\{
\begin{array}{l}
\int d{\theta}'{\hat{\cal R}}^{\imath}_{\alpha}(\theta;
{\theta}')C^{\alpha}(\theta') =0\,,\\
({\cal L}^r_{\imath}S_L)(\theta) -\int d\vec{\theta}_2{\cal
A}^{\ast}_{\jmath}(\theta_1) \displaystyle\frac{\delta {\hat{\cal
R}}^{\jmath}_{\alpha}(\theta_1; {\theta}_2)}{\delta {\cal
A}^{\imath}(\theta)\phantom{xxxx}}
C^{\alpha}(\theta_2)(-1)^{\varepsilon_{\imath}(\varepsilon_{
\alpha}+1)} = 0,\\
(-1)^{\varepsilon_{\imath}}
\int d{\theta}{\cal A}^{\ast}_{\imath}(\theta)
{\hat{\cal R}}^{\imath}_{\alpha}(\theta;
{\theta}')=0\,.
\end{array}\right.
\end{eqnarray}
With regard for Eq.(4.26) trivial solution for irreducible GThGT the system
above is reduced to the form
\begin{eqnarray}
1) & {} & C^{\alpha}(\theta) = 0,\ \; {\cal L}^r_{\imath}(\theta)
S_L(\theta) = 0,\ \; \partial_{\theta}{\cal A}^{\ast}_{\imath}(\theta)
{\hat{\cal R}}^{\imath}_{\alpha}(\theta;{\theta}')=0\,, \nonumber \\
2) & {} & \partial_{\theta}C^{\alpha}(\theta) = 0,\ \;
(\textstyle\frac{\partial}{\partial\theta} + {\stackrel{\circ}{U}}_+(\theta))
S_L,_{\imath}(\theta) = 0,\ \; \partial_{\theta}\partial_{\theta'}{
\cal A}^{\ast}_{\imath}(\theta)
{\hat{\cal R}}^{\imath}_{\alpha}(\theta;{\theta}')=0\,,
\end{eqnarray}
providing the explicit structure for antibracket (5.14)
\begin{eqnarray}
{} & {} &
\hspace{-1.2em}\bigl(\tilde{f}[\Gamma],\tilde{g}[\Gamma]\bigr)
\hspace{-0.2em} = \hspace{-0.2em}\int
\hspace{-0.4em}d\theta\hspace{-0.1em}\left[\hspace{-0.2em}
\frac{\delta_r F[\Gamma]}{\delta {\cal A}^{\ast}_{\imath}(\theta)}
\hspace{-0.2em}\int \hspace{-0.3em}d\theta_1 \frac{\delta_l
\phantom {xxx}}{\delta{\cal A}^{\imath}(\theta)} \frac{\delta_r
Z[{\cal A}]}{\delta{\cal A}^{\jmath}(\theta_1)} \frac{\delta_l
G[\Gamma]}{\delta {\cal A}^{\ast}_{\jmath}(\theta_1)}
(\hspace{-0.1em}-1\hspace{-0.1em})^{\varepsilon_{\imath}}
\hspace{-0.2em}- \hspace{-0.2em}
\left(\hspace{-0.2em}\Bigl(\frac{\delta F[\Gamma]}{\delta {\cal
A}^{\imath}(\theta)} \hspace{-0.2em}\int \hspace{-0.3em} d\theta_1
{\hat{\cal R}}^{\imath}_{\alpha}(\hspace{-0.05em}\theta;
\hspace{-0.05em}{\theta}_1) \frac{\delta_l G[\Gamma]}{\delta
C^{\ast}_{\alpha}(\theta_1)}
\right.\right. \nonumber \\
{} & {} & \hspace{-1.2em} \left.\left. -
\frac{\delta_r F[\Gamma]}{\delta {\cal A}^{\ast}_{\imath}(\theta)}
\hspace{-0.2em}\int \hspace{-0.2em}
d{\theta}_2{\cal A}^{\ast}_{\jmath}(\theta_1)
\displaystyle\frac{\delta
{\hat{\cal R}}^{\jmath}_{\alpha}(\theta_1;
{\theta}_2)}{\delta {\cal A}^{\imath}(\theta)\phantom{xxxx}}
\frac{\delta_l G[\Gamma]}{\delta C^{\ast}_{\alpha}(\theta_2)}
(-1)^{\varepsilon_{\imath}\varepsilon_{\alpha}}\Bigr) -
 (-1)^{(\varepsilon(F)+1)(\varepsilon(G)+1)}(F \leftrightarrow G)
 \hspace{-0.15em}\right)\hspace{-0.15em}\right]_{\mid
{\cal Z}_{Q^1}}\hspace{-0.9em}.
\end{eqnarray}

The validity of the properties (5.11) for brackets on ${\cal Z}_{Q^b}$
directly ensues from master equations (5.3), (5.4) and the same properties
above but for initial brackets (5.5), (5.6). The 1st 2 summands
(1st in square and 1st in round parentheses)
in (5.17), (5.20) permit to define on
${\cal Z}_{Q^0}$,
${\cal Z}_{Q^1}$ correspondingly the nondegenerate 2-forms $\omega_0^2(
\theta)$,
$\omega^2_1(\theta)$ (and the 1st term in (5.17) corresponds for $
\varepsilon_{\imath}=0$ in appropriate basis for superfields ${\cal A}^{
\imath}(\theta)$ of the form (4.15a), specially compatible with other
superfields from $\Gamma^p(\theta)$, to the  symmetric
part $d{\cal A}'{}^*_A(\theta)\wedge d{\cal A}'{}^*_A(\theta)$ of even
2-form $\omega^2_0(\theta)$, $A=\overline{1,n-m}$ [21]). The rest
terms  guarantee the
Jacobi identities validity, i.e. to be closed for  $\omega^2_b(\theta)$
and hence to be for (${\cal Z}_{Q^0}$) ${\cal Z}_{Q^1}$  not only
Poisson but by the (anti)symplectic, at least locally, supermanifolds.

There exists the inverse inclusion of the new even and odd brackets (5.13),
(5.14) given on the corresponding superalgebras of superfunction(al)s modulo
the superfunction(al)s vanishing on ${\cal Z}_{Q^b}$. Really, having
restricted GThGT with $Z_{(1)}[\Gamma]$ (5.3) to GThST with
$Z_{(0)}[\Gamma]$ satisfying to (5.10) we immediately find the relationship,
taking correspondence between superfunction(al)s in (5.13), (5.14) into
account and under condition for $F[\Gamma], G[\Gamma]$ in (5.14) to be
defined through densities only on $T^*_{odd}{\cal M}_{min}\times\{\theta\}$,
\begin{eqnarray}
\bigl(\tilde{f}[\Gamma],\tilde{g}[\Gamma]\bigr)^{Z_{(0)}} & = & -
\int d\theta \bigl\{f(\Gamma(\theta),\theta),
g(\Gamma(\theta),\theta)
\bigr\}_{\theta}\,,\\
\bigl(\tilde{f}[\Gamma],\tilde{g}[\Gamma]\bigr)^{Z_{(0)}} & = &
\bigl\{F[\Gamma],\bigl\{Z_{(0)}[\Gamma], G[\Gamma]\bigr\}\bigr\}_{
\mid{\cal Z}_{\tilde{Q}^1}}, \ {\bf \tilde{Q}}^1=
\bigl\{Z_{(0)}[\Gamma], \ \;\bigr\}\,, 
\end{eqnarray}
in turn based on being easily derived from (5.16) and (5.5), (5.6)
expressions
\begin{eqnarray}
\bigl\{Z_{(0)}[\Gamma], G[\Gamma]\bigr\} & = & - \int d\theta
(S_{(1)}( \Gamma(\theta), \theta), {\cal
G}(\Gamma(\theta),\theta))_{\theta},\; G=\partial_{\theta}{\cal
G}, (G_{\mid{\cal Z}_{\tilde{Q}^1}}, {\cal G}_{\mid{\cal
Z}_{{Q}^0}}) = (\tilde{g}, g),
\nonumber \\
{} & {} & {\cal Z}_{Q^1} \cap T^*_{odd}{\cal M}_{min} =
{\cal Z}_{\tilde{Q}^1} \cap T^*_{odd}{\cal M}_{min} = {\cal Z}_{{Q}^0}
\,.
\end{eqnarray}
In general, even bracket (5.13) is embedded into antibracket (5.14).

The hierarchy of superbrackets permits to suggest a some different ways to
construct the new $\theta$-superfield models starting from initial GThST
and GThGT, firstly, as embedded into ${\cal Z}_{{Q}^b}$, secondly, as
enlarged in more wide special space  than $T_{odd}(T^*_{odd}{\cal M}_{s})$,
$s= min$. These possibilities are based on the interpretation for $Z_{(0)}[
\Gamma_{s}]$ $\subset$ $Z_{(1)}[\Gamma_{s}]$ as the BFV-BRST charges
defined on  $T_{odd}(T^*_{odd}{\cal M}_{s})$ and on the
introduction of an additional earlier hidden  the $\vec{\varepsilon}$-even
{\it first level supertime}
$\Gamma^{1}$ $\equiv$ $\Gamma^{1}_a = (t^1, \theta^1)$, $a=0,1$,
reconstructing therefore the all $\theta$-STF ingridients.

So, the supergroup
$J$ $\equiv$ $J_{(0)}$, superspace ${\cal M}$ $\equiv$ ${\cal M}_{(0)}$
structure and properties of superfields $\Phi^B(\theta)$ and
${\cal M}_{min}^{(0)}$ $\equiv$ ${\cal M}_{min}$ can be extended as follows
\begin{eqnarray}
{} & {} & J_{(1)}  = \bar{J}_{(1)} \times P_{(1)},\;
P_{(1)} = P_0 \times P_1 =
\bigl\{\exp(
\imath\mu_c p^c)\vert \mu_c \in {}^1\Lambda_2(\theta,\theta^1)\bigr\},
[p^c, p^d]_+ =0, c,d=0,1, \nonumber \\
{} & {} & \bar{J}_{(1)}\supset \bar{J}_{(0)}\equiv \bar{J},\ \;
{\cal M}_{(1)} = \tilde{\cal M}_{(1)} \oplus \tilde{P}_{(1)} = \{(
z^a_{(1)}; \theta, \theta^1) = ((x^{\mu}, t^1), \theta^{Aj}; \theta,
\theta^1)\}\,, \\
{} & {} & \Phi^B(\theta, \Gamma^{1}) = \Phi^B(\theta, t^{1}) +
\lambda^B(\theta, t^1)\theta^1,\
\vec{\varepsilon}{}^{1}(\Phi^B(\theta, \Gamma^{1})) =
\vec{\varepsilon}{}^{1}(\lambda^B(\theta, t^{1})) + (1,0,1)
\nonumber \\
{} & {} & \hspace{1em}=
\vec{\varepsilon}{}^{1}(\partial_{\theta}\partial_{\theta^1}
\Phi^B(\theta, \Gamma^{1}))= ((\varepsilon_P^1)_B, (\varepsilon_{
\bar{J}}^1)_B, \varepsilon^1_B),\  \;\vec{\varepsilon}{}^{1}
= (\varepsilon_P^1=
\varepsilon_{P_1}+\varepsilon_{P_0}, \varepsilon_{\bar{J}}^1, \varepsilon^1)
\,.
\end{eqnarray}

The construction of $\theta$-STF model in a space enlarging
$T_{odd}(T^*_{odd}{\cal M}_{min})$ can be realized in 2 directions.
Qualitatively regarding,
firstly, $Z_{(a)}(\Gamma_{min}(\Gamma^1))$, $a=0,1$
now by superfunctions depending upon  $\Gamma^1$ and, secondly, interpreting
them
by virtue of generating equations (5.3), (5.10) as the BFV-BRST
generators in minimal sector further enlarged  in more wide than
$T_{odd}(T^*_{odd}{\cal M}_{min})$
space to superfunction $Z_{(a)}(\Gamma_{tot}(\Gamma^1))$ $\equiv$
$Z_{(a)}(\Gamma^1)$
we must find the corresponding gauge fermion superfunctions
$\Psi_{(a)}(\Gamma_{tot}(\Gamma^1))$ $\equiv$ $\Psi_{(a)}(\Gamma^1)$
 providing the
nondegeneracy of the analogs of Faddeev-Popov supermatrices in the $0$-level
even bracket (5.5) given in enlarged space coordinatized by
$\Gamma_{tot}(\Gamma^1)$ =
$(\Gamma_{min}(\Gamma^1), \Gamma_{add}(\Gamma^1))$ with additional
antighost and Lagrangian multiplier superfields and its superantifields.
Next it is necessary
to define the such $\vec{\varepsilon}{}^{1}$-boson superfunctions
$H_{(a)}(\Gamma_{min}(\Gamma^1))$ $\equiv$ $H_{(a)}(\Gamma^1)$ which
commute with $Z_{(a)}$ w.r.t.
continued $\Gamma^1$-local bracket (5.5) and may be considered as
corresponding Hamiltonians. At last, in correspondence with BFV-prescription
composing the unitarizing Hamiltonians
${\cal H}_{(a)}(\Gamma_{tot}(\Gamma^1))$ we can to obtain the new action
superfunctions $S^1_{H{}(a)}(\Gamma_{tot}(\theta^1))$ depending upon only
odd 1st level time parameter
\begin{eqnarray}
{} & {} &
S^1_{H{}(a)}(\Gamma_{tot}(\theta^1))  =  \hspace{-0.2em}\int
\hspace{-0.2em}dt^1 \left[\hspace{-0.1em}\int \hspace{-0.3em}d\theta
\Phi^*_{B{}tot}(\theta,\Gamma^1)\partial_{t^1}\Phi^{B}_{tot}(\theta,\Gamma^1)
- {\cal H}_{(a)}(\Gamma_{tot}(\Gamma^1))
\right], \nonumber \\
{} & {} &
{\cal H}_{(a)}(\Gamma_{tot}(\Gamma^1)) =
H_{(a)}(\Gamma^1) + \bigl\{
Z_{(a)}(\Gamma^1), \Psi_{(a)}(\Gamma^1)\bigr\}_{\Gamma^1},\ \;
\bigl\{H_{(a)}(\Gamma^1), Z_{(a)}(\Gamma^1)
\bigr\}_{\Gamma^1} = 0\,. 
\end{eqnarray}
Note that this algorithm may be applied directly to $Z[{\cal A}]$ (4.4) with
definite peculiarities.
As the
result we arrive, in fact, at the action superfunction being similar to
superfunction in (5.4) but on the next level of $\theta$-STF model
determination and therefore including
$S_{(1){}L}\bigl(\Gamma_{min}(\theta, \Gamma^1),\theta\bigr)$ for $a=1$,
$S_{(1)}(\theta, \Gamma^1)$ for $a=0$  (5.4) as parts of
the density on $\theta, \theta^1$.

The other variant to obtain the new action is almost similar to so-called
superfield algorithm for generalized Poisson sigma models [31] and results in
the 1st level $\vec{\varepsilon}{}^{1}$-boson superfunctionals being differed
from (5.26)
\begin{eqnarray}
{S}^1_{(a)}[\Gamma_{s}]  =  \hspace{-0.2em}\int
\hspace{-0.2em}d\Gamma^1 \left[\hspace{-0.1em}\int \hspace{-0.2em}d\theta
\Phi^*_{B{}s}(\theta,\Gamma^1)\partial^r_{\theta^1}\Phi^{B}_{s}(
\theta,\Gamma^1) - Z_{(a)}(\Gamma_s(\Gamma^1))\right], a=0,1, s=min, tot.
\end{eqnarray}
The following dynamical equations arise from variational principle for both
actions (5.26), (5.27)
\begin{eqnarray}
\frac{\delta_{l,\theta^1}S^1_{H{}(a)}(\theta^1)}{\delta\Gamma^p_{tot}(
\theta, \Gamma^1)\phantom{x}} = 0 &
\Rightarrow & \partial_{t^1}\Gamma^p_{tot}(\theta,
\Gamma^1) = \bigl\{\Gamma^p_{tot}(\theta, \Gamma^1),
{\cal H}_{(a)}(\Gamma_{tot}(\Gamma^1))\bigr\}_{\Gamma^1}\,, \\ 
\frac{\delta_{l} {S}^1_{(a)}[\Gamma_s]}{\delta\Gamma^p_{s}(
\theta, \Gamma^1)} = 0 & \Rightarrow & \partial^r_{\theta^1
}\Gamma^p_{s}(\theta,
\Gamma^1) = \bigl\{\Gamma^p_{s}(\theta, \Gamma^1),
Z_{(a)}(\Gamma_{s}(\Gamma^1))\bigr\}_{\Gamma^1}, s=min,tot\,, 
\end{eqnarray}
with superfield variational derivatives w.r.t. $\Gamma^p_{tot}(\theta,
\Gamma^1)$ for fixed $\theta^1$ in (5.28) and the  next level
superfunctional one for $s=min, tot$ in (5.29).

Firstly, note the commutation of $H_{(a)}(\Gamma^1)$ with $Z_{(a)}(\Gamma^1)$
w.r.t. bracket (5.26) leads to compatibility of the Eqs.(5.28),(5.29) for
$s=tot$, i.e.
the commutator $[\partial_{t^1}, \partial_{\theta^1}]
\Gamma^p_{s}(\theta, \Gamma^1)$ vanishes on their solutions in view of (5.3),
(5.26) equations fulfilment, antisymmetry and Jacobi identity properties
(5.11a,c) for $k=0$. Besides, the generating equations (5.3), (5.10)
provide the $\theta$-superfield solvability of the Eqs.(5.28), (5.29) in
the sense of (A.6) relation.

Moreover, the actions ${S}^1_{(a)}[\Gamma_s]$ obey to the next level
master equation  with new superfunctional antibracket in the supermanifold
$T^{(1)}_{odd}(T_{odd}(T^*_{odd}{\cal M}_{s}))$
\begin{eqnarray}
{} & {} &
\Bigl({S}^1_{(a)}[\Gamma_s],{S}^1_{(a)}[\Gamma_s]\Bigr)^1 =
 \int d\Gamma^1\Bigl(d\theta \bigl[\partial^r_{\theta^1}
 (\Phi^*_{B{}s}\partial^r_{\theta^1}\Phi^{B}_{s})(
\theta,\Gamma^1) +  \bigl\{Z_{(a)}(\Gamma^1), Z_{(a)}(\Gamma^1)
\bigr\}_{\Gamma^1}\bigr] \nonumber \\
{} & {} &
\phantom{\Bigl({S}^1_{(a)}[\Gamma_s],{S}^1_{(a)}[\Gamma_s]\Bigr)^1}
+ \partial^r_{\theta^1}Z_{(a)}(\Gamma^1)\Bigr)
= 0, \; s=min, tot\,, \\ 
{} & {} & \Bigl({F}^1[\Gamma_s],{G}^1[\Gamma_s] \Bigr)^1 = \int
d\Gamma^1d\theta \left[ \frac{\delta_{r} {F}^1[\Gamma_s]}{\delta
\Phi^B_{s}(\theta, \Gamma^1)} \frac{\delta_{l}
{G}^1[\Gamma_s]}{\delta\Phi^*_{B{}s}(\theta, \Gamma^1)} -
\frac{\delta_{r} {F}^1[\Gamma_s]}{\delta\Phi^*_{B{}s}(\theta,
\Gamma^1)} \frac{\delta_{l}
{G}^1[\Gamma_s]}{\delta\Phi^B_{s}(\theta,
\Gamma^1)}\right] \nonumber \\
{} & {} & \hspace{1em} =
\int d\Gamma^1d\theta
\left[{\cal  L}_{B}^{r}(\theta,\Gamma^1){\cal F}(\Gamma^1)
{\cal  L}^{*{}B}_{l}(\theta,\Gamma^1){\cal G}(\Gamma^1) -
{\cal  L}^{*{}B}_{r}(\theta,\Gamma^1){\cal F}(\Gamma^1){
\cal  L}_{B}^{l}(\theta,\Gamma^1){\cal G}(\Gamma^1)\right]
 \nonumber \\
{} & {} & \hspace{1em} \equiv \int d\Gamma^1 \Bigl\{{\cal
F}(\Gamma^1), {\cal G}(\Gamma^1)\Bigr\}_{\Gamma^1}^{(\Gamma_s,
\partial_{\Gamma^1}\Gamma_s)},\ {D}^1[\Gamma_s] = \int
d\Gamma^1{\cal D}\bigl( \Gamma_s(\Gamma^1),
\partial_{\Gamma^1}\Gamma_s(\Gamma^1), \Gamma^1\bigr),
\end{eqnarray}
where we could not include $t^1$ variable in definition of
antibracket (5.31) to get the type (5.21) connection  and
therefore instead of ${ S}^1_{(a)}[\Gamma_s]$ we should write its
density w.r.t.  $t^1$ in direct analog of the Eq.(5.30). At last,
under Euler-Lagrange operator, for  instance, for
$\Phi^B_{s}(\theta, \Gamma^1)$ in (5.31) we mean the analog of
expression (5.8) written for superfunctional w.r.t. $\theta$:
${\cal F}(\Gamma^1)$,  defined as in (5.31) (with $\delta$-symbol
$\delta_{a1}$)
\begin{eqnarray}
{\cal  L}_{B}^{l(r)}(\theta,\Gamma^1){\cal F}(\Gamma^1) =
\left[\frac{\delta_{l(r),\Gamma^1} \phantom{xx}}{
\delta\Phi^B(\theta,\Gamma^1)}
 -(-1)^{\varepsilon_{B}\delta_{a1}}\partial_{\Gamma^1_a}^{l(r)}\frac{
 \delta_{l(r),\Gamma^1}
\phantom{xxxxx}}{\delta
\left(\partial_{\Gamma^1_a}^{l(r)}\Phi^{B}(\theta, \Gamma^1)
\right)}\right]{\cal F}(\Gamma^1)\,.
\end{eqnarray}
\underline{\bf  Resume}:

\noindent
{\bf 1)} The action ${S}^1_{(a)}$ (5.27) construction defined
on the odd tangent bundle with new antibracket (5.31) starting from $
Z_{(a)}(\Gamma_s(\Gamma^1))$ given on symplectic supermanifold $
T_{odd}(T^*_{odd}{\cal M}_{min})$ with even bracket  (5.5)
appears by none other than the  essense of so-called inverse ZLR problem.

\noindent
{\bf 2)} Together the superfunction(al)s
$S^1_{H{}(a)}(\Gamma_{tot}(\theta^1))$ (5.26), ${S}^1_{(a)}[\Gamma_s]$
(5.27) and their extremals (5.28), (5.29) describe in the new
superfield manner the BFV-BRST classical construction of the corresponding
 dynamical system initially defined on
$T_{odd}(T^*_{odd}{\cal M}_{cl})$ and enlarged into space  parametrized
by $\Gamma_{tot}^p(\theta, \Gamma^1)$ subject to constraints encoded by
$Z_{(a)}(\Gamma_{tot}(\Gamma^1))$ and $Z_{(a)}(\Gamma_{min}(\Gamma^1))$
(5.3), (5.10), in turn,  constructed from $S_L({\cal A}(\theta), \partial_{
\theta}{\cal A}(\theta), \theta)$. The dynamics of this external model is
provided by
the 1st level supertime $\Gamma^1 = (t^1, \theta^1)$ presence in
comparison with initial  $t \in \imath$, $\theta$ describing the original
so-called internal model.

As to the direct ZLR problem  then we can use the
constructed brackets (5.13), (5.14) to define on the corresponding
${\cal Z}_{Q^0}$ the BFV-BRST charge $Z^{(-1)}(\theta)$ and on
${\cal Z}_{Q^1}$ the new superfunction $S^{(-1)}$ which must satisfy to
the corresponding generating equations with appropriate new "$(-1)$-level"
supertime $\Gamma^{-1}$ introduction. Shortly, these new  objects can be
obtained by correlated
with each other as well as their initial analogs $Z_{(a)}[\Gamma]$,
$S_{(1)}(\Gamma(\theta), \theta)$, $S_{(1){} L}(\Gamma(\theta),
\partial_{\theta}\Gamma(\theta), \theta)$  or in correspondence with
anzatz (5.27).

If there exist the special coordinates on ${\cal Z}_{Q^1}$ with values of $
\varepsilon_P^{(-1)} = \varepsilon_{P_{-1}}$ + $\varepsilon_{P_0} = 0$ then
we able to restrict the action superfunction $S^{(-1)}$ to depend only upon
them and therefore to be the classical new action.

Certainly, under appropriate extension of ${\cal Z}_{Q^0}$ in analogy with
superfunctions (5.26) introduction we can to construct $\theta$-superfield
BFV similar triple $\bigl(Z^{(-1)}$, $H^{(-1)}$,
$\Psi^{(-1)}\bigr)(\theta,\Gamma^{-1})$ and $S_{H}^{(-1)}(\Gamma_{
tot}(\theta^{-1}))$ as for inverse ZLR problem if the structure of
${\cal Z}_{Q^0}$ allows to define $Z^{(-1)}(\theta)$, i.e.  ${\cal Z}_{Q^0}$
appears by supermanifold.

The more profound and sufficiently perspective investigation of the produced
schemes for construction of the embebbed and enlarged new
$\theta$-superfield models
and their detailed properties (including the
relationships between the corresponding observables making use the ghost
number prescription) requires  the special
efforts in view of both the models nontrivial connection
and, for instance, the correspondence with  superfield BFV
method [29]. In addition, note on the particular similar features between the
models construction above with the results from ref.[30].
\subsection{$\theta$-STF Component Formulation}
\setcounter{equation}{0}

Let us continue a particular started in Sec.III programm of establishment the
correspondence between superfield and component field quantities and
relations. From representation (2.33)
for
superfunctionals on $T_{odd}{{\cal M}_{cl}} \times \{\theta\}$  find the
expression for their densities in terms of the formers themselves
\begin{eqnarray}
{\cal F}\bigl( {\cal A}(\theta), \partial_{\theta}{\cal
A}(\theta), \theta\bigr)  =  P_0(\theta) {\cal F}(\theta) + \theta
F[{\cal A}] \equiv  {\cal F}\bigl( A, \lambda,
0\bigr) + \theta \bar{F}[A,\lambda] \,, 
\end{eqnarray}
where the bar means the standard change of the form of
dependence  from the component arguments ($F[{\cal A}]$ = $\bar{F}[
P_0{\cal A}, \partial_{\theta}{\cal A}]$ = $\bar{F}[A,\lambda]$).

Taking account of the connection (3.1) for partial superfield
and component derivatives  the component expression for
supermatrix of the 2nd partial superfield derivatives of ${\cal
F}(\theta) \equiv {\cal F}\bigl({\cal A}(\theta), \partial_{\theta}{\cal
A}(\theta), \theta \bigr)$ w.r.t. ${\cal A}^{\imath}(\theta)$,
${\cal A}^{\jmath}(\theta)$ has the form
\begin{eqnarray}
\left\|\frac{\partial_r \phantom{xxx}}{ \partial {\cal
A}^{\imath}(\theta)} \frac{ \partial_l {\cal F}(\theta)}{ \partial {\cal
A}^{\jmath}(\theta)}\right\| =\left\|
\frac{ \partial_r \phantom{xxxx}}{\partial
P_a{\cal A}^{\imath}(\theta)}
\frac{\partial_l {\cal F}(\theta)\phantom{xx}
}{\partial P_b{\cal A}^{\jmath}(\theta)}\right\|, \;
\frac{ \partial_r \phantom{xxxx}}{\partial
P_a{\cal A}^{\imath}(\theta)} = \left(
\frac{\delta_r \phantom{x}}{\delta A^{\imath}},
\frac{\delta_r \phantom{xxx}}{\delta
\bigl(\lambda^{\imath}\theta\bigr)}\right),  
\end{eqnarray}
where in the right-hand side the usual variational derivatives w.r.t.
$A^{\imath}$ and composite
objects $\bigl(\lambda^{\jmath}\theta\bigr)$ are written.
The rank of supermatrix above is determined by one of its
subsupermatrix  for $a=b=0, \theta=0$ with trivial subsupermatrix for
$a=b=1$.

As far as the relations hold
\begin{eqnarray}
\frac{\partial_r \phantom{xxxxxx}}{\partial \left(\partial^r_{\theta}
{\cal A}^{\imath}(\theta)\right)} =
\frac{\delta_r \phantom{x}}{\delta \lambda^{\imath}} \ ,\
\frac{\partial_l \phantom{xxxx}}{\partial (\partial_{\theta}{\cal
A}^{\imath}(\theta))}
= \frac{\delta_l \phantom{x}}{\delta \lambda^{\imath}}
(-1)^{\varepsilon_{\imath} + 1}\,, 
\end{eqnarray}
it is
convenient to introduce the variational component derivatives for
connection with component quantities
\begin{eqnarray}
\frac{\delta_r \phantom{xxx}}{\delta \bigl(\lambda^{\imath}\theta\bigr)} =
\left(\partial^r_{\theta}
\frac{\tilde{\delta}_r \phantom{x}}{\delta
\lambda^{\imath}}\right)(-1)^{\varepsilon_{\imath} + 1}\ , \frac{\delta_l
\phantom{xxx}}{\delta \bigl(\lambda^{\imath}\theta\bigr)} =
\left(\partial_{\theta} \frac{{\tilde{\delta}}_l \phantom{x}}{\delta
\lambda^{\imath}}\right) 
\end{eqnarray}
in view of density ${\cal
F}(\theta)$ dependence  on fields ${\lambda}^{\imath}$ through two arguments
${\cal A}^{\imath}(\theta)$ and $\partial_{\theta}{\cal A}^{\imath}(\theta)$.
The differential operators
$\left(\partial^r_{\theta} \frac{\tilde{\delta}_r
\phantom{x}}{\delta \lambda^{\imath}}\right)$,
$\left(\partial_{\theta} \frac{{\tilde{\delta}}_l\phantom{x}}{
\delta\lambda^{\imath}}\right)$ it should be considered  as the
whole objects.

The connection of  superfield variational derivative on
${\cal A}^{\imath}(\theta)$ with  variational component derivatives w.r.t.
$A^{\imath}$, ${\lambda}^{\imath}$ coincides with proposed in
ref.[25] and follows from (2.34), (6.1), (6.3) and identities (4.6) being
valid on $C^k$
\begin{eqnarray}
\frac{\delta_r
F[{\cal A}]}{\delta{\cal A}^{\imath}(\theta)} & = &
\theta\frac{\delta_r
\bar{F}[A,\lambda]}{\delta A^{\imath} \phantom{xxxx}} + \frac{\delta_r
\bar{F}[A,\lambda]}{\delta \lambda^{\imath}\phantom{xxxx}} (-1) ^{\varepsilon
(F)}\,,\\ 
\left[\partial_{\theta},
\frac{\partial_l \phantom{xxxx}}{\partial (\partial_{\theta}{\cal
A}^{\imath}(\theta))} \right]_s & = & \left[
{\stackrel{\circ}{U}}_{+}(\theta),
\frac{\partial_l \phantom{xxxx}}{\partial (\partial_{\theta}{\cal
A}^{\imath}(\theta))} \right]_s = (-1)^{\varepsilon_{\imath}}
\frac{\partial_l \phantom{xx}}{\partial {\cal
A}^{\imath}(\theta)}\,, 
\end{eqnarray}
with obvious restriction of the last formula to apply in  ${}^{0,0}C^k$.
The differential consequence of the formula (6.5) w.r.t. $\theta$ permits to
express the  superfield
variational derivatives of the form (2.33) superfunctionals  and of
superfunctions  both with
partial superfield derivatives w.r.t. ${\cal A}^{\imath}(\theta)$,
$P_{0}(\theta){\cal A}^{\imath}(\theta)$ and with variational derivatives
w.r.t. $A^{\imath}$
\begin{eqnarray}
\partial_{\theta}\frac{\delta_r F[{\cal
A}]}{\delta{\cal A}^{\imath}(\theta)} & \hspace{-0.5em}= & \hspace{-0.5em}
(-1)^{\varepsilon_{\imath} +
\varepsilon (F)}\frac{\partial F[{\cal A}]}{\partial {\cal A}^{\imath}(
\theta)} =
 (-1)^{\varepsilon_{\imath} + \varepsilon (F)}\displaystyle\frac{\delta
 \bar{F}[A,\lambda]}{ \delta A^{\imath}\phantom{xxxx}}\,, \\ 
\partial_{\theta_1}\frac{\delta_r {\cal F}\bigl(\theta\bigr)}{\delta{\cal
A}^{\imath}(\theta_1)}  & \hspace{-0.5em}= &\hspace{-0.5em}
\partial_{\theta_1}\Bigl(\delta(\theta_1 - \theta) \frac{\partial
{\cal F}(\theta_1)}{\partial {\cal A}^{\imath} (\theta_1)}\Bigr) =
(-1)^{\varepsilon({\cal F}) + \varepsilon_{\imath}}\frac{
\partial {\cal F}(\theta)\phantom{xx}}{\partial
P_0{\cal A}^{\imath} (\theta_1)} \,, 
\end{eqnarray}
with use of the relationships (2.11)
 and trivial identities:
$(P_0(\theta_1),\partial_{\theta_1}){\cal F}(\theta_1) \equiv
(P_0(\theta),\partial_{\theta}){\cal F}(\theta)$
to derive (6.8).

For the operators from {\boldmath${\cal A}_{cl}$} investigated
in Sec.III let us indicate the component expressions  for only basis
operators $\{U_{a}(\theta), {\stackrel{\circ}{U}}_{a}(\theta), a=0, 1\}$
in acting on  $C^{k}$
\begin{eqnarray}
\Bigl(U_0, {\stackrel{\circ}{U}}_0, U_1,
\stackrel{\circ}{U}_1\Bigr)(\theta) =
\left(\lambda^{\imath} \theta \frac{\delta_l \phantom{x}}{\delta A^{\imath}},
(-1)^{\varepsilon_{\imath} + 1}\lambda^{\imath} \frac{\delta_l
\phantom{x}}{\delta A^{\imath}},
  \lambda^{\imath}\theta \left(
\partial_{\theta}\frac{\tilde{\delta}_l \phantom{x}}{\delta \lambda^{\imath}}
\right),
 (-1)^{\varepsilon_{\imath} +
1}\lambda^{\imath} \left(\partial_{\theta}\frac{\tilde{\delta}_l \phantom
{x}}{\delta \lambda^{\imath}}\right)\right). 
\end{eqnarray}

With regard for the formulae above, consider the physically main from the
standard conventional gauge fields theory restriction $\theta=0$, imposed on
the structure of
${\cal M}_{cl}$,
$T_{odd}{\cal M}_{cl}$, classical actions $S_L(\theta)$ (4.1), $S(\theta)$
(4.33),  Euler-Lagrange equations
\renewcommand{\theequation}{\arabic{subsection}.\arabic{equation}\alph{lyter}}
\begin{eqnarray}
\setcounter{lyter}{1}
{} & {} & \hspace{-1em} 1)\ S_L\bigl( {\cal A}(\theta), \partial_{\theta}{\cal
A}(\theta), \theta\bigr)_{\vert\theta=0} = S_L\bigl( A, \lambda,0\bigr)
,\ \ (T_{odd}{\cal M}_{cl})_{\vert\theta=0} =
\{(A^{\imath}, \lambda^{\imath})\}
\,, \\
\setcounter{equation}{10}
\setcounter{lyter}{2}
{} & {} & \hspace{-1em}\phantom{1)}\
S\bigl( {\cal A}(\theta), \theta\bigr)_{
\vert\theta=0} =
{\cal S}(A,0),\  \ {{\cal M}_{cl}}_{\vert\theta=0}= \{\bigl(A^{\imath}\bigr)\}
\,,
\\
\setcounter{equation}{11}
\setcounter{lyter}{1}
{} & {} & \hspace{-1em}2)\ \frac{\delta_l Z[{\cal A} ]}{\delta
{\cal A}^{\imath}(\theta)}_{\vert\theta=0} = -
\frac{\delta_l \bar{Z}[A,\lambda]}{\delta \lambda^{\imath}\phantom{xxxxx}}
(-1)^{\varepsilon_{\imath}}=
- \frac{\delta_l S'_{\theta{}L}\bigl( A, \lambda\bigr)}{\delta
\lambda^{\imath}\phantom{xxxxxx}}(-1)^{\varepsilon_{\imath}}, \\
\setcounter{equation}{11}
\setcounter{lyter}{2}
{} & {} & \hspace{-1em}\phantom{2)\ }
{S},_{\imath}\bigl({\cal A}(\theta), \theta\bigr)_{\vert\theta=0} =
{\cal S},_{\imath}(A, 0) =0\,,
\end{eqnarray}
gauge transformations (4.48), (4.50), their
generators
$\hat{{\cal R}}^{\imath}_{\alpha}
(\theta;{\theta}')$, ${{\cal R}_{0}}^{\imath}_{\alpha}(\theta)$
(for $\xi^{\alpha}(\theta)$ = $\xi^{\alpha}_0 + \xi^{\alpha}_1\theta)$
\begin{eqnarray}
\setcounter{equation}{12}
\setcounter{lyter}{1}
{} & {} & \hspace{-2em}3)\
\delta_g A^{\imath} =
- {\hat{\cal R}_{0}}{}^{\imath}_{\alpha}(A,\lambda,0)\xi^{\alpha}_0
+ (-1)^{\varepsilon_{\imath}}
{\hat{\cal R}_{1}}{}^{\imath}_{\alpha}(A,\lambda,0)\xi^{\alpha}_1\,,
\nonumber \\
{} & {} & \hspace{-2em}\phantom{3)\ }
\delta_g \lambda^{\imath} = - \left({\hat{\cal R}_{0}}{}^{\imath}_{\alpha}(A,
\lambda,0) + {\hat{\cal R}'_{1}}{}^{\imath}_{\theta{}\alpha}(A,\lambda)
\right)\xi^{\alpha}_1 + (-1)^{\varepsilon_{\imath}}
{\hat{\cal R}'_{0}}{}^{\imath}_{\theta{}\alpha}(A,\lambda)
\xi^{\alpha}_0\,, \\
\setcounter{equation}{12}
\setcounter{lyter}{2}
{} & {} &  \hspace{-2em}\phantom{3)\ }
\bigl(\delta A^{\imath}, \delta \lambda^{\imath}\bigr) =
\Bigl({{\cal R}_{0}}{}^{\imath}_{\alpha}(A,0)\xi^{\alpha}_0,\ \;
{{\cal R}_{0}}{}^{\imath}_{\alpha}(A,0)\xi^{\alpha}_1
- (-1)^{\varepsilon_{\imath}} {{\cal R}'_{0}}{}^{\imath}_{\theta{}\alpha}(A)
\xi^{\alpha}_0\Bigr)\,,
\\
\setcounter{equation}{13}
\setcounter{lyter}{1}
{} & {} & \hspace{-2em}4)\ \hat{{\cal R}}^{\imath}_{\alpha}
\bigl({\cal A}(\theta), \partial_{\theta}{\cal A}(\theta),\theta;{\theta}'
\bigr)_{\vert\theta=0} = - \theta'
{\hat{\cal R}_{0}}{}^{\imath}_{\alpha}(A,\lambda,0) +
{\hat{\cal R}_{1}}{}^{\imath}_{\alpha}(A,\lambda,0)\,, \\
\setcounter{equation}{13}
\setcounter{lyter}{2}
{} & {} & \hspace{-2em}\phantom{4)\ }
{{\cal R}_{0}}^{\imath}_{\alpha}({\cal A}(\theta),\theta)_{\vert\theta=0} =
{{\cal R}_{0}}^{\imath}_{\alpha}(A,0)\,, 
\end{eqnarray}
Noether's identities (4.24), (4.30)
\begin{eqnarray}
\setcounter{equation}{14}
\setcounter{lyter}{1}
{} & {} & \hspace{-2em}5)\
\int d\theta \displaystyle\frac{\delta_r Z[{\cal A} ]}{\delta
{\cal A}^{\imath}(\theta)}{}{\hat{{\cal R}}}^{\imath}_{\alpha} \left(\theta;
{\theta}' \right)_{\vert\theta'=0} =
- \frac{\delta_r S'_{\theta{}L}\bigl( A, \lambda\bigr)}{
\delta \lambda^{\imath}\phantom{xxxxxx}}
 \left[{\hat{\cal R}_{0}}{}^{\imath}_{\alpha}(A,\lambda,0) +
{\hat{\cal R}'_{1}}{}^{\imath}_{\theta{}\alpha}(A,\lambda)
\right] \nonumber \\
{} & {} & \hspace{-2em}\phantom{5)\ }
\hspace{1em} + \frac{\delta_r S'_{\theta{}L}(A,\lambda)}{
\delta A^{\imath}\phantom{xxxxx}}
{\hat{\cal R}_{1}}{}^{\imath}_{\alpha}(A,\lambda,0)=0\,, \\
\setcounter{equation}{14}
\setcounter{lyter}{2}
{} & {} & \hspace{-2em}\phantom{5)\ }
S,_{\imath}\bigl({\cal A}(\theta), \theta \bigr) {{\cal
R}}_{0}{}^{\imath}_{\alpha}\bigl( {\cal A}(\theta), \theta\bigr)_{\vert\theta
=0} =
{\cal S},_{\imath}\bigl(A,0\bigr) {{\cal
R}}_{0}{}^{\imath}_{\alpha}\bigl(A, 0\bigr) = 0\,,
\end{eqnarray}
where the notation  was used for arbitrary $f\bigl(
{\cal A}(\theta), \partial_{\theta}{\cal A}(\theta), \theta\bigr)$
\renewcommand{\theequation}{\arabic{subsection}.\arabic{equation}}
\begin{eqnarray}
f'_{\theta}(A,\lambda) \equiv \partial_{\theta} f(\theta)= -
\lambda^{\imath}f,_{\imath}(A,\lambda)(-1)^{\varepsilon_{\imath}
\varepsilon(f)} +
\textstyle\frac{\partial}{\partial\theta}f(\theta)\,.
\end{eqnarray}
Note, the Eqs.(6.11a) in terms of fields $A^{\imath}$,
$\lambda^{\imath}$ have not the form of the usual component Euler-Lagrange
equations for
functional $S_{L}(A,\lambda,0)$ in view of the  different $\theta$-superfield
origin of the arguments $A^{\imath}$, $\lambda^{\imath}$ in $S_L$.

In addition the component form of the identities (4.24) in (6.14a) has the
specific character,  on the one hand showing that for particular case of the
natural system (4.33) and ${\hat{\cal R}_{1}}{}^{\imath}_{\alpha}=0$ this
relation passes into the usual GThST formula (6.14b), and
 on the other hand they represent the extension of the
functional ${\cal S}(A,0)$ invariance w.r.t. only transformations with
$\delta A^{\imath}$ in (6.12b) up to invariance of $\bar{Z}[A,\lambda]$ w.r.t.
transformations (6.12a) with doubled numbers of the generators
${\hat{\cal R}_{k}}{}^{\imath}_{\alpha}(A,\lambda,0)$ and functions
$\xi^{\alpha}_0$, $\xi^{\alpha}_1$ of opposite parities.

At the same time for the special type quantities and relations depending only
upon superfields  ${\cal A}^{\imath}(\theta)$
the component relationships (6.10b)-(6.14b) have the usual
form in view of the purely gauge character of the fields $\lambda^{\imath}$
in this case.

The component form for the transformations (2.18) of $A^{\imath}$,
$\lambda^{\imath}$ w.r.t. $P$ group translations and induced by them the
transformation of superfunction ${\cal F}(\theta)$ (2.21) with taking account
of the rules (4.2), (4.3), (6.15) are defined for $\theta=0$ by the formulae
\begin{eqnarray}
\delta (A^{\imath},\lambda^{\imath}) = (- \lambda^{\imath}\mu,0),\ \;
\delta{\cal F}\bigl( {\cal A}(\theta), \partial_{\theta}{\cal
A}(\theta), \theta\bigr)_{\vert\theta=0} = - \mu{\cal F}'_{\theta}(A,\lambda)
\,. 
\end{eqnarray}

Next, list only a some component expressions for Sec.V objects and relations
devoted to the GAs of GThGT, GThST  and ZLR
problem. So, the superfunction (5.4) being by $\theta$-superfield extension
of BV action and superfunctional (5.3) have the form
\renewcommand{\theequation}{\arabic{subsection}.\arabic{equation}\alph{lyter}}
\begin{eqnarray}
\setcounter{lyter}{1}
{} & {} &S_{(1)}\bigl(\Gamma_{min}(\theta),\theta\bigr)_{\mid\theta =0} =
{\cal S}( A, 0) +
A^{\ast}_{\imath}{\cal R}_{0}{}^{\imath}_{\alpha}(A,0)C^{
\alpha} + O(C^2)\,;  \\ 
\setcounter{equation}{17} \setcounter{lyter}{2} {} & {} &
\bar{Z}_{(1)}[\Gamma_{min}] = S'_{\theta{}L}(A,\lambda) -
A^*_{\imath}\Bigl[{\hat{\cal
R}'_{0}}{}^{\imath}_{\theta{}\alpha}(A,
\lambda)C^{\alpha}(-1)^{\varepsilon_{\imath}} + ({\hat{\cal
R}'_{1}}{}^{\imath}_{\theta{}\alpha}(A,\lambda) + {\hat{\cal
R}_{0}}{}^{\imath}_{\alpha}(A,\lambda,0))\lambda^{\alpha}\Bigr]
\nonumber \\
{} & {} & \phantom{Z_{(1)}[\Gamma_{min}] } -
J_{\imath}\Bigl[{\hat{\cal
R}_{0}}{}^{\imath}_{\alpha}(A,\lambda,0)C^{ \alpha} +
(-1)^{\varepsilon_{\imath}} {\hat{\cal
R}_{1}}{}^{\imath}_{\alpha}(A,\lambda,0)\lambda^{\alpha}\Bigr] +
O(C^{2-l}(\lambda^{\alpha})^l)\,.
\end{eqnarray}
Note, for the simplest case of nondegenerate ThST ($m=0$) for which the only
1st terms in (6.17a,b) survive, so that the extremals ${\cal S},_{\imath}(
A, 0) = 0$ for (6.17a) appear by the 1st class constraints w.r.t.
the 2nd Poisson bracket in (6.18b) for BFV generator $\bar{Z}_{(0)}(A,\lambda)$
= $ \lambda^{\imath}{\cal S},_{\imath}(A, 0)$ $-$ $\textstyle\frac{\partial}{
\partial\theta}S({\cal A}(\theta),\theta)$, revealing the physical
significance of $\lambda^{\imath}$ to be the ghost fields which are different
from $C^{\alpha}, \lambda^{\alpha}$ corresponding to the nontrivial GThST
invariance.

The corresponding $\theta$-local antibracket (5.6) given on the functional
superalgebra of the restricted supermanifold $T^*_{odd}{\cal M}_{min}{}_{
\mid\theta=0}$ $=$ $\{(\phi^B,\phi^*_B)\}$ and superfunctional even bracket
(5.5)
calculated on superfunctionals defined on $T_{odd}(T^*_{odd}{\cal M}_{min})$
with coordinates $\{(\phi^B, \phi^*_B), (\lambda^B, J_B)\}$ forming the flat
phase space structure read as follows with regard for obvious generalization
of  (6.1), (6.3), (6.5) to the case of superfields $\Phi^B(\theta),
\Phi^*_B(\theta)$
\begin{eqnarray}
\setcounter{lyter}{1}
{} &\hspace{-0.5em} {} & \hspace{-0.5em}
({\cal F}(\phi,\phi^*,0), {\cal G}(
\phi, \phi^*,0)) =
\frac{\delta {\cal F}(\phi,\phi^*,0)}{\delta\phi^B\phantom{xxxxxx}}
\frac{\delta_l {\cal G}(\phi,\phi^*,0)}{\delta\phi^{\ast}_{B}\phantom{xxxx
xx}}-
\frac{\delta_r {\cal F}(\phi,\phi^*,0)}{\delta\phi^{\ast}_{B}\phantom{xxx
xxx}}
\frac{\delta_l {\cal G}(\phi,\phi^*,0)}{\delta\phi^B\phantom{xxxxxx}},
\\
\setcounter{equation}{18}
\setcounter{lyter}{2}
{} &\hspace{-0.5em} {} & \hspace{-0.5em}
\bigl\{\bar{F}, \bar{G}\bigr\}  = \bigl\{\bar{F}, \bar{G}\bigr\}^{(
\lambda, \phi^*)} - \bigl\{\bar{F}, \bar{G}\bigr\}^{(J,\phi)} =
\left(\frac{\delta_r \bar{F}}{\delta \lambda^B}
\frac{\delta_l \bar{G}}{\delta \phi^*_{B}} - (-1)^{\varepsilon_A+1}
\frac{\delta_r \bar{F}}{\delta \phi^*_{B}}
\frac{\delta_l \bar{G}}{\delta \lambda^B}\right)\nonumber \\
{} &\hspace{-0.5em} {} & \hspace{-0.5em}\phantom{\bigl\{\bar{F},
\bar{G}\bigr\} } - \left(\frac{\delta \bar{F}}{\delta \phi^B}
\frac{\delta_l \bar{G}}{\delta J_{B}} - (-1)^{\varepsilon_A}
\frac{\delta_r \bar{F}}{\delta J_{B}} \frac{\delta_l
\bar{G}}{\delta \phi^B}\right),\ (\bar{F};\bar{G})=
(\bar{F};\bar{G})[\phi,\lambda,\phi^*,J]\,.
\end{eqnarray}
The last bracket is presented in terms of 2 ones so that the role of
coordinates and momenta play $(\phi^B,\lambda^B)$ and $(J_B,\phi^*_B)$.

It is easy to produce the component form on ${\cal Z}_{Q^0}$
for $\theta$-local even
bracket (5.13) for $\theta =0$ and therefore let us only fulfill it for the
antibracket (5.14) on ${\cal Z}_{Q^1}$ taking the explicit expressions (5.19),
(5.20) for the most important 1st term
 in square and 1st one in round parentheses in (5.20) (being exact for GGTGT
 not  depending upon ${\cal A}^{\imath}, \partial_{\theta}{\cal A}^{\imath}$)
\renewcommand{\theequation}{\arabic{subsection}.\arabic{equation}}
\begin{eqnarray}
{} & {} & \hspace{-1.0em}\bigl(\bar{\tilde{f}\,}[\Gamma],\bar{\tilde{g}}[
\Gamma]\bigr) = (-1)^{\varepsilon_{\imath}}\left(
\frac{\delta_r \bar{F}}{\delta  A^{\ast}_{\imath}}
\left(\frac{\delta_l \phantom {x}}{\delta \lambda^{\imath}}
\frac{\delta_r \bar{Z}}{\delta \lambda^{\jmath}}\right)
\frac{\delta_l \bar{G}}{\delta A^{\ast}_{\jmath}} -
\frac{\delta_r \bar{F}}{\delta  J_{\imath}}
\left(\frac{\delta_l \phantom {x}}{\delta A^{\imath}}
\frac{\delta_r \bar{Z}}{\delta A^{\jmath}}\right)
\frac{\delta_l \bar{G}}{\delta J_{\jmath}}\right)
\nonumber \\
{} & {} & +  \left[\frac{\delta_r \bar{F}}{\delta
A^{\ast}_{\imath}} \left(\frac{\delta_l \phantom {x}}{\delta
\lambda^{\imath}} \frac{\delta_r \bar{Z}}{\delta
A^{\jmath}}\right) \frac{\delta_l \bar{G}}{\delta J_{\jmath}}
(-1)^{\varepsilon_{\imath}+1} + \frac{\delta_r \bar{F}}{\delta
\lambda^{\imath}}\left(
 {\hat{\cal R}'_0}{}^{\imath}_{\theta{}\alpha}
\frac{\delta_l \bar{G}}{\delta J_{\alpha}}(-1)^{\varepsilon_{\imath}}
- \Bigl({\hat{\cal R}'_1}{}^{\imath}_{\theta{}\alpha} +
{\hat{\cal R}_0}{}^{\imath}_{\alpha}\Bigr)
\frac{\delta_l \bar{G}}{\delta C^*_{\alpha}}\right)
\right.
\nonumber \\
{} & {} &   \left.
-\frac{\delta \bar{F}}{\delta A^{\imath}}\left(
 {\hat{\cal R}_0}{}^{\imath}_{\alpha}
\frac{\delta_l \bar{G}}{\delta J_{\alpha}}
-(-1)^{\varepsilon_{\imath}}
 {\hat{\cal R}_1}{}^{\imath}_{\alpha}
\frac{\delta_l \bar{G}}{\delta C^*_{\alpha}}\right)  -
(-1)^{(\varepsilon(F)+1)(\varepsilon(G)+1)}(\bar{F}
\leftrightarrow \bar{G})
 \right]_{\mid{\cal Z}_{Q^1}}.
\end{eqnarray}
Zero locus ${\cal Z}_{Q^1}$ is defined by the equations
\begin{eqnarray}
\left(\hspace{-0.1em}\frac{\delta_l \bar{Z}}{\delta
\lambda^{\imath}}, \frac{\delta_l \bar{Z}}{\delta
A^{\imath}},C^{\alpha}, \lambda^{\alpha}
\hspace{-0.15em}\right)\hspace{-0.1em}=\hspace{-0.1em}
\vec{0}_4,\; \left((-1)^{\varepsilon_{\imath}}A^*_{\imath}\left(
{\hat{\cal R}_0}{}^{\imath}_{\alpha}\hspace{-0.15em} +
\hspace{-0.15em}{\hat{\cal
R}'_1}{}^{\imath}_{\theta{}\alpha}\right)
\hspace{-0.15em}+\hspace{-0.15em} J_{\imath} {\hat{\cal
R}_1}{}^{\imath}_{\alpha}, A^*_{\imath}{\hat{\cal
R}'_0}{}^{\imath}_{\theta{}\alpha}\hspace{-0.15em}+\hspace{-0.15em}
J_{\imath} {\hat{\cal
R}_0}{}^{\imath}_{\alpha}(-1)^{\varepsilon_{\imath}}\right)\hspace{-0.1em}=
\hspace{-0.1em}\vec{0}_2, 
\end{eqnarray}
whose solutions is ambiguous by virtue of the identities (6.14a). In
particular, having considered as $\bar{Z}_{(a)}$ the functional
$\bar{Z}_{(0)}$
from (5.10) we get the first 2 summands in (6.19) vanish together with
${\hat{\cal R}_1}{}^{\imath}_{\alpha}$, ${\hat{\cal R}'_1}{}^{\imath}_{\theta
{}\alpha}$ so, that  we obtain the representation for antibracket
(5.21) $(\ ,\ )^{Z_{(0)}}$  with definition therefore the even bracket (5.13)
as well but in terms of component superfunctionals (5.23).
\subsection{Lagrangian $\theta$-STF Models}
\subsubsection{Massive Complex Scalar Superfield Models}
\setcounter{equation}{0}

Let us choose as the supergroups $\bar{J}, \overline{M}, {\bar{J}}_{
\tilde{A}}$ (2.3) the following Lie groups
\begin{eqnarray}
\bar{J} =
\Pi(1,3)^{\uparrow},\; \overline{M} = T(1,3),\; {\bar{J}}_{\tilde{A}} =
SO(1,3)^{\uparrow}\,,
\end{eqnarray}
to be respectively by proper
Poincare group, group of space-time translations and proper Lorentz group. As
the group ${\bar{J}}_{\tilde{A}}$ one can take  $SL(2,{\bf C})$
being by the universal covering group
for $SO(1,3)^{\uparrow}$ . The corresponding
quotient superspace  has the form
\begin{eqnarray}
{\cal M} = {\bf
R}^{1,3} \times \tilde{P}=\{(x^{\mu}, \theta)\},\,\;{\rm diag}\ \eta_{\mu\nu}
= (1,-1,-1,-1)\,.
\end{eqnarray}

An action of $\Pi(1,3)^{\uparrow} \times P$  has the standard
character of Poincare transformations on ${\bf R}^{1,3}$ (with identical
action of $P$) and on $\tilde{P}$ is given as in (2.7).

Choose as the Lorentz type superfields ${\cal A}^{\imath}(\theta)$ the
complex scalar superfield $\varphi(x,\theta)$ $\in$
$\tilde{\Lambda}_{4\mid 0+1}(x^{\mu},\theta;{\bf C})$
\begin{eqnarray}
{} & {} & \varphi(x,\theta) = \varphi_1(x,\theta) +
\imath\varphi_2(x,\theta)= \varphi(x) +\lambda(x)\theta,\ \varphi_j(x,\theta)
= \varphi_j(x) + \lambda_j(x)\theta\,,  \nonumber \\
{} & {} & \phantom{\varphi(x,\theta) = }
\varphi_j(x,\theta)\in\tilde{\Lambda}_{4\mid 0+1}(x^{\mu},\theta;{
\bf R}),\
j=1,2,\  \imath^2=-1\,. 
\end{eqnarray}
The index $\imath$ condensed contents  (2.12) and vector of
Grassmann gradings $\vec{\varepsilon}$ = $(\varepsilon_P,
\varepsilon_{\Pi}, \varepsilon)$ for $\varphi(x,\theta)$ and its complex
components on $\theta$  are written in the form
\begin{eqnarray}
\imath = ([\varphi],[\overline{\varphi}],x),\ n=(2,0),\
\vec{\varepsilon}\varphi(x) = \vec{\varepsilon}\varphi(x,\theta) =
\vec{\varepsilon}\lambda(x)+(1,0,1)=\vec{0}\,.
\end{eqnarray}
Superfield $\varphi(x,\theta)$ and its $\theta$-component  fields are
transformed in a standard way w.r.t. restriction onto
$\Pi(1,3)^{\uparrow}$ of the supergroup
$J$ $\theta$-superfield representation $T$ as the spin 0 and mass $m$
elements of Poincare group representation [38].  As to restricted
representation $T_{\mid P}$, then only $(\varphi,\overline{\varphi})(
x,\theta)$ are transformed nontrivially
according to the general rule (2.18)
\begin{eqnarray}
\delta(\varphi,\overline{\varphi})(x,\theta) = (\varphi,
\overline{\varphi})'(x,\theta)
- (\varphi, \overline{\varphi})(x,\theta) = -\mu
(\partial_{\theta}{\varphi}, \partial_{\theta}\overline{\varphi})(x,\theta)
 = \mu (\lambda,\overline{\lambda})(x)\,. 
\end{eqnarray}
As the classical action $S_{L}\Bigl(\bigl(\varphi, \overline{\varphi},
\partial_{\theta}{\varphi}, \partial_{\theta}\overline{\varphi}\bigr)(\theta)
\Bigr)$ $\in$ $\Lambda_1(\theta;{\bf R})$ for free
superfields $(\varphi,\overline{\varphi})(x,\theta)$, describing 2
opposite charged spinless massive particles,  let us construct
the superfunction
having the type (4.33) almost natural system
with $g(\theta)= {\rm antidiag}(\nu,\nu)$ ($\vec{\varepsilon}\nu = (1,0,1)$)
in (4.34) and real dimensional in
the units of length (for $\hbar = c =1$) constants $a_{\varphi}, b_{\varphi}$
($[a_{\varphi}]_l$
= $[\varphi]_l +1$ = $[b_{\varphi}]_l$ = $-2$) in order to $T(\theta)$ and
therefore $S_L(\theta)$ would be by dimensionless
\begin{eqnarray}
{} &\hspace{-0.5em} {} & \hspace{-0.5em}
S_L(\theta) \equiv S_L\bigl(\bigl(\varphi,\overline{\varphi},
\partial_{\theta}{\varphi}, \partial_{\theta}\overline{\varphi}
\bigr)(\theta)\bigr)
= T\bigl(\bigl({\varphi}, \overline{\varphi}, \partial_{\theta}{\varphi},
\partial_{\theta}\overline{\varphi}\bigr)(\theta)\bigr) -
S_{0}\bigl(\bigl(\varphi, \overline{\varphi}\bigr)(\theta)\bigr)
\nonumber \\
{} &\hspace{-0.7em} {} & \hspace{-0.7em}\phantom{S_L}
= \int d^4 x \Bigl[\Bigl(-\imath a_{\varphi} \partial_{\theta}
\overline{\varphi}\partial_{\theta}{\varphi}
+ b_{\varphi}\bigr(\partial_{\theta}\overline{\varphi}\nu \varphi  +
\partial_{\theta}{\varphi}\nu\overline{\varphi} \bigr)\Bigr) -
\Bigl(\partial_{\mu}\overline{\varphi}\partial^{\mu}\varphi - m^2
\overline{\varphi}\varphi\Bigr)\Bigr](x,\theta)\,.
\end{eqnarray}
Note, the requirement on $a_{\varphi}, b_{\varphi}, T(\theta)$ above are
naturally
derived from realization of $J$ as direct product of $\bar{J}, P$ and
therefore from condition $[\partial_{\theta}]_l = [\theta]_l =0$ and,
secondly,
the reality of $S_{L}(\theta)$ is provided by the natural continuation
of the complex conjugation from the $P_0(\theta)$ real component fields
$\varphi_j(x)$ (in general $A^{\imath}$) up to the same property fulfilment
for $P_1(\theta)$ component $\lambda_j(x)$ (in general $\lambda^{\imath}$)
written in the form
\begin{eqnarray}
(\lambda_{j}(x))^{\ast} = \lambda_{j}(x),\ \theta^{\ast} = \theta\,.
\end{eqnarray}
Superfunction  $S_{L}(\theta)$ being considered as the more fundamental
object than $Z[\varphi,\overline{\varphi}]$ (see remark after (2.34)) is
invariant w.r.t. Poincare
transformations, but  w.r.t. $P$ group ones (7.5)
is transformed  according to (4.3) as follows  simultaneously with
operator ${\stackrel{\circ}{U}}_{+}(\theta)$ realization
\renewcommand{\theequation}{\arabic{subsection}.\arabic{equation}}
\begin{eqnarray}
\delta
S_L(\theta) & = &  \mu \displaystyle\int d^4x \Biggl[
{\stackrel{\circ}{\varphi}}(x,\theta) P_0(\theta)\frac{\partial
S_L(\theta)\phantom{x}}{\partial \varphi(x,\theta)}
+ {\stackrel{\circ}{\overline{\varphi}}}(x,\theta)P_0(\theta) \frac{\partial_l
S_L(\theta)\phantom{x}}{\partial\overline{\varphi}(x,\theta)}\Biggr]
\nonumber \\
{} &  = &\mu\int d^4x \Bigl[\partial_{\theta}{\varphi}(x,\theta)
\bigl(\Box + m^2\bigr)\overline{\varphi}(x,0) + c.c.
\Bigr],\ \Box=\eta_{\mu\nu}\partial^{\mu}
\partial^{\nu}\,. 
\end{eqnarray}
The invariance of $S_{L}(\theta)$ w.r.t. above transformations is restored
on the solutions for following independent
Euler-Lagrange equation of the form (4.5)
appearing by virtue of (4.8), (4.9) by  one from HCLF
$\Theta_{\varphi}\bigl(\bigl(\overline{\varphi},
\Box\overline{\varphi}\bigr)(x,\theta)\bigr)=0$,
$\Theta_{\overline{\varphi}}
\bigl(\bigl(\varphi, \Box\varphi\bigr)(x,\theta)\bigr)=0$
\begin{eqnarray}
{} \hspace{-1em}\frac{\delta_l Z[\varphi,\overline{\varphi}]}{\delta
\overline{\varphi}(x,\theta)\phantom{x}} =  - \frac{\partial_l S_0(\theta)
\phantom{x}}{\partial
\overline{\varphi}(x,\theta)} - \partial_{\theta} \frac{\partial_l T(\theta)
\phantom{xxx}}{\partial (\partial_{\theta}\overline{\varphi}(x,\theta))} =
\bigl(\imath a_{\varphi}\partial^2_{\theta} +
(\Box +m^2)\bigr)\varphi(x,\theta) = 0\,,
\end{eqnarray}
where we preserve the form for the 1st summand in $T(\theta) \in {\rm Ker}\{
\partial_{\theta}\}$ by fixed that does not influence on
$Z[\varphi,\overline{\varphi}]$ value and therefore on LS (7.9)  structure
as in (4.7) but demonstrate the nondegeneracy of (4.17) in question.

Really, the supermatrix (4.17) has the form in this case
\renewcommand{\theequation}{\arabic{subsection}.\arabic{equation}}
\begin{eqnarray}
K(\theta,x,y)=\left\|\frac{\partial_r \phantom{xxxxx}}{
{\partial(\partial_{\theta}{E}_a(x,\theta))}}\frac{\partial_l
S_L(\theta)\phantom{xxx}}{\partial(\partial_{\theta}{E}_b(y,\theta))}
\right\| =
\imath\left\|
\begin{array}{cc}
0 & a_{\varphi} \\
-a_{\varphi} & 0
\end{array}
\right\|\delta(x-y),\ E_b = (\varphi,\overline{\varphi})\,,
\end{eqnarray}
being by the usual matrix w.r.t. ${\varepsilon}_{\Pi}$ grading and by
the supermatrix  w.r.t. $\varepsilon$ one $K(\theta,x,y)$
with only odd-odd nontrivial block by virtue of (7.4).

Solutions for Eq.(7.9) being by the superfield (on $\theta$) generalization
of the Klein-Gordon equation exist providing the assumption (4.13) fulfilment
in question.
The rank of supermatrix (4.15) is calculated by the rule (4.21) according to
Corollary 2 and  formulae (4.27), (4.36)
\begin{eqnarray}
{\rm rank}\left\|\frac{\partial_r \phantom{xxxxx}}{
\partial{E}_a(x,\theta)}\frac{\partial_l
S_0(\theta)\phantom{x}}{\partial{E}_b(y,\theta)}\right\| = {\rm rank}
\left\|
\begin{array}{cc}
-(\Box + m^2) & 0 \\
0 & -(\Box + m^2)
\end{array}
\right\|\delta(x-y)=2\,. 
\end{eqnarray}
From the last formula being valid almost everywhere in
${\cal M}_{cl}$ it follows that there are not  differential identities among
two HCLF and  the number of physical degrees of freedom is equal to 2.
As the Cauchy problem for the 2nd  order
w.r.t. $x^{\mu}, \theta$ independent part of the LS (7.9)
one can choose the initial conditions
\renewcommand{\theequation}{\arabic{subsection}.\arabic{equation}}
\begin{eqnarray}
\Bigl(\varphi, \partial_0\varphi,
{\stackrel{\circ}{\varphi}},\partial_0{\stackrel{\circ}{\varphi}
}\Bigr)(x,\theta)_{\mid x^0=\theta=0}=
\Bigl(\varphi_0, \varphi_1, \mbox{\boldmath${\stackrel{\circ}{
\varphi}}$}, \mbox{\boldmath$\lambda_1$}\Bigr)(x^i),\
x^{\mu}=(x^0,x^i)\,.
\end{eqnarray}
Therefore according to terminology introduced in Sec.IV, given
$\theta$-STF model belongs to the class of nondegenerate ThSTs.

A generalization of the  model (7.6) onto interacting theory case
 may be realized in terms of the local superfunction, for
instance, by means of addition to  $S_0(\theta)$
at least the cubic  w.r.t. $(\varphi,  \overline{\varphi})(x,\theta)$
polynomial $V(\theta)$  without derivatives on $(x^{\mu},\theta)$  with real
constants $\zeta$, $\eta$
\begin{eqnarray}
S_{0{}M}(\theta)= S_{0}(\theta) - V((\varphi,
\overline{\varphi})(\theta)),\ V(\theta) = \hspace{-0.3em}\int\hspace{-0.3em}
d^4x\Bigl(\frac{\zeta}{3}
\overline{\varphi}\varphi(\overline{\varphi}+\varphi) + \frac{\eta}{2}(
\overline{\varphi}\varphi)^2 + \ldots\Bigr)(x,\theta). 
\end{eqnarray}
The corresponding independent dynamical equation
for $Z_M[\varphi,\overline{\varphi}]$ = $Z[\varphi, \overline{\varphi}]$ +
$\int d\theta V(\theta)$ appears by nonlinear
\begin{eqnarray}
\frac{\delta_l Z_M[\varphi,\overline{\varphi}]}{\delta
\varphi(x,\theta)\phantom{xx}} =
\Bigl(-\imath a_{\varphi}\partial^2_{\theta}+\Box + m^2 + \frac{\zeta}{3}(
\overline{\varphi} + 2\varphi) +
{\eta}(\overline{\varphi}\varphi) + \ldots\Bigr)\overline{\varphi}(x,\theta)
= 0\,, 
\end{eqnarray}
so that, if the $(P_0,P_1)(\theta)$ components $\varphi(x)$,
$\lambda(x)$ for free superfield  had satisfied to the same
Klein-Gordon equation (7.9) respectively, in fact, (formally for $\lambda(x)$)
describing 2 identical of the same name
charged spinless massive particles then component (on $\theta$) equations
following from (7.14) lead to nontrivial interaction for $\overline{\lambda
}(x)$ with fields $\varphi(x)$, $\overline{\varphi}(x)$
\renewcommand{\theequation}{\arabic{subsection}.\arabic{equation}\alph{lyter}}
\begin{eqnarray}
\setcounter{lyter}{1}
{} & {} & (\Box + m^2)\overline{\varphi}(x) = -
\frac{\delta_l
P_0(\theta)V(\theta)}{\delta\varphi(x)\phantom{xxxxx}}\;, \\ 
\setcounter{equation}{15}
\setcounter{lyter}{2}
{} & {} &
(\Box + m^2)\overline{\lambda}(x) = - \displaystyle\int d^4y\Bigl[
\overline{\lambda}(y)\displaystyle\frac{\delta_l\phantom{xxx}}{\delta
\overline{\varphi}(y)} + \lambda(y)\displaystyle\frac{\delta_l\phantom{xxx}}{
\delta\varphi(y)}\Bigr]\displaystyle\frac{\delta_l
P_0(\theta)V(\theta)}{\delta\varphi(x)\phantom{xxxxx}}\,, 
\end{eqnarray}
resulting in different dynamics for the particles corresponding to the fields
$\overline{\varphi}(x)$,  $\overline{\lambda}(x)$. In deriving of
Eqs.(7.15) the formulae (6.5) have been taken into account and
the component
form for ${\stackrel{\circ}{U}}_0(\theta)$ (6.9) in question is defined by
the  last equation.

The requirement of invariance for $S_{0{}M}(\theta)$ w.r.t.
transformations from $U(1)$ group results in restriction $\zeta=0$  in (7.13).
The only relationships (7.8), (7.11) above  are
changed in an obvious way with allowance made for Eq.(7.14) so that the rank
condition (7.11) remains invariable together with classification for given
interacting model as ThST.

The nondegeneracy, for instance, of  free model (7.6) with
BRST similar  charge (5.10)
\renewcommand{\theequation}{\arabic{subsection}.\arabic{equation}}
\begin{eqnarray}
Z_{(0)}[\varphi,\overline{\varphi},\varphi^{\ast},\overline{\varphi}{}^{\ast}]
= \bar{Z}[\varphi,\overline{\varphi}, \lambda,\overline{\lambda}] = -\int d^4x
\Bigl(\overline{\lambda}(x)(\Box + m^2){\varphi}(x) + c.c.\Bigr) 
\end{eqnarray}
and $S_{(1)}\bigl((\varphi,\overline{\varphi},\varphi^{\ast},
\overline{\varphi}{}^{\ast})(\theta)\bigr)$
$\equiv$ $S_{0}\bigl((\varphi,\overline{\varphi})(\theta)\bigr)$ (5.4) leads
to trivial of the form (5.4), (5.10) generating equations given by means of
standard superbrackets (5.6), (5.5) defined in question on the antisymplectic
$T^{\ast}_{odd}{\cal M}_{cl}$ $=$ $\{\bigl((\varphi,\varphi^{\ast})$, $(
\overline{\varphi},\overline{\varphi}{}^{\ast})\bigr)(x,\theta)\}$ $\equiv$
$\{\Gamma^p_{cl}(x,\theta)\}$ and symplectic  supermanifolds $T_{odd}(
T^{\ast}_{odd}{\cal M}_{cl})$ $=$ $\{(\Gamma^p_{cl},\partial_{\theta}
\Gamma^p_{cl})(x,\theta)\}$, $p=\overline{1,4}$ based on the complex
superantifields introduction
\begin{eqnarray}
{} & {} & (\varphi^{\ast},\overline{\varphi}{}^{\ast})(x,\theta) =
(\varphi^{\ast},\overline{\varphi}{}^{\ast})(x) - \theta
(J_{\varphi},J_{\overline{\varphi}})(x) =
(\varphi_1^{\ast},\varphi_1^{\ast})(x,\theta) +
\imath(\varphi_2^{\ast},-{\varphi}{}_2^{\ast})(x,\theta)\,, \nonumber\\
{} & {} & \hspace{1em}{\varphi}{}_j^{\ast}(x,\theta) \in
\tilde{\Lambda}_{4\mid 0+1}(x^{\mu},\theta;{\bf R}),\
\vec{\varepsilon}(\varphi^{\ast},\overline{\varphi}{}^{\ast}) =
\vec{\varepsilon}(J_{\varphi},J_{\overline{\varphi}}) + (1,0,1) = (1,0,1)\,.
\end{eqnarray}
Corresponding ZLR odd and even brackets are given by the particular
$\theta$-superfield relationships (5.22), (5.13) with fulfilment of the
property (5.21) (for $z=(x,\theta)$)
\begin{eqnarray}
{} &\hspace{-1.5em} {} & \hspace{-1.5em}
\bigl(\tilde{f}[\varphi^{\ast},\overline{\varphi}{}^{\ast}],
\tilde{g}[\varphi^{\ast},\overline{\varphi}{}^{\ast}]\bigr)^{Z_{(0)}} = -
\int dz
\Bigl[
\frac{\delta_r \tilde{f}\phantom{xx}}{\delta\varphi^{\ast}(z)}(\Box + m^2)
\frac{\delta_l \tilde{g}\phantom{xx}}{\delta\overline{\varphi}{}^{\ast}(z)} +
\frac{\delta_r \tilde{f}\phantom{xx}}{\delta\overline{\varphi}{}^{\ast}(z)}(
\Box + m^2)
\frac{\delta_l \tilde{g}\phantom{xx}}{\delta{\varphi}{}^{\ast}(z)}\Bigr]
, \nonumber \\
{} & {} & \hspace{0em} {\cal Z}_{\tilde{Q}^1} =
\bigl\{\bigl(
\varphi^{\ast},\overline{\varphi}{}^{\ast},
\partial_{\theta}(\varphi^{\ast},\overline{\varphi}{}^{\ast})\bigr)(z)
\bigr\}
\supset {\cal Z}_{{Q}^0} =
\bigl\{\bigl(\varphi^{\ast},\overline{\varphi}{}^{\ast}\bigr)(z)\bigr\}
\,,\\
{} &\hspace{-1.5em} {} & \hspace{-1.5em}
\bigl\{{f}((\varphi^{\ast},\overline{\varphi}{}^{\ast})(\theta)),
{g}((\varphi^{\ast},\overline{\varphi}{}^{\ast})(\theta))\bigr\}_{
\theta}  =  \hspace{-0.2em}\int\hspace{-0.2em} d^4 x
\hspace{-0.1em}\Bigl( \frac{\partial_r
{f}(\theta)}{\partial\varphi^{\ast}(z)}(\Box + m^2)
\frac{\partial_l
{g}(\theta)}{\partial\overline{\varphi}{}^{\ast}(z)} +
(\overline{\varphi}{}^{\ast} \leftrightarrow {\varphi}^{\ast})
\Bigr), 
\end{eqnarray}
where ${\cal Z}_{\tilde{Q}^1}$, ${\cal Z}_{{Q}^0}$ are determined on the
solution $(\varphi_0, \overline{\varphi}_0)(z)$ of Eqs.(7.9).

Since ${\cal Z}_{{Q}^0}$ is parametrized by only  $\vec{\varepsilon}$-odd
variables being by coordinates in the fiber over
$(\varphi_0, \overline{\varphi}_0)(z)$ in $T^{\ast}_{odd}{\cal M}_{cl}$ then
the corresponding  $Z^{(-1)}((\varphi^{\ast},\overline{\varphi}{}^{\ast})
(\theta))$ is trivial. As the classical physical action given on
${\cal Z}_{\tilde{Q}^1}$ for direct ZLR problem by virtue of $\vec{
\varepsilon}$-vector distribution in (7.17), satisfying to the master
equation with antibracket (7.18), one can choose the superfunction by
introducing of only $\theta^{-1}$ odd time
\begin{eqnarray}
S^{(-1)}\Bigl((J_{\varphi},J_{\overline{\varphi}})(\theta^{-1})\Bigr) =
\int d^4x \Bigl(\partial_{\mu}
J_{\varphi}\partial^{\mu}J_{\overline{\varphi}} -
m^2J_{\varphi}J_{\overline{\varphi}}\Bigr)(x,\theta^{-1})\,.
\end{eqnarray}

Having added the kinetic term $T(\theta^{-1})$ with derivatives on
$\theta^{-1}$ to the quantity above one can obtain the coincidence of
$S^{(-1)}_L(\theta^{-1})$ = $T(\theta^{-1})$ $-$ $S^{(-1)}(\theta^{-1})$ with
$S_L(\theta)$ (7.6) under identification $\theta \leftrightarrow \theta^{-1}$,
$(J_{\varphi},J_{\overline{\varphi}})$ $\leftrightarrow$ $(\varphi, \overline{
\varphi})$, reflecting the fact of the special duality for these
superfunctions in solving the ZLR problem.

The similar construction may be carried out for interacting model
with obvious change of the operator $(\Box + m^2)$, for instance,
for $\overline{\varphi}(z)$ in (7.16)--(7.20) onto $\Bigl[(\Box +
m^2)+ \textstyle\frac{\partial_r
\phantom{xx}}{\partial\overline{\varphi}(z)}
\textstyle\frac{\partial_l V(\theta)}{\partial\varphi(z)}\Bigr]$.

At last, setting $\theta=0$ in $S_L(\theta)$ (7.6) or (7.13) we get
for $\partial_{\theta}\bigl({\varphi}, \overline{\varphi}\bigr)(x,\theta)
= 0$ the standard action
$S_{0}(\varphi,\overline{\varphi})$ for complex scalar fields whereas without
the latter restriction we shall have the additional (of the topological
nature) presence of the background nonpropagating
fields $\lambda_j(x)$ in accordance with (6.10a).
\subsubsection{Massive Spinor Superfield of Spin $\frac{1}{2}$ Models}

In the framework of the Subsec.VI.1 representations for
$J$, ${\cal M}$  choose as
${\cal A}^{\imath}(\theta)$ the Dirac spinor superfields
$\Psi (x,\theta)$ $\in$
$\tilde{\Lambda}_{4\mid 0+1}(x^{\mu},\theta;{\bf C})$
and its Dirac conjugate $\overline{\Psi} (x,\theta)$  in the 4- and
2-component spinor formalisms
\begin{eqnarray}
\setcounter{lyter}{1} {} & {} & \hspace{-2em} \Psi (x,\theta)  =
\bigl(\psi_{\gamma}(x,\theta), \chi^{\dot{\gamma}}(x,\theta)
\bigr)^T,\; \Psi (x,\theta) = \psi(x) + \psi_1(x)\theta,\ \gamma =
1, 2,\; \dot{\gamma} = \dot{1}, \dot{2}\,, \nonumber \\
{} & {} &
\hspace{-2em} \overline{\Psi}(x,\theta) =
\bigl(\overline{\chi}^{\beta}(x,\theta),
\overline{\psi}_{\dot{\beta}}(x,\theta)\bigr),\;
\overline{\Psi}(x,\theta) = \overline{\psi}(x) +
\overline{\psi}_1(x)\theta ,\; \beta = 1, 2,\;
\dot{\beta} = \dot{1}, \dot{2}\,,
\end{eqnarray}
being by elements of $(\frac{1}{2},0)\bigoplus (0,\frac{1}{2})$
Lorentz group reducible  representation. The index $\imath$ condensed
contents  and Grassmann gradings for superfields
$(\Psi,\overline{\Psi})(x,\theta)$ and its $\theta$-component
fields $((\psi,\overline{\psi}), ({\psi}_{1}, \overline{\psi}_1))(x)$ read in
the form
\renewcommand{\theequation}{\arabic{subsection}.\arabic{equation}}
\begin{eqnarray}
\imath = (\gamma, \dot{\gamma}, \beta, \dot{\beta}, x),\ n=(0,8),\
\vec{\varepsilon}(\psi,\overline{\psi}) =
\vec{\varepsilon}(\Psi,\overline{\Psi}) =
 \vec{\varepsilon}(\psi_1,\overline{\psi}_1) + (1,0,1)= (0,1,1).
\end{eqnarray}

\begin{sloppypar}
If  $({\Psi},  \overline{\Psi})(x,\theta)$ and their
$\theta$-components are transformed in the standard
way w.r.t. $\theta$-superfield  representation  $T_{\mid\Pi(1,3)^{\uparrow}}$
as   spin $\frac{1}{2}$ and mass $m$ elements of Poincare group representation
[38], then w.r.t. $T_{\mid P}$ operators the only $({\Psi}, \overline{\Psi})
(x,\theta)$ have the nontrivial  transformation law following from  (2.18)
\begin{eqnarray}
\delta(\Psi, \overline{\Psi})(x,\theta) = (\Psi', \overline{\Psi}')(
x,\theta)
- (\Psi, \overline{\Psi})(x,\theta) = -\mu \partial_{\theta}(
{\Psi}, {\overline{\Psi}}\bigr)(x,\theta)
= - \mu (\psi_1, \overline{\psi}_1)(x)\,. 
\end{eqnarray}
\end{sloppypar}

As the classical actions $S^l_{L}\Bigl(({\Psi},
\overline{\Psi}, \partial_{\theta}({\Psi}, {\overline{
\Psi}}))(\theta)\Bigr)$, $l=1,2$ for free spinor superfields  describing
massive particle and its antiparticle with spin $\frac{1}{2}$, let us
construct the superfunctions consisting only of quadratic w.r.t.
$\bigl({\Psi}, \overline{\Psi}, \partial_{\theta}({\Psi},
{\overline{\Psi}})\bigr)(x,\theta)$ parts without writing of the special
dimensional constants,
one from which appears by the 1st order w.r.t. derivatives on $(x^{\mu},
\theta)$ superfunction  with $g_{\imath \jmath}(\theta)$ = $-(-1)^{
\varepsilon_{\imath}\varepsilon_{\jmath}}g_{\jmath\imath }(\theta)$ in (4.34)
and another with $g_{\jmath\imath }(\theta)=0$
\renewcommand{\theequation}{\arabic{subsection}.\arabic{equation}\alph{lyter}}
\begin{eqnarray}
\setcounter{lyter}{1}
{} \hspace{-1em}S_L^{(1)1}(\theta) &
\hspace{-0.5em}\equiv & \hspace{-0.5em}S_L^1\Bigl((\Psi,
\overline{\Psi}, \partial_{\theta}({\Psi}, {\overline{
\Psi}}))(\theta)\Bigr) = T\Bigl(\partial_{\theta}({\Psi},
{\overline{\Psi}})(\theta)\Bigr) - S_{0}\bigl((
\Psi, \overline{\Psi})(\theta)\bigr) \nonumber \\
{}  & \hspace{-0.5em} = &
\hspace{-0.5em}
\int d^4 x  \Bigl(\partial_{\theta}{\overline{\Psi}}
\partial_{\theta}{\Psi}-
\bigl(\overline{\Psi}\left(\imath
\Gamma^{\mu}\partial_{\mu}- m\right){\Psi}\bigr)
\Bigr)(x,\theta)
 \,, \\ 
\setcounter{equation}{24} \setcounter{lyter}{2} {}
\hspace{-1em}S_L^{(1)2}(\theta) & \hspace{-0.5em}\equiv &
\hspace{-0.5em}S_L^2\Bigl((\Psi, \overline{\Psi},
\partial_{\theta}({\Psi}, {\overline{ \Psi}}))(\theta)\Bigr) =
- \int d^4x \Bigl(\overline{\Psi}\bigl(\imath
\Gamma^{A}\partial_{A}- m\bigr){\Psi}\Bigl)(x,\theta)\,,\\
\setcounter{equation}{24}
\setcounter{lyter}{3}
{} & {} & \Gamma^A=(\Gamma^{\nu}, \hat{\mu}{\bf 1}_{4}),\ A=(\nu,
\theta),\ \partial_{A}=(\partial_{\nu}, \partial_{\theta}),
\vec{\varepsilon}\hat{\mu} = (1,0,1), (\Gamma^{\theta})^{+} =
\Gamma^{\theta}\,,  \nonumber \\
{} & {} & \Gamma^{A}\Gamma^{B} + (-1)^{\varepsilon_A\varepsilon_B}\Gamma^{B}
\Gamma^{A} = 2 \eta^{AB}{\bf 1}_4,\ \eta^{\nu \theta}{\bf 1}_4= \hat{
\mu}\Gamma^{\nu}, \eta^{\theta\theta} = 0\,, 
\end{eqnarray}
where ${\Gamma}^{\mu}$ are  the Dirac matrices in the ref.[39] notations.
The    superfunctions $S^{(1)l}_{L}(\theta)$ are  invariant w.r.t.
Poincare transformations whereas w.r.t. $P$ group ones
$S^{(1)l}_{L}(\theta)$ are transformed according to (4.3) respectively
\begin{eqnarray}
\setcounter{lyter}{1}
\delta
S^{(1)1}_L(\theta) & = &
\mu \displaystyle\int d^4x \Biggl[P_0(\theta)\displaystyle\frac{\partial
S^{(1)}_0(\theta)}{\partial {\Psi}(x, \theta)} {\stackrel{\circ}{\Psi}}(x,
\theta)
+ {\stackrel{\circ}{\overline{\Psi}}} (x,\theta)P_0(\theta) \frac{\partial_l
S^{(1)}_0(\theta)}{\partial \overline{\Psi}(x, \theta)}\Biggr]
\nonumber \\
{} & = & \mu \displaystyle\int d^4x \Bigl[\overline{\Psi}(x,0) \bigl(\imath
\Gamma^{\mu}\partial_{\mu} - m\bigr){\stackrel{\circ}{\Psi}} (x,\theta) +
{\stackrel{\circ}{\overline{\Psi}}}(x,\theta) \bigl(\imath
\Gamma^{\mu}\partial_{\mu} - m\bigr){\Psi}(x,0)\Bigr], \\ 
\setcounter{equation}{25} \setcounter{lyter}{2} \delta
S^{(1)2}_L(\theta) & = & \delta S^{(1)1}_L(\theta) - \mu\int d^4x
\Bigl(\partial_{\theta}{\overline{\Psi}} \imath\hat{\mu}
\partial_{\theta}{\Psi}\Bigr)(x,\theta)\,, 
\end{eqnarray}
with writing in (7.25a) the expression for operator
${\stackrel{\circ}{U}}_+(\theta)$.

The reality of $S^{(1)l}(\theta)$
follows from the complex conjugation relations for the case of
bispinors having the form (without writing of arguments)
\renewcommand{\theequation}{\arabic{subsection}.\arabic{equation}}
\begin{eqnarray}
{} & {} & \hspace{-2em}
\Bigl({\overline{\Psi}}_1
{\Psi}_2,  \overline{\Psi}_1
{\stackrel{\circ}{\Psi}}_2,
{\stackrel{\circ}{\overline{\Psi}}}_1{\stackrel{\circ}{{\Psi}}}_2,
{\stackrel{\ \ \circ}{\overline{\theta\Psi}}},
\hat{\mu},
{\overline{\Psi}} \imath\Gamma^{\theta}{\stackrel{\circ}{\Psi}}\Bigr)^{\ast} =
\Bigl(\overline{\Psi}_2 {\Psi}_1,
{\stackrel{\circ}{\overline{\Psi}}}_2{\Psi}_1,
 {\stackrel{\circ}{\overline{\Psi}}}_2{\stackrel{\circ}{{\Psi}}}_1,
\theta{\stackrel{\circ}{\Psi}},
 \hat{\mu},
- {\stackrel{\circ}{\overline{\Psi}}} \imath\Gamma^{\theta}{\Psi}\Bigr)\,,
\end{eqnarray}
with accuracy up to total derivative w.r.t. $\theta$ in
$S^{(1)2}_L(\theta)$ and w.r.t. $x^{\mu}$ for $S^{(1)}_0(\theta)$.

Euler-Lagrange equations (4.5) have the form, with the same
comments on $T^{(1)}(\theta)$ role  as after Eq.(7.9),  in terms
of the corresponding superfunctionals $Z^l[\Psi,\overline{\Psi}]$
= $\int d\theta S^{(1)l}_L(\theta)$
\renewcommand{\theequation}{\arabic{subsection}.\arabic{equation}\alph{lyter}}
\begin{eqnarray}
\setcounter{lyter}{1}
{} \hspace{-1em}\frac{\delta_l Z^1[\Psi,\overline{\Psi}]}{\delta
\Psi(x,\theta)\phantom{x}} & = & - \frac{\partial_l S^{(1)}_0(\theta)}{
\partial
\Psi(x,\theta)} + \partial_{\theta}\frac{\partial_l T^{(1)}(\theta)
\phantom{xx}}{\partial (\partial_{\theta}{\Psi}(x,\theta))} = -\Bigl(
\imath\partial_{\mu} \overline{\Psi}\Gamma^{\mu} + m\overline{\Psi}-
\partial_{\theta}^2{ \overline{\Psi}}\Bigr)(x,\theta) = 0\,,
\\
\setcounter{equation}{27}
\setcounter{lyter}{2}
{} \hspace{-1em}\frac{\delta_l
Z^1[\Psi,\overline{\Psi}]}{\delta \overline{\Psi}(x,\theta)\phantom{x}} & = & -
\frac{\partial_l S^{(1)}_0(\theta)}{\partial \overline{\Psi}(x,\theta)} +
\partial_{\theta}\frac{\partial_l T^{(1)}(\theta) \phantom{x}}{\partial(
\partial_{\theta}{ \overline{\Psi}}(x,\theta))} =
\Bigl(\partial_{\theta}^2 - \bigl(\imath
\Gamma^{\mu}\partial_{\mu} - m\bigr)\Bigr)\Psi (x,\theta) = 0\,,\\ 
\setcounter{equation}{28}
\setcounter{lyter}{1}
{} \hspace{-1em}\frac{\delta_l Z^2[\Psi,\overline{\Psi}]}{\delta
\Psi(x,\theta)\phantom{x}
} & = & \frac{\partial_l S^{(1)2}_L(\theta)\phantom{}}{\partial
\Psi(x,\theta)} + \partial_{\theta}\frac{\partial_l S^{(1)2}_L(\theta)
}{\partial(\partial_{\theta}{\Psi}(x,\theta))} = -\bigl(
\imath\partial_{A} \overline{\Psi}\Gamma^{A} + m\overline{\Psi}
\bigr)(x,\theta) = 0\,, \\
\setcounter{equation}{28}
\setcounter{lyter}{2}
{} \hspace{-1em}\frac{\delta_l
Z^2[\Psi,\overline{\Psi}]}{\delta \overline{\Psi}(x,\theta)\phantom{x}} &
= &
\frac{\partial_l S^{(1)2}_L(\theta)}{\partial \overline{\Psi}(x,\theta)}
= - \bigl(\imath
\Gamma^{A}\partial_{A} - m\bigr){\Psi} (x,\theta) = 0\,, 
\end{eqnarray}
and represent, by virtue of (4.8), (4.9) the HCLF
$\Theta^1\bigl(\bigl({\Psi}, \overline{\Psi},
{\partial}_{\mu}{\Psi}, {\partial}_{\mu}{\overline{\Psi}})\bigr(x,\theta)
\bigr) = 0$ and DCLF(!)
$\Theta^2\bigl(\bigl({\Psi}, \overline{\Psi},
{\partial}_{A}{\Psi}, {\partial}_{A}{\overline{\Psi}})\bigr(x,\theta)
\bigr) = 0$,  containing 2 equations in terms of
bispinors, respectively.

The   2nd derivatives
of $S^{(1)1}_{L}(\theta)$ w.r.t $\partial_{\theta}{\Psi}(x,\theta)$,
$\partial_{\theta}{\overline{\Psi}}(x,\theta)$
supermatrix (4.17) has the
following block form in 4-component spinor formalism
\renewcommand{\theequation}{\arabic{subsection}.\arabic{equation}}
\begin{eqnarray}
K^{(1)}(\theta,x,y)  =  \left\|
\frac{\partial_r \phantom{xxxx}}{ \partial
(\partial_{\theta}E_a(x,\theta))}\frac{\partial_l
T^{(1)}(\theta)\phantom{x}}{\partial (\partial_{\theta}E_b(y,\theta))}
\right\| = \left\|
\begin{array}{ll}
{\mbox{\large{\bf 0}}}_4 & {\mbox{\large{\bf 1}}}_4 \\
{\mbox{\large{\bf 1}}}_4 & {\mbox{\large{\bf 0}}}_4
\end{array}
\right\| \delta(x - y),\; (E_1, E_2)=(\Psi, {\overline{\Psi}}), 
\end{eqnarray}
appearing by the usual nondegenerate matrix
w.r.t. $\varepsilon$ grading and by the supermatrix with only odd-odd
nonvanishing block for the case of $\varepsilon_{\Pi}$ parity. The
 corresponding  supermatrix for $S^{(1)2}_L(\theta)$  will be
trivial $(n = {\stackrel{\circ}{m}})$ complicating the analysis of DCLF (7.28)
in comparison with HCLF (7.27).

Solutions for Eqs.(7.27), (7.28) (for instance, trivial) exist, providing
the fulfilment of assumption (4.13) in question.
Since the Eqs.(7.27) are the HCLF, then in view  of Corollary 2
and  formula (4.36) validity the
rank (4.15b) for $S^{(1)1}_L(\theta)$ is calculated by the
rule (4.21)
\begin{eqnarray}
\left\| \frac{\partial_r \phantom{xxx}}{ \partial
E_a(x,\theta)}\frac{\partial_l
S^{(1)1}_L(\theta)\phantom{}}{\partial E_b(y,\theta)} \right\| =
\left\|
\begin{array}{cc}
{\mbox{\large{\bf 0}}}_4 &
- \bigl(\imath \Gamma^{\mu}\partial_{\mu} - m
{\mbox{\large{\bf 1}}}_4\bigr) \\
-\bigl(\imath \Gamma^{\mu}\partial_{\mu} + m{\mbox{\large{\bf 1}}}_4
\bigr) & {\mbox{\large{\bf 0}}}_4
\end{array} \right\| \delta(x - y)\,.  
\end{eqnarray}
It is well known fact that on mass-shell  $\Sigma^{(1)}_{\Psi}$
the ranks of the $4 \times 4$ matrices $\bigl(\imath
\Gamma^{\mu}\partial_{\mu} - m {\mbox{\large{\bf 1}}}_4\bigr)$ and
$\bigl(\imath \Gamma^{\mu}\partial_{\mu} + m {\mbox{\large{\bf
1}}}_4\bigr)$ are equal to 2 so that supermatrix (7.30) contains
only odd-odd block with rank being equal to $4$ in terms of Weyl
spinor's components. From the  equality
$\dim{\Sigma^{(1)}_{\Psi}}=0$  it follows  that $m=0$ in (4.14b)
and  therefore there are not differential identities among
Eqs.(7.27).

Although the same conclusion can be derived from the investigation
results for the  model with $S^{(1)2}_L(\theta)$, the structure of
Eqs.(7.28) as DCLF leads to the definition for given theory status
from the general grounds developed in Sec.IV. So, the rank of the
supermatrix (4.15b) for $Z^2[\Psi,\overline{\Psi}]$  may be
calculated by the rule (4.20) with regard for notations in (7.29)
in the form ($z_a=(x_a,\theta_a), a=1,2$)
\begin{eqnarray}
\left\|\frac{\delta_r \phantom{xx}}{
\delta {E}_a(z_1)}\frac{\delta_l Z^2[\Psi,\overline{\Psi}]
}{\delta {E}_b(z_2)\phantom{x}}\right\| =
\left\|
\begin{array}{cc}
{\mbox{\large{\bf 0}}}_4 & -
 \bigl(\imath \Gamma^{A}\partial_{A} - m
{\mbox{\large{\bf 1}}}_4\bigr) \\ - \bigl(\imath
\Gamma^{A}\partial_{A} + m{\mbox{\large{\bf 1}}}_4 \bigr) &
{\mbox{\large{\bf 0}}}_4
\end{array} \right\|\delta(z_1 - z_2)\,.
\end{eqnarray}
The rank value for supermatrix (7.31) on mass-shell $\Sigma^{(2)}_{\Psi}$ for
Eqs.(7.28) coincides with rank for supermatrix (7.30) on $\Sigma^{(
1)}_{\Psi}$ (i.e. equal to $4$) in view of nilpotent character for term with
$\hat{\mu}{\bf 1}_4$, therefore not affecting on rank value.
As in the previous case the supermatrix (7.31) appear by nondegenerate in
any neighbourhood of $\Sigma^{(2)}_{\Psi}$ reflecting the fact of the type
(4.24) differential identities absence for this model.

By  independent initial conditions for the 1st order w.r.t. derivatives on
$x^{\mu}$ and the 2nd order w.r.t. derivative on $\theta$
($\partial^2_{\theta}{\Psi}(x,\theta)=0$
, $\partial^2_{\theta}{\overline{\Psi}}(x,\theta)=0$) partial
differential equations (HCLF) (7.27) and
for the DCLF (7.28) being by the 1st order w.r.t. derivatives on
$(x^{\mu}, \theta)$ superfield differential equations one can choose the
expressions respectively
\begin{eqnarray}
\bigl(\Psi, \partial_{\theta}{\Psi}\bigr)(x,\theta)_{\mid x^0=\theta=0} =
\bigl({\bf \Psi},{\bf {\stackrel{\circ}{\Psi}}}\bigr)(x^i),\
\Psi(x,\theta)_{\mid x^0=\theta=0} =
\underline{\bf \Psi}(x^i)\,. 
\end{eqnarray}
In the framework of Sec.IV terminology the $\theta$-superfield models
described by the actions $S_L^{(1)1}(\theta)$ (7.24a) and
$S_L^{(1)2}(\theta)$ (7.24b) belong to the classes of
nondegenerate ThSTs and nondegenerate ThGTs with vanishing supermatrix (4.17)
respectively. Really, given theories for
$\partial_{\theta}{\Psi}(x,\theta) = \partial_{\theta}{\overline{\Psi}}(x,
\theta) = \theta = 0$ have
2 second-class constraints, in terms of Dirac spinors leading to survival
of only 4 physical degrees of freedom in terms of Weyl spinor's components
for every model.

The interacting $\theta$-superfield spinor ThST and ThGT may be constructed
in the framework of local theory
by means of addition to $S_L^{(1)l}(\theta)$ (7.24a,b) at least quadratic
combinations
w.r.t. product $(\overline{\Psi}\Psi)(x,\theta)$ without derivatives
w.r.t. $(x^{\mu},\theta)$
\begin{eqnarray}
S_{L{}M}^{(1)l}(\theta) & = & S_L^{(1)l}(\theta) +
V(\bigl(\Psi, \overline{\Psi}\bigr)(\theta)),\ V^{(1)}(\theta)\equiv
V(\bigl(\Psi, \overline{\Psi}\bigr)(\theta))\nonumber \\
{} & = &  \int d^4x\Bigl[
\frac{\lambda_1}{2}(\overline{\Psi}\Psi)^2 +
\frac{\lambda_2}{2}(\overline{\Psi}\Gamma^{\mu}\Psi)(
\overline{\Psi}\Gamma_{\mu}\Psi)\Bigr](x,\theta),\
\lambda_1,\lambda_2\in {\bf R}\,, 
\end{eqnarray}
including the $\theta$-superfield  generalization of the
Fermi interaction term (polynomial at $\lambda_1$).

Corresponding nonlinear independent Euler-Lagrange equations have the form
for $Z_M^l[\Psi,\overline{\Psi}]$ = $Z^l[\Psi,\overline{\Psi}]$ +
$\int d\theta V^{(1)}(\theta)$
\begin{eqnarray}
\frac{\delta_l Z_M^l[\Psi,\overline{\Psi}]}{\delta
\overline{\Psi}(x,\theta)\phantom{xx}} = -
\Bigl(\Bigl[\imath
\partial_{\mu}\Gamma^{\mu} + \delta_{l2}\imath\hat{\mu}\partial_{\theta}
- m -\delta_{l1}\partial^2_{\theta} -
\lambda_1(\overline{\Psi}{}\Psi) -
\lambda_2(\overline{\Psi}\Gamma^{\mu}\Psi)\Gamma_{\mu}\Bigr]{
\Psi}\Bigr)(x,\theta) = 0. 
\end{eqnarray}
While in view of (4.8), (4.9)
the linear HCLF (7.27) describe 2 pairs of the opposite
charged particles (electrons $e^-$ and positrons $e^+$) corresponding to
$\psi(x)$ and formally to $\psi_1(x)$, the nonlinear HCLF for $l=1$
in (7.34) contains the following $P_a(\theta)$  components
\renewcommand{\theequation}{\arabic{subsection}.\arabic{equation}\alph{lyter}}
\begin{eqnarray}
\setcounter{lyter}{1}
{} & {} & \hspace{-3.0em}
- [\imath\partial_{\mu}\Gamma^{\mu} - m]{\psi}(x) = -
\displaystyle\frac{\delta_l P_0 V^{(1)}(\theta)}{\delta\overline{
\psi}(x)\phantom{xxxx}}\;,
\\ 
\setcounter{equation}{35}
\setcounter{lyter}{2}
{} & {} & \hspace{-3.0em}
- [\imath\partial_{\mu}\Gamma^{\mu} - m]{\psi}_1(x) = - \hspace{-0.3em}
\displaystyle\int \hspace{-0.3em}d^4y
\Bigr[\overline{\psi}_1(y)\displaystyle\frac{\delta_l\phantom{xxx}}{\delta
\overline{\psi}(y)}\displaystyle\frac{\delta_l P_0 V^{(1)}(\theta)}{
\delta\overline{\psi}(x)\phantom{xxxx}} +
\displaystyle\frac{\delta_l\phantom{xxx}}{\delta{\psi}(y)}\displaystyle\frac{
\delta_l P_0 V^{(1)}(\theta)}{\delta\overline{\psi}(x)\phantom{xxxx}}
{\psi}_1(y)\Bigr], 
\end{eqnarray}
with simultaneous definition of the component form for
${\stackrel{\circ}{U}}_0(\theta)$ (6.9).
One can consider that  Eq.(7.35b) for
${\psi}_1(x)$  is given in an external field being determined by
a solution of the Eq.(7.35a) for ordinary spinor $\psi(x)$.

The structure of linear DCLF (7.28) is more difficult in view of  the
superfields
$\partial_{\theta}\bigl({\Psi}, {\overline{\Psi}}
\bigr)(x,\theta)$  nontrivial occurrence which complicates the
$P_0(\theta)$  component  of Eqs.(7.34) for $l=2$ in comparison with HCLF.
On the other hand the $P_1(\theta)$  component  of Eqs.(7.34) has the form
(7.35b).
The relationships (7.25), (7.30), (7.31) are changed taking
(7.33), (7.34) into account in an evident way for interacting models.

It is not difficult to repeat here the all computations for ZLR and its direct
problem made for preceding model in (7.16)--(7.20).
Let us demonstrate a some moments, for only free ThGT with $S_L^{(1)2}(
\theta)$
(7.24b) with taking notations (7.25) into account. So, the superfunctional
\renewcommand{\theequation}{\arabic{subsection}.\arabic{equation}}
\begin{eqnarray}
Z_{(1)}[\Psi, \overline{\Psi}, \Psi^{\ast}, \overline{\Psi}{}^{\ast}]
= Z^2[\Psi, \overline{\Psi}] = \textstyle\frac{\partial_l}{\partial\mu}
\delta S_L^{(1)2}(\theta)\,
\end{eqnarray}
and superfunction $S_{(1)}\Bigl((\Psi, \overline{\Psi}, \Psi^{\ast},
\overline{\Psi}{}^{\ast})(\theta)\Bigr)$ $\equiv$ $S_0\Bigl((\Psi,
\overline{\Psi})(\theta)\Bigr)$ lead to the trivial generating equations (5.3)
(for $\Gamma^{\theta} = 0$ to the Eqs.(5.10)), (5.4) defined on the symplectic
$T_{odd}(T^{\ast}_{odd}{\cal M}_{cl})$ $=$ $\{(\Gamma_{cl},\partial_{\theta}
\Gamma_{cl})(z)\}$ and antisymplectic $T^{\ast}_{odd}{\cal M}_{cl}$ $=$ $\{
\bigl((\Psi,\Psi^{\ast})$, $(\overline{\Psi},\overline{\Psi}{}^{\ast})\bigr)
(z)\}$ $\equiv$ $\{\Gamma_{cl}(z)\}$ supermanifolds, where we have introduced
the spinor superantifields (for $z=(x,\theta)$)
\begin{eqnarray}
(\Psi^{\ast}, \overline{\Psi}{}^{\ast})(z)
= (\psi^{\ast}, \overline{\psi}{}^{\ast})(x) - \theta (J_{\Psi},
J_{\overline{\Psi}})(x),\ \;\vec{\varepsilon}(\Psi^{\ast}, \overline{\Psi}{
}^{\ast})=
\vec{\varepsilon}(J_{\Psi}, J_{\overline{\Psi}}) +(1,0,1) = (1,1,0).
\end{eqnarray}
Let us only find here the more general than in (7.18) form for antibracket
(5.14) on ${\cal Z}_{Q^1}$ $=$ $\{(\Psi^{\ast}, \overline{\Psi}{}^{\ast}$,
$\partial_{\theta}(\Psi^{\ast}, \overline{\Psi}{}^{\ast}))(z)\}$
defined by ${\bf Q}^1$ $=$ $\{Z^2[\Psi,\overline{\Psi}],\ \;\}$ over a some
solution $(\Psi_0,\overline{\Psi}_0)(z)$ for Eqs.(7.28) (for $Z_{(0)}$ =
$Z^1[\Psi,\overline{\Psi}]$)
\begin{eqnarray}
{} &\hspace{-1em} {} &\hspace{-1em}
\bigl(\tilde{f}[\Psi^{\ast},\overline{\Psi}{}^{\ast}],
\tilde{g}[\Psi^{\ast},\overline{\Psi}{}^{\ast}]\bigr) = \int dz \Bigl[
\frac{\delta_r \tilde{f}\phantom{xx}}{\delta\overline{\Psi}{}^{
\ast}(z)}(\imath\Gamma^A\partial_A - m)
\frac{\delta_l \tilde{g}\phantom{xx}
}{\delta{\Psi}{}^{\ast}(z)} - (-1)^{(\varepsilon(f)+1)(
\varepsilon(g)+1)}(\tilde{f}\leftrightarrow \tilde{g})\Bigr] \nonumber \\
{} &\hspace{-1em} {} & \hspace{-0.5em} = \int dz\Bigl[
\frac{\delta_r \tilde{f}\phantom{xx}}{\delta\overline{\Psi}{}^{\ast}(z)}
\imath\mu\partial_{
\theta}\frac{\delta_l \tilde{g}\phantom{xx}}{\delta{\Psi}{}^{\ast}(z)}
- (-1)^{(\varepsilon(f)+1)(
\varepsilon(g)+1)}(\tilde{f}\leftrightarrow \tilde{g})\Bigr] +
\bigl(\tilde{f}[\Psi^{\ast},\overline{\Psi}{}^{\ast}],
\tilde{g}[\Psi^{\ast},\overline{\Psi}{}^{\ast}]\bigr)^{Z_{(0)}}
\hspace{-0.2em}. 
\end{eqnarray}

Again the corresponding BRST charge $Z^{(-1)}(\theta)$ on ${\cal Z}_{Q^0}$
$=$ $\{(\Psi^{\ast}, \overline{\Psi}{}^{\ast})\}$ is trivial whereas the
superfunction $S^{(-1)2}((J_{\Psi}$, $J_{\overline{\Psi}})(\theta^{-1}))$
being analogous to (7.20) continued up to $S^{(-1)2}_L(\theta^{-1})$ exists
\begin{eqnarray}
S^{(-1)2}_L\bigl((J_{\Psi}, J_{\overline{\Psi}},
\partial_{\theta^{-1}}(J_{\Psi}, J_{\overline{\Psi}}))(
\theta^{-1})\bigr) = -\int d^4x\bigl(
J_{\Psi}(\imath\Gamma^A\partial_A -
m)J_{\overline{\Psi}}\bigr)(x,\theta^{-1})
\,,
\end{eqnarray}
satisfying to the master equation with antibracket (7.38) and being dual to
$S^{(1)2}_L(\theta)$ (7.24b).
\subsubsection{Free Vector Superfield Models}

Setting for supergroups and quotient space (2.3)
\begin{eqnarray}
\bar{J}&\hspace{-0.5em} =
& \hspace{-0.5em}\Pi(1,D-1)^{\uparrow},\; \overline{M} = T(1,D-1),\;
{\bar{J}}_{\tilde{A}} = SO(1,D-1)^{\uparrow} ,\;D\geq 2,\; D \in
\mbox{\boldmath$N$}\,, \\
{\cal M}&\hspace{-0.5em} =&\hspace{-0.5em}
{\bf R}^{1,D-1} \times \tilde{P}=\{(x^{\mu}, \theta)\},\;{\rm
 diag}\ \eta_{\mu\nu} = (1,-1,\ldots,-1),\;\mu,\nu = 0,1,\ldots,D-1\,,
\end{eqnarray}
we consider as ${\cal A}^{\imath}(\theta)$  the real vector superfield
${\cal A}^{\mu}(x,\theta)$
\begin{eqnarray}
{\cal A}^{\mu}(x,\theta) = {A}^{\mu}(x) + {A}_1^{\mu}(x)\theta,\
{\cal A}^{\mu}(x,\theta)\in \tilde{\Lambda}_{D\mid 0+1}(x^{\mu},
\theta;{\bf R})\,,
\end{eqnarray}
being by element of $\Pi(1,D-1)^{\uparrow}$ group massless irrep
space and encoding $n=n_{+}=D$ real degrees of freedom. The index $\imath$
contents,  Grassmann vector values for quantities above and the type (7.7)
obvious properties of  conjugation  read as follows
\begin{eqnarray}
\imath = (\mu,x),\
 \vec{\varepsilon}({A}^{\mu}(x), {\cal A}^{\mu}(x,\theta)) =
\vec{\varepsilon} ({A}_1^{\mu}(x)) + (1,0,1) = \vec{0},\
\bigl(\partial_{\theta}{\cal A}^{\mu}\bigr)^{\ast}(x,\theta) =
\partial_{\theta}{\cal A}^{\mu}(x,\theta)\,.
\end{eqnarray}
The superfields $\bigl({\cal
A}^{\mu}, \partial_{\theta}{\cal A}^{\mu}\bigr)(x,\theta)$ are
transformed, in a standard way, as Lorentz vectors
w.r.t.  $T_{\mid\Pi(1,D-1)^{\uparrow}}$ $\theta$-superfield  representation
whereas the only ${\cal A}^{\mu}(x,\theta)$ have nontrivial transformation
law w.r.t. $T_{\mid P}$ action of the form (2.18)
\begin{eqnarray}
\delta{\cal A}^{\mu}(x,\theta) = {{\cal
A}'}^{\mu}(x,\theta) -{\cal A}^{\mu}(x,\theta)= - \mu
\partial_{\theta}{\cal A}^{\mu}(x,\theta)= \mu{A}_{1}^{\mu}(x)\,.
\end{eqnarray}
As the classical $\Lambda_1(\theta, {\bf R})$-valued   action
$S_{L}\Bigl( ({\cal A}^{\mu},
\partial_{\theta}{\cal A}^{\mu})(\theta) \Bigr)$ $\equiv$
$S_{L}^{(2)}(\theta)$ for free vector superfield, describing massless particle
(helicity $\lambda = \pm 1$ for $D=4$), choose the local
superfunction in the natural system form  with $g_{\imath
\jmath}(\theta)=0$ in (4.34)  not explicitly depending upon $\theta$,
without dimensional constants as in (7.6) and with antisymmetric
$\varepsilon_{\mu\nu}$
\begin{eqnarray}
 S_{L}^{(2)}(\theta) =
T\bigl(\partial_{\theta}{\cal A}^{\mu}(\theta)\bigr) - S_{0}\bigl({
\cal  A}^{\mu}(\theta)\bigr) =
\int\hspace{-0.2em} d^D x\Bigl(
\frac{1}{2}\varepsilon_{\mu\nu} \partial_{\theta}{\cal
A}^{\nu}\partial_{\theta}{\cal A}^{\mu}- \bigl(
- \frac{1}{4}F_{\mu\nu}F^{\mu\nu}\bigr)\Bigr)(x,\theta)
\,. 
\end{eqnarray}
The transformation law for $\Pi(1,D-1)^{\uparrow}$-scalar
$S_{L}^{(2)}(\theta)$ w.r.t.  $T_{\mid P}$ action has the form in agreement
with (4.3) realizing among them the superfield structure for operator
${\stackrel{\circ}{U}}_+(\theta)$
\begin{eqnarray}
\delta S_L^{(2)}(\theta) = \mu\int d^Dx
{\stackrel{\ \circ}{\cal
A}}{}^{\nu}(x,\theta)P_0(\theta) \frac{\partial_l S_{0}\bigl( {{\cal
A}}^{\mu}(\theta)\bigr)}{\partial {{\cal A}}^{\nu}(x,\theta)\phantom{xx}} =
\mu\int d^Dx {\stackrel{\ \circ}{\cal
A}}{}^{\nu}(x,\theta) \partial^{\mu}F_{\mu\nu}(x,0)\,. 
\end{eqnarray}
The invariance of $S_L^{(2)}(\theta)$ w.r.t. $T_{\mid P}$ are restored on the
Euler-Lagrange equations
\begin{eqnarray}
\frac{\delta_l Z[{\cal A}^{\mu}] \phantom{x}}{\delta {{\cal
A}}^{\nu}(x,\theta)}  =  - \frac{\partial_l S_{0}\bigl( {{\cal
 A}}^{\mu}(\theta)\bigr)}{\partial {{\cal A}}^{\nu} (x,\theta)\phantom{x}} -
\partial_{\theta}\frac{\partial_l T\bigl(\partial_{\theta}{\cal
A}^{\mu}(\theta)\bigr)}{ {\partial(\partial_{\theta}{\cal
A}^{\nu}(x,\theta))}\phantom{x}} =
(\varepsilon_{\nu\mu} \partial^2_{\theta}{\cal A}^{\mu}
- \partial^{\mu}F_{\mu\nu}\bigr)(x,\theta) = 0\,,
\end{eqnarray}
containing in view of identical fulfilment $\partial^2_{\theta}{\cal
A}^{\mu} = 0$  the HCLF ${\Theta}_{\nu}\bigl({\cal
A}^{\mu}(x,\theta)\bigr)=0$ being by the $2$nd order on $(x^{\mu}, \theta)$
$D$ linear homogeneous partial differential equations.

The supermatrix (4.12) for $S_{L}^{(2)}(\theta)$ has the form in question
\begin{eqnarray}
K^{(2)}(\theta,x,y)=\left\|\frac{\partial_r \phantom{xxxxxx}}{
{\partial(\partial_{\theta}{\cal A}^{\mu}(x,\theta))}} \frac{\partial_l
S_L\bigl({\cal A}^{\rho}(\theta), \partial_{\theta}{\cal A}^{\rho
}(\theta)\bigr) }{ {\partial(\partial_{\theta}{\cal
A}^{\nu}(y,\theta))}\phantom{xxxxxxx}}\right\| = \left\|
\varepsilon_{\mu\nu}\right\|\delta(x - y)\,,
\end{eqnarray}
being by the usual matrix w.r.t. ${\varepsilon}_{\Pi}$ grading and by the
supermatrix with only nontrivial odd-odd block w.r.t. $\varepsilon$ parity.
Rank of
$K^{(2)}(\theta,x,y)$ depends on values of $D=\dim{{\bf R}^{1,D-1}}$.  So, for
odd $D$, this supermatrix is always degenerate
$({\stackrel{\circ}{m}} > 0)$ in view
of skew-symmetry  for ${\varepsilon}_{\mu \nu}$ whereas the choice
for ${\varepsilon}_{\mu \nu}$ for even $D$, for instance,   in the form
\begin{eqnarray}
\left\|\varepsilon_{\mu\nu}\right\| = {\rm antidiag}(-{1}_k, {1}_k),\
 D = 2k, \; k \in \mbox{\boldmath$N$}\,,  
\end{eqnarray}
yields the nondegenerate $K^{(2)}(\theta,x,y)$.

The solutions for Eqs.(7.47) exist, providing the fulfilment of assumption
(4.13) in question. In its turn the rank of supermatrix (4.15)
for given model is equal to
\begin{eqnarray}
{\rm rank}\left\|\frac{\partial_r \phantom{xxxxx}}{\partial {\cal
A}^{\mu}(x,\theta)} \frac{\partial_l  S_0\bigl({\cal
A}^{\rho}(\theta)\bigr)}{\partial {\cal
A}^{\nu}(y,\theta)\phantom{xx}}\right\|_{\mid\Sigma_{\cal A}} = {\rm
rank}\left\|(\Box \eta_{\mu\nu} -
\partial_{\mu}\partial_{\nu})\right\|_{\mid\Sigma_{\cal A}} \delta(x - y)
= D-1\,,
\end{eqnarray}
being always strictly less than $n$ in the
whole ${\cal M}_{cl}$ = $\{{\cal A}^{\mu}(x,\theta)\}$.

Therefore there is only one ($m=1$) differential identity among Eqs.(7.47)
with  standard choice for linear independent generator compatible with the
conclusions from Corollary 1
\begin{eqnarray}
{\cal R}^{\mu}(x,y) =
\partial^{\mu}\delta(x - y),\;\alpha = y \,.
\end{eqnarray}
Given $\theta$-STF model is the GThST with GGTST above in the Sec.IV
terminology.
GTST being invariance transformation for only $S_0\bigl({\cal A}^{\mu}(
\theta)\bigr)$ has the form
of standard gradient transformation
$\delta{\cal A}^{\mu}(x,\theta)$ = $\partial^{\mu}{\xi}(x,\theta)$
with $\vec{\varepsilon}$-boson arbitrary
superfield ${\xi}(x,\theta)$.
As consequence, not all from the following initial conditions for LS (7.47)
are independent
\begin{eqnarray}
\Bigl({\cal A}^{\mu}, \partial_0{\cal A}^{\mu},
\partial_{\theta}{\cal A}^{\mu},\partial_0\partial_{\theta}{
\cal A}^{\mu}\Bigr)(x,\theta)_{\mid x^0=\theta=0}=
\Bigl(\mbox{\boldmath${\cal A}$}^{\mu}, \mbox{\boldmath${\cal A}$}^{\mu}_1,
\mbox{\boldmath${\stackrel{\ \circ}{
{\cal A}}}$}{}^{\mu}, \mbox{\boldmath$\lambda$}_1^{\mu}\Bigr)(x^i),\
i = \overline{1,D-1}\,.
\end{eqnarray}

As the another example of a vector model
consider the theory of free complex massive vector superfield  for
arbitrary $D\geq 2$ (for $D=4$ describing two mass $m$ charged particles of
spin 1). In this case the configuration space ${\cal M}_{cl}$
coordinatized by $\bigl({\cal A}^{\mu}, \overline{\cal A}{}^{\mu}\bigr)(x,
\theta)$ $\in$ $\tilde{\Lambda}_{D\mid 0+1}(x^{\mu},\theta;{\bf C})$
describing $2D$ real degree of freedom
\begin{eqnarray}
{\cal A}^{\mu}(x,\theta)& = &{\cal A}^{\mu}_1(x,\theta) +
\imath{\cal A}^{\mu}_2(x,\theta)= A^{\mu}(x) +  \lambda^{\mu}(x)\theta,\
{\cal A}^{\mu}_j(x,\theta)
=  A^{\mu}_j(x) + \lambda^{\mu}_j(x)\theta\,, \nonumber \\
{} & {} &  {\cal A}^{\mu}_j(x,\theta)
\in\tilde{\Lambda}_{D\mid 0+1}(x^{\mu},\theta;{\bf R}),\
j=1,2\,. 
\end{eqnarray}
The condensed index $\imath$ contents  is extended up to
 $\imath  = (\mu, \nu, x)$ $\mapsto$ $(
[{\cal A}^{\mu}], [\overline{\cal A}{}^{\nu}], x)$ in comparison with
(7.43) whereas the Grassmann
gradings appear by the same but for complex (super)fields.
The properties (7.7) for  scalar
$P_1(\theta)$-component fields  remain valid for  $\lambda^{\mu}_j(x)$.

The  $\Lambda_1(\theta, {\bf R})$-valued  superfunction
\begin{eqnarray}
 S_{L{}m}^{(2)}(\theta) &\hspace{-0.6em} = &\hspace{-0.6em}
S_{L{}m}^{(2)}\Bigl(\bigl({\cal  A}^{\mu}, \overline{\cal  A}{}^{\mu},
\partial_{\theta}({\cal A}^{\mu}, {\overline{\cal A}}{
}^{\mu})\bigr)(\theta)\Bigr) =
 T\bigl(\partial_{\theta}({\cal A}^{\mu},
{\overline{\cal A}}{}^{\mu})(\theta)\bigr) -
S_{0{}m}\bigl(\bigl({\cal  A}^{\mu},\overline{\cal A}{}^{\mu}\bigr)(
\theta)\bigr)  \nonumber \\
{} &\hspace{-0.6em} = &\hspace{-0.6em}\int d^D x\Bigl(
\bigl(\varepsilon_{\mu\nu}
\partial_{\theta}{\overline{
\cal A}}{}^{\nu}\partial_{\theta}{\cal A}^{\mu} -
\imath \partial_{\theta}{\overline{\cal A}}{}^{\mu}\partial_{\theta}
{\cal A}_{\mu}\bigr)-
\Bigl(-\frac{1}{2}
\overline{F}_{\mu\nu}F^{\mu\nu} + m^2\overline{\cal A}_{\mu}{\cal A}^{
\mu}\Bigr)\Bigr)(x,\theta)
\end{eqnarray}
may be chosen as the classical action leading to the $2$nd order w.r.t.
$(x^{\mu}, \theta)$ complex linear partial differential equations (LS)
written by means of superfunctional $Z_m[{\cal A}^{\mu},\overline{\cal A}{}^{
\mu}]$ = $\int d\theta S_{L{}m}^{(2)}(\theta)$
\begin{eqnarray}
\frac{\delta_l Z_m[{\cal A}^{\mu},\overline{\cal A}{}^{\mu}]}{\delta
\overline{\cal A}{}^{\nu}(x,\theta)\phantom{xx}}  =
\bigl(\varepsilon_{\nu\mu}  + \imath
\eta_{\nu\mu}\bigr)\partial^2_{\theta}{\cal A}^{\mu}(x,\theta)
- (\partial^{\mu}F_{\mu\nu} + m^2{\cal A}_{\nu})(x,\theta) =0 \,,
\end{eqnarray}
providing the property for  $S_{L{}m}^{(2)}(\theta)$ to be integral w.r.t.
$T_{\mid P}$ global transformations with simultaneous realization of
${\stackrel{\circ}{U}}_+(\theta)$ for this model
\begin{eqnarray}
\delta S_{L{}m}^{(2)}(\theta) &
= & \mu\int d^Dx \Biggr[
{\stackrel{\ \circ}{\cal
A}}{}^{\nu}(x,\theta)P_0(\theta) \frac{\partial_l \phantom{xxxx}}{\partial
{{\cal A}}^{\nu}(x,\theta)} +
{\stackrel{\ \circ}{\overline{\cal
A}}}{}^{\nu}(x,\theta)P_0(\theta) \frac{\partial_l \phantom{xxxx}}{\partial {
\overline{\cal A}}{}^{\nu}(x,\theta)}\Biggr] S_{0{}m}(\theta) \nonumber \\
{} &= {} &
 \mu\Bigl[\partial_{\theta}{\cal
A}^{\nu}(x,\theta) (\partial^{\mu}\overline{F}_{\mu\nu} + m^2\overline{
\cal A}{}_{\nu})(x,0) + (c.c.)\Bigl]\,. 
\end{eqnarray}
As the independent initial conditions for complex LS (7.55) one can take the
complexified Cauchy problem (7.52). Really, in first, the solution for
HCLF in (7.55) exists, in second, the supermatrix (4.12) in question
for $E_b\in\{{\cal A}, \overline{\cal A}\}$
\begin{eqnarray}
K^{(2)}_m(\theta,x,y)=\left\|
\frac{\partial_r \phantom{xxxxxx}}{
{\partial(\partial_{\theta}{E}_a^{\mu}(x,\theta))}}
\frac{\partial_l
S_{L{}m}^{(2)}(\theta)\phantom{xx}}{{\partial(
\partial_{\theta}{E}_b^{\nu}(y,\theta))}}
\right\| =
\left\|
\begin{array}{cc}
0 & \varepsilon_{\mu\nu} + \imath\eta_{\mu\nu}\\
\varepsilon_{\mu\nu} - \imath\eta_{\mu\nu} & 0
\end{array}\right\|\delta(x - y)\,, 
\end{eqnarray}
may be chosen  by nondegenerate for any $D$ and, in third, the rank of
supermatrix (4.15) is calculated by the rule (4.21) in the form, being
differed from the double value for rank of the previous vector model
supermatrix (7.50),
\begin{eqnarray}
\left\|\frac{\partial_r \phantom{xxxxx}}{
\partial E_a^{\mu}(x,\theta)}
\frac{\partial_l  S_{0{}m}(\theta)}{\partial E_b^{\nu}(y,\theta)}
\right\| =
\left\|
\begin{array}{cc}
0 & (\Box + m^2)\eta_{\mu\nu} - \partial_{\mu}\partial_{\nu} \\
(\Box + m^2)\eta_{\mu\nu} - \partial_{\mu}\partial_{\nu} & 0
\end{array}\right\|
\delta(x - y)\,, 
\end{eqnarray}
and equal to $2D$ almost everywhere in ${\cal M}_{cl}$ in view of
zero-dimensionality of the mass-shell
$\Sigma_{{\cal A}{}m}$ $(m=0)$ in question.

In contrast to massless case the model has 6 physical degrees of freedom for
$D=4$ and appears by singular [39] nondegenerate (i.e. nongauge) ThST
because of the 2 second class constraints presence in applying
of Dirac-Bergmann algorithm.

The dynamical equations  (7.47), (7.55) appear by the same for both their
$P_a(\theta)$ components $a=0,1$,
thus describing formally the similar dynamics  for
corresponding  to the fields  $A^{\mu}(x)$, $A_1^{\mu}(x)$ particles.

The generalization of the massive complex vector model to interacting theory
appears by evident as it was made for the examples with
scalar and spinor superfields.

Since the ZLR brackets and objects construction for superfunction
(7.54), in fact, repeats the scalar superfield properties then
we consider only the
analogouos problem for GThST with $S_L^{(2)}(\theta)$ (7.45). Corresponding
exact $\theta$-superfield BRST charge as in (5.10) and, in fact, BV action
(5.4) have the form in $T_{odd}(T^{\ast}_{odd}{\cal M}_{min})$ $=$ $\{(
\Gamma^p_{min}, \partial_{\theta}\Gamma^p_{min})(z)\}$, $p=\overline{1,2(4+1)
}$ and
$T^{\ast}_{odd}{\cal M}_{min}$ $=$ $\{({\cal A}^{\mu},C)$, $({\cal A}^{
\ast}_{\mu},C^{\ast})(z)\}$ $\equiv$ $\{\Gamma^p_{min}(z)\}$ respectively
\begin{eqnarray}
{}&\hspace{-1em} {} & \hspace{-1em} Z_{(0)}[\Gamma_{min}] = -
\partial_{\theta} S_{(1)}(\Gamma_{min}(\theta)) =
\hspace{-0.2em}\int \hspace{-0.2em}d^Dx
\Bigl(A_1^{\nu}\partial^{\mu}F_{\mu\nu} + J_{\mu}\partial^{
\mu}C + A^{\ast}_{\mu}\partial^{\mu}\lambda\Bigr)(x),\\
{}&\hspace{-1em} {} & \hspace{-1em}
S_{(1)}(\Gamma_{min}(\theta))  =  \int d^Dx \left(-\textstyle\frac{1}{4}
F_{\mu\nu}F^{\mu\nu} + {\cal A}_{\mu}^{\ast}\partial^{\mu}C\right)(x,\theta).
\end{eqnarray}
The above quantities satisfy to the generating equations (5.10),
(5.4) with simple superbrackets (5.5), (5.6). Whereas the
corresponding new even bracket (5.13)  on ${\cal Z}_{Q^0}$ ($\dim
{\cal Z}_{Q^0} = (2,3)$) and odd (5.22) on ${\cal
Z}_{\tilde{Q}^1}$ ($\dim {\cal Z}_{\tilde{Q}^1} = (2+3,3+2)$) are
written as follows (for $z=(x^{\mu},\theta)$)
\begin{eqnarray}
{} &\hspace{-1.2em} {} & \hspace{-1.2em}
\{f, g\}_{\theta} \hspace{-0.2em}= \hspace{-0.2em}\int
\hspace{-0.2em}d^Dx\Bigl[\hspace{-0.1em}-
\frac{\partial_r {\cal F}(\theta)}{\partial{\cal A}_{\nu}^{\ast}(z)}
(\Box \eta_{\mu\nu}\hspace{-0.1em} -\hspace{-0.1em} \partial_{\mu}
\partial_{\nu})
\frac{\partial_l {\cal J}(\theta)}{\partial{\cal
A}_{\mu}^{\ast}(z)} \hspace{-0.1em}+ \hspace{-0.1em}
\Bigl(\frac{\partial {\cal F}(\theta)}{\partial{\cal A}^{\mu}(z)}
\partial_{\mu}
\frac{\partial {\cal J}(\theta)}{\partial C^{\ast}(z)}\hspace{-0.1em}
+\hspace{-0.1em} (C^{\ast} \hspace{-0.25em}
\leftrightarrow \hspace{-0.25em}
{\cal A}^{\mu})\hspace{-0.1em}\Bigr)\hspace{-0.1em}
\Bigr]_{\mid{\cal Z}_{Q^0}}\hspace{-0.3em}, \\ 
{} &\hspace{-1.2em} {} &\hspace{-1.2em}
\bigl(\tilde{f}, \tilde{g}\bigr) \hspace{-0.2em}=
\hspace{-0.2em}
\hspace{-0.2em}\int\hspace{-0.2em} dz \Bigl[\hspace{-0.1em}
\frac{\delta_r F[\Gamma]}{\delta{\cal A}_{\nu}^{\ast}(z)}
(\Box \eta_{\mu\nu}\hspace{-0.1em} - \hspace{-0.1em}
\partial_{\mu}\partial_{\nu})
\frac{\delta_l G[\Gamma]}{\delta{\cal A}_{\mu}^{\ast}(z)}\hspace{-0.1em}
- \hspace{-0.1em}\Bigl(\frac{\delta F[\Gamma]}{\delta{\cal A}^{\mu}(z)}
\partial_{\mu}
\frac{\delta G[\Gamma]}{\delta C^{\ast}(z)}\hspace{-0.1em}
+\hspace{-0.1em} (C^{\ast}\hspace{-0.2em} \leftrightarrow \hspace{-0.2em}
{\cal A}^{\mu})\Bigr)\hspace{
-0.1em}\Bigr]_{\mid{\cal Z}_{\tilde{Q}^1}}\hspace{-0.2em}, 
\end{eqnarray}
where ${\cal Z}_{Q^0}$ may be parametrized by $C^{\ast}(z)$, 3 antifields
$\tilde{\cal A}{}_{k}^{\ast}(z)$ from ${\cal A}_{\mu}^{\ast}(z)$ (determined
by equation $\partial^{\mu}{\cal A}_{\mu}^{\ast}(z)$ = $0$) and 1 from
superfield ${\cal A}^{\mu}(z)$ not being conjugate w.r.t. initial antibracket
to
$\tilde{\cal A}{}_{k}^{\ast}(z)$. In turn, the ${\cal Z}_{\tilde{Q}^1}$
therefore may be coordinatized by the variables above and their derivatives on
$\theta$.

In view of   ${\cal Z}_{{Q}^0}$ structure the new BRST charge $Z^{(-1)}(
\theta)$ vanishes again.
The new dual classical action representing the GThST on
${\cal Z}_{\tilde{Q}^1}$ may be
determined by the formula with introduction of new $\theta^{-1}$
\begin{eqnarray}
S_{L}^{(-1)2}\Bigl(\partial_{\theta}({\cal A}_{\mu}^{\ast},
\partial_{\theta^{-1}}{\cal A}_{\mu}^{\ast})(\theta^{-1})\Bigr) =
\int\hspace{-0.2em} d^D x\Bigl(
\frac{1}{2}\varepsilon^{\mu\nu} \partial_{\theta}\partial_{\theta^{-1}}{\cal
A}_{\nu}^{\ast}\partial_{\theta}\partial_{\theta^{-1}}{\cal A}_{\mu}^{\ast}
+ \frac{1}{4}F^{\ast}_{\mu\nu}F^{\mu\nu{}{\ast}}
\Bigr)(x,\theta^{-1}), 
\end{eqnarray}
where the superantifield strength $
F^{\ast}_{\mu\nu}(x,\theta^{-1})$ = $[\partial_{\mu},
\partial_{\theta}{\cal A}_{\nu}^{\ast} (z,\theta^{-1})]$ is invariant
w.r.t. standard gradient transformation for $\partial_{\theta}{\cal A}_{
\mu}^{\ast} (z,\theta^{-1})$ with arbitrary $\xi(z,\theta^{-1})$.
\subsubsection{$\theta$-superfield $U(1\vert 0)\times U(0\vert 1)$
Abelian GThGT Model}

As the initial model we choose the any from the models above with complex
superfields for $D=4$ admitting the realization of the global
transformations generated by two-parametric supergroup (without sum on $a$ in
(7.64))
\begin{eqnarray}
{U}^{1\vert 1} & = &U(1\vert 0)\times U(0\vert 1),\ U(\delta_{a0}\vert
\delta_{a1}) = \{\exp\{\imath  \xi^a e^a\}\vert\, e^0,\xi^0 \in {\bf R},\,
e^1, \xi^1 \in {}^{1}\Lambda_1(\theta;{\bf R})\},\nonumber \\
{} & {} & (e^a,\xi^a)^{\ast}= (e^a, \xi^a),\
\vec{\varepsilon}(e^a,\xi^a) =
(a,0,a), a=0,1\,,
\end{eqnarray}
for composite superfields $(\Phi, \overline{\Phi})$ $\in$ $\{
(\varphi, \Psi, {\cal A}^{\mu})$, $(\overline{\varphi}, \overline{\Psi},
\overline{\cal A}^{\mu})\}$ in the form
\begin{eqnarray}
(\Phi, \overline{\Phi})(x,\theta)  \mapsto  (\Phi',
\overline{\Phi}{}')(x,\theta)=(\exp\{\imath  \xi^ae^a\}\Phi,
\overline{\Phi}\exp\{-\imath \xi^a e^a\})(x,\theta) = (g \Phi,
\overline{\Phi} g^{\ast})(x,\theta)\,. 
\end{eqnarray}
The set of transformations (7.65) leaves the actions (7.6), (7.24a,b), (7.54)
by invariant.

The realization of the Yang-Mills type gauge principle [40], for
instance, for the case of spinor superfields is based on the
change of superparameters $\xi^a$ onto arbitrary superfields
$\xi^a(x,\theta)$ in such a way that the resultant action
$S_{LG}(\theta)$, which we shall seek now, must be invariant
w.r.t. GTGT for all the superfields parametrizing the new extended
configuration space ${\cal M}_{cl}$ $(n=(n_+,n_-)$ = $(4 + 1, 8+ 4
+1)$ = $([{\cal A}^{\mu 0}]+ [C^1_{(0)}], [\Psi,\overline{\Psi}] +
[{\cal A}^{\mu 1}]+ [C^0_{(0)}]))$:
\begin{eqnarray}
{} & {} &
(\Psi,\overline{\Psi}, {\cal A}^{A a})(x,\theta)
\mapsto \hspace{-0.4em} (\Psi',
\overline{\Psi}{}', {\cal A}'{}^{A a})(x,\theta) = \bigl(g\Psi,
\overline{\Psi}g^*, {\cal A}^{A a} + \partial^{A}\xi^a\bigr)(x,\theta)\,,
\\ 
{} & {} & {\cal A}^{A a}(x,\theta)  =
 A^{A a}(x) + A_1^{A a}(x)\theta = ({\cal A}^{\mu a}, C^a_{(0)})(x,\theta)
 \in  \tilde{\Lambda}_{4\vert 0+1}(x^{\mu},\theta;{\bf R})\,,\\ 
{} & {} & \vec{\varepsilon}({\cal A}^{A a}(x,\theta),A^{A a}(x)) =
 \vec{\varepsilon}(A_1^{A a}(x)) + (1,0,1) = ((\varepsilon_P)_A +a, 0,
a + \varepsilon_A)\,. 
\end{eqnarray}
In contrast to the models above the $\varepsilon_P$ Grassmann parity spectrum
is nontrivial even for the case of standard $U(1\vert 0)\equiv U(1)$
transformations in view of ghost superfield $C_{(0)}(x,\theta) \equiv
C^0_{(0)}(x,\theta)$
inclusion (being differed from the role of $C(x,\theta)$ in Sec.V) on
the initial level of the model formulation.

Adapting the general form of infinitesimal GTGT (4.48) and GGTGT (4.25)
one can write their realization together with specification
of the index $\imath, \alpha$
contents as follows
\begin{eqnarray}
\delta_g{\cal A}^{\imath}(\theta) &\hspace{-0.5em} = & \hspace{-0.5em}
\int d\theta'
\hat{\cal R}^{\imath}_b({\cal A}(\theta),\theta;\theta')\delta\xi^b(\theta')
= \displaystyle\int d\theta'd^4y
\hat{\cal R}^{\tilde{\imath}}_b({\cal A}(x,\theta),x,\theta;y,\theta')
\delta\xi^b(y,\theta'), \,\nonumber \\
{\cal A}^{\imath}(\theta) &\hspace{-0.5em} = &\hspace{-0.5em}\bigl({\cal
A}^{A a}, \overline{\Psi}, {\Psi}\bigr)(x,\theta),\ \imath = ((A,a),
\beta, \dot{\beta}, \gamma, \dot{\gamma},x) = (\tilde{\imath}, x),\;
\alpha = (b,y)\,,\\
{} & {} & \hspace{-3.5em}
\hat{\cal R}^{\tilde{\imath}}_b({\cal A}(x,\theta),x,\theta;y,\theta') =
\sum\limits_{k\geq 0} \left(\left(\partial_{\theta}
\right)^k\delta(\theta - \theta')\right)
\hat{\cal R}_k^{\tilde{\imath}}{}_b({\cal A}(x,\theta))\delta(x-y)\,,
\nonumber \\
{} & {} & \hspace{-3.5em}
\hat{\cal R}_0^{\tilde{\imath}}{}_b({\cal A}(x,\theta)) = \left\{
\begin{array}{ll}
- \partial^{\mu}_{x}\delta^a{}_b,& \hspace{-0.3em} \tilde{\imath}=(\mu, a) \\
\imath e^b \overline{\Psi}(x,\theta), & \hspace{-0.3em}
\tilde{\imath}=(\beta,\dot{\beta}) \\
- \imath e^b {\Psi}(x,\theta), & \hspace{-0.3em}
\tilde{\imath}=(\gamma,\dot{\gamma},x)
\end{array}\right.\hspace{-0.7em}, \nonumber \\
{} & {} & \hspace{-3.5em}
\hat{\cal R}_1^{\tilde{\imath}}{}_b({\cal A}(x,\theta))= \delta^a{}_b, \tilde{
\imath} =
(\theta, a),\;
(\varepsilon_P)_{\alpha}=\varepsilon_{\alpha} = b,\  m=(m_+,m_-) = (1,1)\,.
\end{eqnarray}
The easily obtained only trivial solution  for superfunctions
$u^b\bigl( {\cal A}(\theta), \partial_{\theta}{\cal A}(\theta), \theta\bigr)$
in Eq.(4.26) implies the set of  GGTGT above is the  functionally
independent and forms the gauge
algebra of GTGT with abelian gauge supergroup $U^{1\vert 1}$.

To construct  the classical action, realizing the minimal inclusion of
interaction for spinor superfields by means of connectedness coefficients
${\cal A}^{A a}(x,\theta)$, let us consider the not Lorentz type covariant
derivatives
\begin{eqnarray}
{\cal D}^A \equiv \partial^A - \imath  {\cal A}^{A
a}(x,\theta)e^a =
({\cal D}^{\mu}, {\cal D}^{\theta})\,,
\end{eqnarray}
whose supercommutator permit to obtain the GTGT-invariant $\theta$-superfield
strength in the almost standard manner
\begin{eqnarray}
{\cal F}_{AB}{}^a(x,\theta) & =&  {\imath}\frac{d_r}{d {e}^a}[{\cal D}_A,{\cal
D}_B]_s = \bigl(\partial_A{\cal A}_{B}^a - (-1)^{\varepsilon_A\varepsilon_B}
\partial_B{\cal A}_{A}^a\bigr)(x,\theta)\,, \nonumber \\
{\cal A}^{A a}(x,\theta) & = & \hat{\eta}^{AB}{\cal A}_B^{a}(x,\theta), \
\hat{\eta}^{AB} =  {\rm diag}(1, -1, -1, -1\vert 1) = (\eta^{\mu\nu},
\hat{\eta}^{\theta\theta})\,, \\
{\cal F}_{AB}{}^a(x,\theta) & = &
\left\|
\begin{array}{lr}
F_{\mu\nu}{}^a & F_{\mu \theta}{}^a\\
F_{\theta\nu}{}^a & F_{\theta\theta}{}^a
\end{array} \right\|(x,\theta) = \left\|
\begin{array}{cc}
\partial_{[\mu}{\cal A}_{\nu]}^a &
\partial_{\mu}C^a_{(0)} - \partial_{\theta}{\cal A}_{\mu}^a\\
\partial_{\theta}{\cal A}_{\nu}^a - \partial_{\nu}C^a_{(0)}  &
2\partial_{\theta}{C}^a_{(0)}
\end{array} \right\|(x,\theta) \nonumber \\
{} & = &
 -(-1)^{\varepsilon_A\varepsilon_B}{\cal F}_{BA}{}^a(x,\theta),\ \;
(A,B)= ((\mu, \theta),(\nu, \theta))\,. 
\end{eqnarray}
In view of special structure for gauge algebra the following
$\Lambda_1(x^{\mu}, \theta;{\bf R})$-valued quadratic w.r.t.
${\cal F}_{AB}{}^a(x,\theta)$ superfunctions possess by the properties of
Poincare and GTGT invariances
\begin{eqnarray}
{}\hspace{-1.5em} 1) &  {} &
({\cal F}_{AB}{}^a{\cal F}^{AB a})(x,\theta)= \left(F_{\mu\nu}{}^0F^{\mu\nu 0}
+ 2 F_{\mu \theta}{}^1 F^{\mu \theta 1} + 4\bigl(\partial_{\theta}{C}^0_{(0)}
\bigr)^2\right)(x,\theta), \nonumber \\
{}\hspace{-1.5em}  &  {} & \hspace{2em}
F_{\mu \theta}{}^1 F^{\mu \theta 1}(x,\theta) = \Bigl(\partial_{\mu
}C^1_{(0)}\partial^{\mu}C^1_{(0)} + \partial_{\theta}{\cal A}^1_{\mu}
\partial_{\theta}{\cal A}^{\mu 1} -
2(\partial_{\theta}{\cal A}^{\mu 1})\partial_{\mu}{C}^1_{(0)}\Bigr)(x,
\theta), \\
{} \hspace{-1.5em} 2) & \hspace{-1.5em}{} & \hspace{-1.5em}
\varepsilon_{ABCD}({\cal F}^{AB a}{\cal F}^{CD a})(x,\theta) = \left(
\varepsilon_{\mu\nu\rho\sigma}F^{\mu\nu 0}F^{\rho\sigma 0} +
2\varepsilon_{\mu\nu \theta\theta}\bigl(F^{\mu\nu 0}F^{\theta\theta 0} -
2F^{\mu\theta 0}F^{\nu\theta 0}\bigr)\right. \nonumber \\
{}\hspace{-1.5em}
& {} & \left.\hspace {3em}+ \varepsilon_{\theta\theta\theta\theta}(
F^{\theta\theta 0})^2 +
 4(\varepsilon_{\mu\nu\rho \theta}F^{\mu\nu a}F^{\rho \theta a} +
\varepsilon_{\mu \theta\theta\theta}F^{\mu \theta a}F^{\theta\theta a})
\right)(x,\theta)\,,\\
{} \hspace{-1.5em}3) & \hspace{-1.5em}{} &\hspace{-1.5em}
\bar{\varepsilon}_{ABCD}({\cal F}^{AB a}{\cal F}^{CD b}\varepsilon_{ab})(x,
\theta) = 4\left\{(
\bar{\varepsilon}_{\mu\nu\rho \theta}F^{\mu\nu 1} - 2
\bar{\varepsilon}_{\mu\rho\theta\theta}F^{\mu\theta 1}+
\bar{\varepsilon}_{\rho \theta\theta\theta}F^{\theta\theta 1})
F^{\rho\theta  0}\right\}(x,\theta),\\
{} & {} & \hspace {2em}
(\varepsilon, \bar{\varepsilon})_{ABCD} =
-(-1)^{\varepsilon_A
\varepsilon_B} (\varepsilon, \bar{\varepsilon})_{BACD} = \ldots =
-(-1)^{\varepsilon_C\varepsilon_D}(\varepsilon, \bar{\varepsilon})_{ABDC}
\,,
\end{eqnarray}
with $\varepsilon_{ab}$ defined in (3.15b). The such choice  for
quantities above is stipulated by the general solutions for 3 independent
equations, following from the trivial requirement for
${\cal F}_{AB}{}^a{\cal F}^{AB a}$,
$\varepsilon_{ABCD}{\cal F}^{AB a}{\cal F}^{CD a}$,
$\bar{\varepsilon}_{ABCD}{\cal F}^{AB a}{\cal F}^{CD b}\varepsilon_{ab}$ to
be invariant w.r.t. permutation of their comultipliers ${\cal F}^{AB a}$
 taking the gradings distribution (7.68) into account
\renewcommand{\theequation}{\arabic{subsection}.\arabic{equation}\alph{lyter}}
\begin{eqnarray}
\setcounter{lyter}{1}
1) & {} & \varepsilon_A + \varepsilon_B +a = 0 \hspace {2em} \Leftrightarrow
\hspace {2em} \left\{
\begin{array}{ll}
\varepsilon_A = \varepsilon_B, & a=0\\
\varepsilon_A = \varepsilon_B + 1, & a=1
\end{array}\right. , \\ 
\setcounter{equation}{78}
\setcounter{lyter}{2}
2) & {} & a(\varepsilon_A + \varepsilon_B + \varepsilon_C + \varepsilon_D
+ 1)  = 0 \hspace {0.7em} \Leftrightarrow \hspace {0.7em}
\left\{
\begin{array}{ll}
\forall A, B, C, D, & a=0\\
\varepsilon_A +\varepsilon_B=\varepsilon_C + \varepsilon_D + 1, & a=1
\end{array}\right. , \\ 
\setcounter{equation}{78}
\setcounter{lyter}{3}
3) & {} & a(\varepsilon_A + \varepsilon_B + \varepsilon_C + \varepsilon_D)
+ \varepsilon_A + \varepsilon_B  = 1  \Leftrightarrow
\left\{
\begin{array}{ll}
\varepsilon_A =\varepsilon_B + 1, \forall C, D, & a=0\\
\varepsilon_C = \varepsilon_D  +1,\forall A, B, & a=1
\end{array}\right..  
\end{eqnarray}
Let us restrict for simplicity now  the choice for components of totally
superantisymmetric constant tensors
$(\varepsilon,\bar{\varepsilon})_{ABCD}$ to be only with even
w.r.t. $(\varepsilon_{P})_A$ = $\varepsilon_A$ parities for the 1st
tensor and with vanishing for the 2nd tensor values  respectively
\renewcommand{\theequation}{\arabic{subsection}.\arabic{equation}}
\begin{eqnarray}
\varepsilon_{0123}=
\varepsilon_{\theta\theta\theta\theta}=1,\;(\varepsilon,\bar{\varepsilon})_{
\mu\nu\rho \theta}=
(\varepsilon,\bar{\varepsilon})_{\mu\theta\theta\theta }=
\bar{\varepsilon}_{\mu\nu\theta\theta }=0,\;
\varepsilon_{\mu\nu \theta\theta}= -
\varepsilon^{(1)}_{\mu\nu}=  \varepsilon^{(1)}_{\nu\mu}\,.
\end{eqnarray}
That representation permit to simplify the only nonzero terms in
(7.75) as follows
\begin{eqnarray}
\Bigl(\varepsilon_{\mu\nu\rho\sigma}F^{\mu\nu 0}F^{\rho\sigma 0} +
4\varepsilon^{(1)}_{\mu\nu}\Bigl(F^{\nu\mu {}0}
\partial_{\theta}{C}^0_{(0)}  + F^{\mu\theta 0}F^{\nu \theta 0}\Bigr) +
4(\partial_{\theta}{C}^0_{(0)})^2\Bigr)(x,\theta)\,. 
\end{eqnarray}
Note,  firstly, the density $\bigl(\partial_{\theta}
{C}^0_{(0)}\bigr)^2(x,\theta)$ appears by the self-dual, secondly, the
summands  with $\varepsilon^{(1)}_{\mu\nu}$ factor are reduced with
accuracy up to total derivatives w.r.t. $(x^{\mu}, \theta)$ to the form
\begin{eqnarray}
\varepsilon^{(1)}_{\mu\nu}\left(F^{\nu\mu 0}
\partial_{\theta}{C}^0_{(0)}+ F^{\mu \theta 0}F^{\nu \theta 0}\right)(x,
\theta)
= \varepsilon^{(1)}_{\mu\nu}\bigl(\partial_{\theta}{\cal A}^{\mu 0}
\partial_{\theta}{\cal A}^{\nu 0}\bigr)(x,\theta)\,,
\end{eqnarray}
and, thirdly, we transform the 3rd term in $F_{\mu \theta}{}^1
F^{\mu \theta 1}$ (7.74) up to the same accuracy to $2{\cal A}^{\mu 1}
\partial_{\theta}\partial_{\mu}{C}^1_{(0)}$.

Now we have all means in order to construct the GTGT invariant
superfunction $S_{L G}(\theta)$ defining the GThGT  with
incorporation both the ghost superfield with its real even scalar
superpartner $C^1_{(0)}(x,\theta)$ and electromagnetic superfield
${\cal A}^{\mu 0}(x,\theta)$ with its real odd vector superpartner
${\cal A}^{\mu 1}(x,\theta)$ into $\theta$-superfield multiplet of the gauge
classical superfields ${\cal A}^{\imath}(\theta)$. Besides, making use of the
inclusion  into $S_{L G}(\theta)$ the quantities (7.80) by means of
the "$\tilde{\theta}$-term" (vacuum angle) addition that leads to
application in the electromagnetic duality theory, see for instance ref.[41],
we choose the action in the form (for $
{\cal D}^{\ast}_{A}\overline{\Psi} = \bigl(\partial_{A} +
\imath   {\cal A}_A^ae^a\bigr)\overline{\Psi}$)
\begin{eqnarray}
{} &\hspace{-1.2em} {} &\hspace{-1.2em} S_{L G}(\theta) \hspace{-0.2em}
=  \hspace{-0.2em}S_{L G}\bigl(\bigl({
\cal A}^a_A,
\partial_{\theta}{\cal A}^a_A, \Psi, \overline{\Psi},
\partial_{\theta}{\Psi}, \partial_{\theta}{\overline{\Psi}}\bigr)
(\theta)\bigr) =
T_{\rm inv}\bigl(\bigl({\cal D}_{\theta}\Psi,{\cal D}^{\ast}_{\theta}
\overline{\Psi}\bigr)(\theta)\bigr)  \nonumber \\
{} &\hspace{-1.2em} {} &\hspace{-1.2em} \phantom{S_{L G}(\theta)}
\hspace{-0.2em} + \hspace{-0.2em}
S^{(1)}_{L{}{\rm inv}}\bigl(\bigl(\Psi,\overline{\Psi},{\cal
A}^a_{A}\bigr)(\theta)\bigr)
+ S_{L{}{\rm inv}}\bigl(\bigl({\cal A}^a_{A},
\partial_{\theta}{\cal A}^a_{A}\bigr)(\theta)\bigr)  \nonumber \\
{} &\hspace{-1.0em} {} &\hspace{-1.0em} \phantom{S_{L G}(\theta)}
 \hspace{-0.2em}= \hspace{-0.2em}
\int d^4x\Bigl[\bigl({\cal D}^{\ast}_{\theta}\overline{\Psi}\bigr)
\bigl({\cal D}_{\theta}\Psi\bigr)
 - \left(\overline{\Psi}(\imath \Gamma^A{\cal D}_{A} - m )
\Psi\right) +
\Bigl(\frac{1}{4}{\cal F}_{AB}{}^a{\cal F}^{AB a}+{\cal L}_{
\tilde{\theta}}\Bigr)\Bigr](x,\theta) \,, \\ 
{} &\hspace{-1.2em} {} &\hspace{-1.2em}
{\cal L}_{\tilde{\theta}} \hspace{-0.2em}
= \hspace{-0.2em}
 \frac{\tilde{\theta}(e^0)^2}{
32\pi^2}\varepsilon_{ABCD}{\cal F}^{AB 0}{\cal F}^{CD 0}\hspace{-0.2em} =
\hspace{-0.2em}\frac{\tilde{\theta}(e^0)^2}{32\pi^2}\Bigl(\hspace{-0.1em}
\varepsilon_{\mu\nu\rho\sigma}F^{\mu\nu 0}F^{\rho\sigma 0}\hspace{-0.1em} +
\hspace{-0.1em}4\varepsilon^{
(1)}_{\mu\nu}\partial_{\theta}{\cal A}^{\mu 0}
\partial_{\theta}{\cal A}^{\nu 0}\hspace{-0.1em}  + \hspace{-0.1em}
4\bigl(\partial_{\theta}{C}^0_{(0)}\bigr)^2\hspace{-0.1em}\Bigr).
\end{eqnarray}
Every summand in (7.82)  is invariant w.r.t. GTGT (7.66).
Omitting the expression for $\delta S_{LG}(\theta)$ under
$T_{\vert P}$ transformations of the type (4.3), having the more complicated
form than in formulae (7.8), (7.25), (7.46) for the models above and now
nonvanishing on  one's mass-shell, we write
down the   Euler-Lagrange equations  for
$Z_{LG}[ {\cal A}_A^a, \Psi,\overline{\Psi}]$ = $\int d\theta
S_{L G}(\theta)$ taking account of the
notation (2.34) (by omitting  $(x,\theta)$ on the right)
\begin{eqnarray}
\hspace{-0.5em}\frac{\delta_l Z_{L G}\phantom{xx}}{\delta
\Psi(x,\theta)} & \hspace{-0.5em} = & \hspace{-0.5em}
{\cal L}^l_{\Psi}(x,\theta)(T_{\rm inv} +  S^{(1)}_{L{}{\rm inv}})(
\theta) = -\Bigl[ \imath
{\cal D}^{\ast}_{A}\overline{\Psi}\Gamma^{A} -(\partial_{\theta}^2 -m
 + \imath e^a \partial_{\theta}{C}^a_{(0)})\overline{\Psi}\Bigr]
= 0, \\
\hspace{-0.5em}\frac{\delta_l Z_{L G}\phantom{xx}}{\delta
\overline{\Psi}(x,\theta)} & \hspace{-0.5em} = &\hspace{-0.5em}
{\cal L}^l_{\overline{\Psi}}(x,\theta)(T_{\rm inv} +
S^{(1)}_{L{}{\rm inv}})(\theta) = -
\Bigl[\bigl(\imath \Gamma^A {\cal D}_{A} -
m +  \imath e^a \partial_{\theta}{C}^a_{(0)} -\partial_{\theta}^2\bigr)
{\Psi}\Bigr]= 0\,, \\
\hspace{-0.5em}\frac{\delta_l
Z_{L G}\phantom{xx}}{\delta {\cal A}_{\mu}^0(x,\theta)} & \hspace{-0.5em} = &
\hspace{-0.5em} {\cal L}^{l{}{}\mu 0}_{{\cal A}^0}(x,\theta)S_{L G}(\theta) =  -
\Bigl[\jmath^{\mu 0} +
\partial_{\nu}F^{\nu\mu 0} +
\textstyle\frac{\tilde{\theta}(e^0)^2}{4\pi^2}
\varepsilon^{(1){}\mu\nu}\partial_{\theta}^2{\cal A}^{0}_{\nu}\Bigr] = 0\,,
\\ 
\hspace{-0.5em}\frac{\delta_l
Z_{L G}\phantom{xx}}{\delta {\cal A}_{\mu}^1(x,\theta)} & \hspace{-0.5em}
= & \hspace{-0.5em}
{\cal L}^{l{}{}{\mu 1}}_{{\cal A}^1}(x,\theta)
S_{L G}(\theta) = -
\Bigl[\jmath^{\mu 1} + \partial^{\mu}\partial_{\theta}{C}^1_{(0)} -
\partial_{\theta}^2{\cal A}^{\mu 1}\Bigr] = 0\,,  \\
\hspace{-0.5em}\frac{\delta_l Z_{L G}\phantom{xx}}{\delta C^0_{(0)}(x,\theta)}
& \hspace{-0.5em} = & \hspace{-0.5em}
{\cal L}^{l{}{}{\theta 0}}_{C^0}(x,\theta)S_{L G}(\theta) = -
\Bigl[\jmath^{\theta 0}  - (2 +
\textstyle\frac{\tilde{\theta}(e^0)^2}{4\pi^2})\partial_{\theta}^2{
C}^0_{(0)} \Bigr) = 0\,,\\ 
\hspace{-0.5em}\frac{\delta_l Z_{L G}\phantom{xx}}{\delta C^1_{(0)}(x,\theta)}
& \hspace{-0.5em}= & \hspace{-0.5em} {\cal L}^{l{}{}{\theta 1}}_{
C^1}(x,\theta)S_{L G}(\theta) = -
\Bigl[\jmath^{\theta 1}  + \Box C^1_{(0)} - \partial_{\mu}\partial_{\theta}{
\cal A}^{\mu 1}\Bigr] = 0\,. 
\end{eqnarray}
In (7.86)-(7.89) we have introduced  the $\theta$-superfield generalization
of the standard
electromagnetic current $\jmath^{\mu}(x)$ $\equiv$ $\jmath^{\mu{}0}(x)$,
extended to (5+5) even and odd superfield components w.r.t.
$\vec{\varepsilon}$ parity and following from the 1st Noether's theorem
analog applied to the $S_{LG}(\theta)$ invariance w.r.t. global $U^{1\vert
1}$ transformations
\begin{eqnarray}
\jmath^{A a}(x,\theta) = - \frac{\partial_l (S^{(1)}_{L{}{\rm inv}} +
T_{\rm inv})(\theta)}{\partial {\cal A}_{A}^a(x,\theta)\phantom{xxxxxx}}
= (e^a\overline{\Psi}\Gamma^A \Psi + \imath \hat{\eta}^{A\theta}e^a
\partial_{\theta}(\overline{\Psi} \Psi))(x,\theta)\,,
\end{eqnarray}
which are conserved on the solutions for corresponding dynamical equations
\begin{eqnarray}
\bigl(\partial_{\mu}\jmath^{\mu 0}(x,\theta)_{\vert
{\cal L}^{l{}\mu 0}_{{\cal A}^0}S_{L G}=0},\
\jmath^{\theta 0}(x,\theta)_{\vert
{\cal L}^{l{}\theta 0}_{{C}^0}S_{L G}=0},\
(-1)^{\varepsilon_A}\partial_A \jmath^{A 1}(x,\theta)_{\vert
{\cal L}^{l{}{A} 1}_{{\cal A}^1}S_{L G}=0}\bigr) = \vec{0}_3\,. 
\end{eqnarray}
The solution for the 2nd algebraic equation w.r.t. $x^{\mu}$ exists
with use of (7.24c) in the form
\begin{eqnarray}
\psi_1(x)=\imath\psi(x)\hat{\mu}\,,
\end{eqnarray}
permitting to express $(\psi_1, \overline{\psi}_1)(x)$ in all the
equations in LS in terms of only
$(\psi, \overline{\psi})(x)$.

Given LS contains after representation (4.8) application  the
18 (8+5 odd and 4+1 even w.r.t. $\varepsilon$) of the
2nd order w.r.t. derivatives on
$(x^{\mu},\theta)$, of all the superfields,  nonlinear  equations
including the 2nd (1st)
order w.r.t. $x^{\mu}$ ($\theta$) DCLF. So, subject to
assumption $T_{\rm inv}(\theta)=0$ the spinor subsystem (7.84), (7.85) will
pass to the 1st
order w.r.t. $(x^{\mu}, \theta)$ $\theta$-superfield generalization of
Dirac equations in presence of dynamical composite superfields ${\cal A}^{A
a}(x,\theta)$.

The both supermatrices (4.17) and (4.15b) are degenerate (the former in sector
of $C^1_{(0)}(x,\theta)$) complicating the analysis of DCLF in
(7.84)--(7.89), in particular, restricting the Cauchy problem setting,
therefore providing the classification for the theory as the irreducible GThGT
with degenerate supermatrix (4.17). Really there are 2
Noether's identities in correspondence with general formula (4.24) among DCLF
\renewcommand{\theequation}{\arabic{subsection}.\arabic{equation}}
\begin{eqnarray}
{} & {} &
\partial_{\mu}\Bigl(
{\cal L}^{l{}{}\mu b}_{{\cal A}^b}(y,\theta)S_{L G}(\theta)\Bigr) -
\partial_{\theta}\Bigl(
{\cal L}^{l{}{}{\theta b}}_{C^b}(y,\theta)S_{L G}(\theta)\Bigr)(-1)^b
\nonumber \\
{} & {} & - \imath(-1)^b e^b \Bigl(\overline{\Psi}(y,\theta){\cal L}^l_{
\overline{\Psi}}(y,\theta)S_{L G}(\theta) +
\bigl({\cal L}^l_{\Psi}(y,\theta)S_{L G}(\theta)\bigr)\Psi(y,\theta)\Bigr)
\equiv 0,\ b=0,1\,. 
\end{eqnarray}
The number of real physical degrees of freedom is equal to
\begin{eqnarray*}
[\Psi] + \textstyle\frac{1}{2}[{\cal A}^{\mu 0}] + [C^1] = 4 + 2 + 1 = 7\,,
\end{eqnarray*}
that is greater on one degree than for standard
quantum electrodynamics in view of the structure for the action
$S_{L{}{\rm inv}}(\theta)$ in (7.82) and Eq.(7.89).

In turn, under reduction of the given  ${\cal M}_{cl}$ onto
hypersurface ${\cal A}^{A 1}(x,\theta) = 0$, the theory  become by the
irreducible GThGT
with $U(1)$ gauge group ($\partial_A\xi^1 =0$) and nondegenerate supermatrix
(4.17) for $T_{\rm inv}(\theta) \ne 0$,
therefore representing now the $\theta$-superfield generalization of quantum
electrodynamics at least on the classical level.
The GTGT, GGTGT, classical action, Euler-Lagrange equations,
$\theta$-superfield current $\jmath^{A 0}(x,\theta)$  and its conservation,
Noether's identity can be easily obtained from the formulae (7.66)--(7.93)
subject to conditions ${\cal A}^{A{}1}(x,\theta) = 0$.

The $\theta$-superfield free  GThGT with $U^{1\vert 1}$  gauge group
described by  only  ${\cal A}^{A{}a}(x,\theta)$ and,
in particular, the $\theta$-superfield free electrodynamics with $U(1)$ gauge
group for ${\cal A}^{A{}1}(x,\theta) = 0$ are yielded from the formulae
(7.82)--(7.89) in the form of superfunctionals respectively
\renewcommand{\theequation}{\arabic{subsection}.\arabic{equation}\alph{lyter}}
\begin{eqnarray}
\setcounter{lyter}{1}
Z_{F{}L{}G}[{\cal A}^a_A] & = & Z_{L{}G}[{\cal A}^a_A,\Psi,\overline{\Psi}]_{
\vert\Psi=\overline{\Psi}=0} =  \int d\theta S_{L{}{\rm inv}}\bigl(\bigl({\cal A}^a_{
A}, \partial_{\theta}{\cal A}^a_{A}\bigr)(\theta)\bigr)\,,\\
\setcounter{equation}{94}
\setcounter{lyter}{2}
Z_{F{}E{}D}[{\cal A}^0_A] & = & Z_{F{}L{}G}[{\cal A}^a_A]_{
\vert {\cal A}^1_A=0} =  \int d\theta S_{L{}{\rm inv}}\bigl(\bigl({\cal A}^0_{
A}, \partial_{\theta}{\cal A}^0_{A}\bigr)(\theta)\bigr)\,.
\end{eqnarray}
Subject to constraints $C^a_{(0)}(x,\theta)=0$ the every from  free
GThGTs become
by the irreducible GThSTs representing now the examples of the type (4.33)
natural systems  and revealing here the validity of the general Theorem
and its corollary 2 application  in these cases.
Evidently, that with accuracy up to total derivative w.r.t. $\theta$, that
is the term $(\partial_{\theta}{\cal A}^{\mu{}1}\partial_{\theta}{\cal A}_{
\mu}^1)$, the classical actions above coincide for
$C^1_{(0)}(x,\theta) = 0$.

The presence of boson scalar superfield $C^1_{(0)}(x,\theta)$ increases
the number
of physical degrees of freedom on 1 for the $1$st model (7.94a) in contrast to
(7.94b)
reflecting the nontrivial character of $U(0\vert 1)$ group realization as
gauge one in spite of $e^1$ charge nilpotency.

Next, setting $C^0_{(0)}(x,\theta) = 0$ together with
absence of the topological summand
$\varepsilon_{\mu\nu\rho\sigma}$ $\times$ $({\cal F}^{\mu\nu{}0}{\cal F}^{\rho
\sigma{}0})(x,\theta)$, we can derive from (7.94b) the
GThST described for $D=4$ by massless
vector superfield with action (7.45)
under identification  $\varepsilon_{\mu\nu} = -\frac{\tilde{\theta}
(e^0)^2}{4\pi^2}\varepsilon^{(1)}_{\mu\nu}$.

As the consequence, consider in the model (7.82) the only potential terms
having singled out them by means of the constraints
$\partial_{\theta}\bigl({\Psi}, {\overline{\Psi}}, {\cal A}^a_A\bigr)(
x,\theta)$ = $0$
\renewcommand{\theequation}{\arabic{subsection}.\arabic{equation}}
\begin{eqnarray}
{} & {} &
S\bigl(\bigl(\Psi, \overline{\Psi}, {\cal A}^a_{\mu}, C^a_{(0)}\bigr)(
\theta)\bigr)
 = - \int  d^4x\Bigl[ \overline{\Psi}(\imath \Gamma^{\mu}\partial_{
\mu}  -  m )
\Psi +  {\cal A}_{A}^a\jmath^{A{}a} \nonumber \\
{} & {} & \hspace{2em} -
\Bigl(\textstyle\frac{1}{4}F_{\mu\nu}{}^0F^{\mu\nu 0}
 +  \textstyle\frac{\tilde{\theta}(e^0)^2}{
32\pi^2}\varepsilon_{\mu\nu\rho\sigma}F^{\mu\nu 0}F^{\rho\sigma 0}
 + \textstyle\frac{1}{2}
\partial_{\mu}C^1_{(0)}\partial^{\mu}C^1_{(0)}\Bigr)\Bigr](x,\theta)\,.
\end{eqnarray}
As it follows from Corollary 2 application and formula (4.51) the
superfunction (7.95) is invariant w.r.t. (7.66) GTST now with
nongauge ghost superfields $C^a_{(0)}(x,\theta)$, whose
infinitesimal form  written by means of GGTST have the form  as
independent ${\hat{\cal R}{}_0^{\tilde{\imath}}}_b({\cal
A}(x,\theta))$ with opposite sign in (7.70) and ${\hat{\cal
R}{}_1^{\tilde{\imath}}}_b({ \cal A}(x,\theta)) = 0$, providing
the fulfilment of the relation (4.31) in question
\begin{eqnarray}
\hat{\cal R}{}^{\tilde{\imath}}_b({\cal A}(x,\theta),x, \theta; y,
\theta') = -\delta(\theta - \theta') {{\cal
R}_0^{\tilde{\imath}}}_b({\cal A}(x,\theta))
\delta(x-y)\,. 
\end{eqnarray}

One exist the another invariant possibility to extract the potential term in
action (7.82) based on the superfield BRST symmetry realization for
Yang--Mills type theories [23,24]. To this end one serve the independent
constraints onto matter superfields and strength components (pointed out in
[23,24] for $a=0$)
\begin{eqnarray}
\Bigl(F_{\theta\theta}{}^a, F_{\mu\theta}{}^a, {\cal D}_{\theta}\Psi
\Bigr)(x,\theta)
= 0 \Longleftrightarrow \partial_{\theta}\Bigl(C^a_{(0)}, {\cal A}^a_{\mu},
\Psi\Bigr)(x,\theta) = (0,\partial_{\mu}C^a_{(0)}, \imath C^a_{(0)}e^a)
(x,\theta)\,, 
\end{eqnarray}
satisfying to the type (A.6) solvability conditions and permitting to write
the BRST transformations for all superfields on the solutions $\tilde{\cal
A}^{\imath}(\theta)$ for Eqs.(7.97) in the form of translations along
$\theta$ on odd constant $\mu$ as in (2.18)
\begin{eqnarray}
\delta_{\mu}{\cal A}^{\imath}(\theta) = \mu \partial_{\theta}
\tilde{\cal A}^{\imath}(\theta)\,.
\end{eqnarray}
These superfield transformations become by the component BRST ones for $
\theta = 0$.

In limiting onto "BRST surface" (7.97), the action (7.82), with
allowance made for strength's transformations (7.80), (7.81) and
remark after (7.81), passes into superfunction (7.95) depending
upon $\tilde{\cal A}^{\imath}(\theta)$ with vanishing
$\tilde{j}{}^{\theta a}$, with regards of the fact that
$(\varepsilon^{(1)}_{\mu\nu}\partial^{\mu}C^0_{(0)}
\partial^{\nu}C^0_{(0)}) (x,\theta)$ is equal to the total
derivative w.r.t. $x^{\mu}$ (therefore omitted) and without
$\frac{1}{2}$ factor before $(\partial_{\mu}C^1_{(0)}
\partial^{\mu}C^1_{(0)})$.
Moreover the purely electromagnetic terms with ${\cal A}^a_{\mu}$
do not depend upon $\theta$ by virtue of (7.97)
($\tilde{F}_{\mu\nu}{}^0(x,\theta)$ = $\partial_{[\mu}\tilde{\cal
A}{}^0_{\nu]}(x,\theta)$ = ${F}_{\mu\nu}{}^0(x))$.

At last, for $\theta =0$ we obtain from (7.95) the spinor
electrodynamics extended by the topological $\tilde{\theta}$-term with not
interacting scalar nongauge field $C^1_{(0)}(x)$ and additional current's
summand $ A_{\mu}^1e^1\overline{\psi}\Gamma^{\mu}\psi$. Note that
fermion nature for the vector field $A_{\mu}^1(x)$ is not so unusual
from physical viewpoint if
to recollect that to describe the (super)particle models it is widely used
the twistor variables being by boson spinors.

In turn, for massive nongauge models from Secs.VII.1,2,3 the restriction
$\partial_{\theta}(\varphi,\overline{\varphi}, \Psi$, $\overline{\Psi}, {\cal
A}^{\mu}, \overline{\cal A}{}^{\mu})$ = $0$ to get the standard action
functionals is associated with superfield form of the trivial BRST
transformations considered as in (7.97) for the matter superfields.

The embedding of GA GTST given by classical action (7.95) $S(\theta)$ into
GA GTGT with classical $Z_{L{}G}[{\cal A}^a_A,\Psi,\overline{\Psi}]$ is
realized by means of the type (5.4), (5.3) real quantities
construction
\begin{eqnarray}
{} & {} & Z_{(1)}[\Gamma_{min}] =
\hspace{-0.2em}\int\hspace{-0.2em} d \theta \Bigl(S_{LG}(\theta) -
\hspace{-0.2em}\int \hspace{-0.2em}d^4x\Bigl[ {\cal
A}^{a{}\ast}_{\mu}\partial^{\mu}C^a + \imath e^a (\Psi^{\ast}\Psi
- \overline{\Psi}\overline{\Psi}{}^{\ast})C^a \nonumber \\
{} & {} & \phantom{Z_{(1)}[\Gamma_{min}]}
- C^{a{}\ast}_{(0)}\partial_{
\theta}C^a\Bigr](x,\theta)\Bigr) = \partial_{\theta}S_{(1){}LG}(\Gamma_{min}
(\theta), \partial_{\theta}\Gamma_{min}(\theta))\,,\\
{} & {} & S_{(1)}(\Gamma_{min}(\theta)) = S(\theta) +
\hspace{-0.2em}\int \hspace{-0.2em}d^4x\Bigl[ {\cal
A}^{a{}\ast}_{\mu}\partial^{\mu}C^a + \imath e^a (\Psi^{\ast}\Psi
- \overline{\Psi}\overline{\Psi}{}^{\ast})C^a \Bigr](x,\theta)
\,,
\end{eqnarray}
defined on $T^{\ast}_{odd}{\cal M}_{min}$ $=$ $\{({\cal A}^{\mu{}a},C^a_{(
0)}, \Psi, \overline{\Psi}, C^a)$,
$({\cal A}^{a{}\ast}_{\mu}, C^{a{}\ast}_{(0)}$,
$\Psi^{\ast}, \overline{\Psi}{}^{\ast}, C^{a{}\ast}))(x,\theta)\}$ $\equiv$
$\{\Gamma^p_{min}(x,\theta)\}$ and
$T_{odd}(T^{\ast}_{odd}{\cal M}_{min})$ $=$ $\{(
\Gamma^p_{min}, \partial_{\theta}\Gamma^p_{min})(x,\theta)\}$
respectively. The configuration space ${\cal M}_{min}$ in question is
enlarged by  ghost superfields $C^a(x,\theta)$ with the same Grassmann
gradings as for the classical ghosts $C^a_{(0)}(x,\theta)$.

By construction, the superfunction(al)  above appear by the
exact solutions of corresponding master equations (5.4), (5.3) with
easily defined for this case superfield brackets (5.6), (5.5).

The problems of ZLR brackets and new models construction appear by the
more complicated than  for the GThST and ThSTs from Sec.VII.1,2,3 and allows,
in particular, the existence of the nontrivial BRST charge $Z^{(-1)}(\theta^{
-1})$ defined on the corresponding zero locus ${\cal Z}_{Q^0}$ (${\bf Q}^0(
\theta) = (S_{(1)}(\theta),\ )_{\theta}$), in  turn, being the local
supermanifold with nontrivial Bose--Fermi distribution for one's coordinates
(for $C^{a\ast}$ and, for instance, for two from ${\cal A}^a_{\mu}$).
It is interesting to more carefully investigate this problem for $U^{1\vert 1}
$ model independently.
\subsection{Conclusion}

The programm of Lagrangian formulation of $\theta$-STF realized on the
whole  in the paper constitutes the 1st
step in order to construct the general superfield quantization
method for gauge theories in the usual Lagrangian formalism. By the next
large effort to resolve the last problem one will appear the construction
a so-called {\it Hamiltonian formulation of $\theta$-STF} (whose elements,
in part, have been used in Secs.V--VII for ZLR problems and to describe the
GAs structure) based on the
powerful use on the classical level the superantifields
${\cal A}^{\ast}_{\imath}(\theta)$ permitting to reformulate, not
always equivalently, the
$\theta$-superfield models given in the Lagrangian $\theta$-STF.

The noncontradictory  possible description of an arbitrary superfield model,
being by natural extension of one from the usual field theory, is
guaranteed by a number of mathematical tools, leading to the classical
action superfunction $S_L(\theta)$ construction defined on the odd tangent
bundle $T_{odd}{\cal M}_{cl} \times \{\theta\}$ over configuration space
${\cal M}_{cl}$ of classical superfields ${\cal A}^{\imath}(\theta)$.

Concepts for the Lagrangian  description of the GThGT, GThST and
nondegenerate
theories also together with the statements and their corollaries on
structure of dynamical equations (the so-called constraints) allow to extend
the notions from standard gauge fields theory onto $\theta$-STF.
Simultaneously the concept of  gauge invariance is prolonged onto superfield
case,  based on the embedding of the local on $\theta$
special type objects and relations (being equivalent for $\theta=0$ to
the ones from usual  field theory) into their general type
analogs.

As it have been waited the $\theta$-STF provides the iterative algorithms to
construct the new $\theta$-superfield models by means of the introduction
the earlier hidden supertimes $\Gamma^{l} = (t^l,\theta^l)$, $l=\ldots,-1,0,1,
\ldots$ in the framework
of so-called direct and inverse ZLR problems with revealing the nontrivial
role for such variables as $\lambda^{\imath}, J_{\imath}$. Additionally,
it is demonstrated the
possibility of the another superfield form for BFV method
construction.

The component formulation for Lagrangian $\theta$-STF and ZLR problems,
in fact, completely
establishes the connection of treatment of the new $\theta$-superfield
models  with their description in terms of component objects.

The statements and consequences of $\theta$-STF have found one's
confirmation of their validity and interconnection with each other on the
example of the $\theta$-superfield models construction in Sec.VII. A number
from
them have sufficiently simple illustrative character, for instance, for the
models of scalar and vector superfields realizing the nondegenerate and
gauge ThSTs. At the same time, the one from the spinor models with
action $S^{(1)2}_L(\theta)$  appears now by nondegenerate
general type theory.

The gauge principle have allowed to construct from these models
the abelian GThGT with $U^{1\vert 1}$ gauge group  leading, in
first, to ghost superfield $C^0_{(0)}(x,\theta)$ inclusion into
model on the classical level, in second, the scalar boson
superpartner $C^1_{(0)}(x,\theta)$ for $C^0_{(0)}(x,\theta)$,
connected with scalar parameter $\xi^1$, have provided the
appearance of the additional physical degree of freedom within
interacting theory. This model contains both the
$\theta$-superfield generalization of quantum electrodynamics and
the $\theta$-superfield QED itself in complete accordance with
Sec.IV general statements. The generalization of the last model
onto case with nonabelian gauge group together with possible
interpretations appear by very perspective.

As to the ways to
construct the $\theta$-superfield arbitrary model starting from usual
relativistic field theory  in the form  of
natural system then the following algorithm may be applied. It is sufficient
to extend the fields $A^{\imath}$ up to superfields
${\cal A}^{\imath}(\theta)$, having built the superfunction
$S(\theta)$. Next, we must add the "kinetic" $T_{\mid\bar{J}}$-invariant
term in superfield form
$T\bigl(\partial_{\theta}{\cal A}(\theta)\bigr)$. Resultant action
$S_{L}\bigl({{\cal A}}(\theta), \partial_{\theta}{\cal A}(\theta)\bigr)$
appears by their
difference and defines GThST or nondegenerate ThST, which leads to the HCLF,
 in turn, being by the nontrivial (as it have
been shown for interacting models(!)) $\theta$-superfield
extension of the initial $P_0(\theta)$-component  dynamical equations.
Original field theory
model is obtained from superfield one, in first, under restriction
from $C^{k}\bigl(T_{odd}{\cal M}_{cl} \times \{\theta\}\bigr)$ down to
$C^{k}\bigl( {\cal M}_{cl} \times \{\theta\}\bigr)$, and then
the ordinary functionals are singled out from the superfunctions, in the
invariant way, by means of involution $\ast$ (3.6), (3.8) or, equivalently,
by setting $\theta= 0$. The another way to do this is the imposing
of the type (7.97) constraints realizing simultaneously the
$\theta$-superfield BRST transformations.
At last, the models from the class of GThGT may be obtained,
for example, by means of gauge principle application [40] to the
corresponding ThST or GThST.

These models appear by appropriate proving ground for demonstration of the
general   constructions validity and efficiency both of Hamiltonian
$\theta$-STF formulation and next the quantization procedure itself.

As it was mentioned in introduction we have intensively used
the following analogy between quantities and
relations of the Lagrangian $\theta$-STF and classical
mechanics in the
usual Lagrangian formulation with ${\varepsilon}_P$-even
objects organized in the table form
\begin{eqnarray*}
\begin{array}{|l|l|} \hline
\phantom{xxxxxxx}
{\bf  usual \phantom{x} classical \phantom{x} mechanics} & \phantom{xxxxxxx}
 {\bf \theta- STF} \\ \hline
1.\;  t\in {\bf R} - {\rm time}\phantom{x}
({\varepsilon}_{\bar{J}}, \varepsilon)t = (0,0) & \theta \in
{}^{1}{\Lambda}_{1}(\theta) - {\rm odd \phantom{x} time} \phantom{x}
({\varepsilon}_{P}, {\varepsilon}_{\bar{J}}, \varepsilon)\theta =  \\
\phantom{xxxxx} & \phantom{x}(1,0,1)\\ \hline
2.\;  q^{a}(t) - {\rm
generalized \phantom{x}coordinates} & {\cal A}^{\imath}(\theta) - {\rm
 superfields} \phantom{x} (\imath \phantom{x} {\rm contains} \phantom{x}t) \\
\phantom{2.\;} ({\varepsilon}_{\bar{J}} , \varepsilon)q^{a}(t) =
({\varepsilon}_{a},{\varepsilon}_{a}) & \phantom{xxx} \\ \hline
3.\;
\dot{\;q^{a}}(t) - {\rm generalized \phantom{x} velocities}  &
{\stackrel{\ \circ}{\cal A}}{}^{\imath}(\theta) - {\rm odd \phantom{x}
generalized \phantom{x} velocities} \\
\phantom{3.\;}
\varepsilon(\dot{\;q^{a}}(t)) = {\varepsilon}_{a} & \phantom{xxx}\\ \hline
4.\;  {\cal M}=\bigl\{q^{a}\bigr\}, T{\cal
M}=\bigl\{(q^{a},\dot{\;q^{a}})\bigr\} - & {\cal M}_{cl}=\bigl\{{\cal
A}^{\imath}(\theta)\bigr\}, T_{odd}{\cal M}_{cl} = \\
\phantom{4.\;} {\rm
configuration \phantom{x} space \phantom{x} and} & \bigl\{\bigl( {\cal
A}^{\imath}(\theta), {\stackrel{\ \circ}{\cal
A}}{}^{\imath}(\theta)\bigr)\bigr\} - {\rm configuration \phantom{x} space}\\
\phantom{4.\;}  {\rm tangent \phantom{x} bundle} &  {\rm and \phantom{x} odd
 \phantom{x} tangent \phantom{x} bundle}\\ \hline
5.\;  L(q, \dot{q}, t)
\equiv L(t):T{\cal M} \times \{t\} \to {\bf R} - & S_{L}\bigl({\cal A}
(\theta), {\stackrel{\ \circ}{\cal A}}(\theta), \theta \bigr) \equiv
S_{L}(\theta):  T_{odd}{\cal M}_{cl} \times \\
\phantom{5.\;} {\rm Lagrange
\phantom{x} function},\phantom{x}
 &  \{\theta\} \to {\Lambda}_{1}(\theta, {\bf R}) - {\rm superfunction
\phantom{x} of} \\
\phantom{5.\;}
({\varepsilon}_{\bar{J}}, \varepsilon)L(t)= (0,0) & \phantom{x} {\rm
Lagrangian \phantom{x} classical \phantom{x} action,} \\
\phantom{5.\;}
\phantom{xxxxxxxxxxxxxxxxxxxxxxxxxxx} & \phantom{x}(\varepsilon_{P},
\varepsilon_{\bar{J}}, \varepsilon ) S_{L}(\theta) = (0, 0, 0)\\ \hline
6.\;
S[q]=\int dt L(q, \dot{q}, t) - {\rm classical} & Z[{\cal A}]= \int d\theta
S_{L}(\theta), (\varepsilon_{P}, \varepsilon_{\bar{J}}, \varepsilon )Z = (1,
0, 1)\\ \phantom{6.\;} {\rm action \phantom{x} functional}, \phantom{x}
\varepsilon (S) = 0 & \phantom{xxx} \\ \hline
7.\; \displaystyle\frac{\delta_l
S\phantom{x}}{\delta q^{a}(t)}= \left( \frac{\partial_l
\phantom{xx}}{\partial q^{a}(t)} - \frac{d}{dt}\frac{ \partial_l
\phantom{xx}}{\partial \dot{q^{a}}(t)}\right)L(q, \dot{q}, t) &
\displaystyle\frac{{\delta}_{l}Z[{\cal A}]}{\delta {\cal
A}^{\imath}(\theta)}= {{\cal L}}^{l}_{\imath}(\theta) S_{L}(\theta) = 0 - \\
\phantom{7.\;} = 0 -  {\rm usual \phantom{x} \mbox{Euler-Lagrange}}   & {\rm
\mbox{Euler-Lagrange} \phantom{x} equations \phantom{x} for \phantom{x}}
Z[{\cal A}]
\\
\phantom{7.\;\ } {\rm equations } & \phantom{xxxxxxxxxxxxxxxxxxxx}\\
\hline
\end{array}
\end{eqnarray*}
{\bf Acknowledgments:}
Author is grateful to referees,  ${\rm \ddot{O}}$.Dayi, A. Nersessian for
useful crititical comments and to B.R.Mishchuk for
some stimulating discussions of the present results and
for help in preparation of the paper to publication.
\appendix
\renewcommand{\thesubsection}{\Alph{subsection}}
\subsection{ODE with odd operator $\frac{d}{d\theta}$}
\setcounter{equation}{0}

\underline{\bf Statement 1} (Existence and Uniqueness theorem for solutions
of the 1st order on $\theta$ systems of $N$  ODE) \\
{\bf 1.} The general solutions of $N$ ODE systems  in  normal form
(NF) of the 1st order w.r.t. unknowns ${g}^{j}(\theta), j = 1, \ldots
, N = (N_{+}, N_{-})$ in the domain $U$ of the supermanifold ${\cal N}$
locally coordinatized by ${g}^{j} (\theta)$, for $\theta \in {\Gamma}_{(0,1)}
\subset {}^{1}\Lambda_{1}(\theta)$
\renewcommand{\theequation}{\Alph{subsection}.\arabic{equation}}
\begin{eqnarray}
{} \hspace{-6em}{\rm a})\hspace{0,5cm}
\partial_{\theta}{g}^j(\theta) & = &
P_0(\theta)f^j\bigl(g(\theta),\theta\bigr),\ \;f^j
 \bigl(g(\theta),\theta\bigr),\;g^j (\theta) \in C^2 \bigl({\cal N}\times
 \{\theta\}\bigr)\ , \\ 
{} \hspace{-6em}{\rm b})\hspace{0.5cm}
\partial_{\theta}{g}^j(\theta) & = & h^j \bigl( g(\theta),\theta\bigr)\
,\ h^j\bigl( g(\theta),\theta\bigr) \in C^2 \bigl({\cal N}\times
\{\theta\}\bigr)\ 
\end{eqnarray}
exist and have the form respectively
\begin{eqnarray}
{} \hspace{-2em} {\rm a})\hspace{0.5cm}
{g}^j(\theta)  & = & P_0(\theta)k^j(\theta) + \theta
f^j\bigl(g(\theta),\theta\bigr),\
\varepsilon(g^j(\theta)) = \varepsilon(k^j(\theta)) =
\varepsilon(f^j(\theta)) + 1\,, \\ 
{} \hspace{-2em} {\rm b})\hspace{0.5cm} {g}^j(\theta)
& =&  P_0(\theta)l^j(\theta) + \theta h^j\bigl(g(\theta),\theta\bigr),\
\varepsilon(g^j(\theta)) = \varepsilon(l^j(\theta))
= \varepsilon(h^j(\theta)) + 1\,,
\end{eqnarray}
with arbitrary superfunctions ${k}^{j}(\theta), {l}^{j}(\theta) \in
C^{2}({\cal N})$ restricted on $U$.

\noindent
{\bf 2.} The integral curve $\check{g}{}^{j}(\theta)$ for system  (A.1), (A.2)
satisfying to Cauchy problem for $\theta={0}$\footnote{having represented
solution for (A.2) in the form
${g}^{j}(\theta - {\theta}_{0}) = (\theta - {\theta}_{0}) {h}^{j}\bigl(
g(\theta - {\theta}_{0}), \theta - {\theta}_{0} \bigr) + {P}_{0}(\theta -
{\theta}_{0}) {l}^{j}(\theta - {\theta}_{0})$  one can set up the Cauchy
problem for arbitrary $\theta = {\theta}_{0} \in {\Gamma}_{(0,1)}$}
in the domain $U \subset {\cal N}$
\begin{eqnarray}
P_0(\theta)\check{g}^j(\theta)=\check{g}^j(\theta)_{\mid \theta=0}
= \bar{g}^j,\ 0 \in
{\Gamma}_{(0,1)} 
\end{eqnarray}
is unique.

To be solvable in explicit superfield form it is necessary, in first, to
impose a so-called necessary solvability
conditions  for system
(A.2) in view of the identity (4.6) validity for ${g}^{j}(\theta)$.
The condition above has the form  of the 1st order on $\theta$
$2N$ ODE system
\begin{eqnarray}
\partial_{\theta}{g}^j(\theta) =
h^j\bigl(g(\theta),\theta\bigr),\ \
\partial_{\theta}{h}^j\bigl(g(\theta),\theta\bigr)  =  0\,, 
\end{eqnarray}
so that the properly solvability conditions written in the 2nd subsystem means
that superfunctions $h^j\bigl(g(\theta),\theta\bigr)$ on the possible
solutions for Eqs.(A.2) appear by the integrals for this system.

\noindent
\underline{\bf Statement 2} (Existence and Uniqueness theorem for solution
of the 2nd order on $\theta$  $N$ ODE system) \\
{\bf 1.} The general
solution for the 2nd order on $\theta$  $N$ ODE system  in NF
\begin{eqnarray}
\partial^2_{\theta}{g}^j(\theta) =
0\;,\;j=1,\ldots,N\,, 
\end{eqnarray}
in the domain $U$ of supermanifold ${\cal N}$ with unknowns
${g}^{j}(\theta) \in C^{2}\bigl({\cal
N} \times \{\theta\}\bigr)$ exists in the form
\begin{eqnarray}
g^j(\theta) = k^j(\theta),\ \varepsilon(g^j(\theta)) =
\varepsilon(k^j(\theta))\,, 
\end{eqnarray}
with arbitrary superfunctions ${k}^{j}(\theta) \in C^{2}\bigl({\cal N} \times
\{\theta\}\bigr)$.\\
{\bf 2.} Particular solution $\check{g}^j(\theta)$ for system  (A.7)
satisfying  to the initial conditions for $\theta = 0$
\begin{eqnarray}
\bigl(\check{g}{}^j(\theta), \partial_{\theta}{\check{g}}}{}^j(\theta)
\bigr)_{\mid\theta=0} = \bigl(\bar{g}{}^j, {\overline{\partial_{\theta}g}{
}^j\bigr) 
\end{eqnarray}
is unique.

By  more complicated of the 2nd order on $\theta$ $N$ ODE system in NF one
appear the equations
\begin{eqnarray}
\partial^2_{\theta}{g}^j(\theta) = f_1^j\bigl(g(\theta),
\partial_{\theta}{g}(\theta), \theta\bigr)\;,\; f_1^j(\theta) \in
C^{1}\bigl( T_{odd}{\cal N} \times \{\theta\} \bigr)\,,
\end{eqnarray}
being equivalent to the following  $2N$ ODE system
\begin{eqnarray}
{} & {} & \partial^2_{\theta}{g}^j(\theta) = 0\;,
\\ 
{} & {} & f_1^j\bigl(g(\theta), \partial_{\theta}{g}(\theta), \theta\bigr)
= 0\,. 
\end{eqnarray}
Eqs.(A.12) appear by the 1st order on $\theta$ differential constraints both
for the initial conditions values (A.9) and onto possible values of
${g}^{j}(\theta), \partial_{\theta}{g}^j(\theta)$
obtained by resolution of Eqs.(A.11) for all $\theta \in {\Gamma}_{(0,1)}$.

It should be noted the $N$ ODE system with given superfunctions
\begin{eqnarray}
m^j\bigl(g(\theta), \partial_{\theta}{g}(\theta), \theta\bigr) = 0,\
m^{j}\bigl( {g}(\theta),
\partial_{\theta}{g}(\theta), \theta \bigr) \in C^{1}\bigl( T_{odd}{\cal
N} \times \{\theta\} \bigr) 
\end{eqnarray}
may be written in the equivalent form of $2N$ ODE system by means of
projectors $P_{a}(\theta)$ action
\begin{eqnarray}
P_0(\theta)m^j\bigl(g(\theta), \partial_{\theta}{g}(\theta), \theta\bigr)
= m^j\bigl( P_0 g(\theta), \partial_{\theta}{g}(\theta), 0\bigr) = 0,\ \
\partial_{\theta}{m}^j\bigl(g(\theta), \partial_{\theta}{g}(\theta),
\theta\bigr) = 0. 
\end{eqnarray}
Therefore a solution for the 2nd subsystem (A.14) must belong to a set of
solutions for the 1st one, if the solvability conditions is true for Eqs.(A.13)

It is convenient to
introduce the following single-valued functions of degree and least degree
w.r.t. any from elements $C(\theta)\in \{g(\theta)$, $\partial_{\theta}{
g}(\theta)$, $g(\theta)\partial_{\theta}{g}(\theta)$, $\ldots\}$
respectively
\begin{eqnarray}
\bigl( {\rm deg}_{C(\theta)},
\min{\rm
deg}_{C(\theta)}\bigr){}:{}C^{k}\bigl( T_{odd}{\cal N}
\times \{\theta\} \bigr) {}\rightarrow \mbox{\boldmath$N$}_0\,,
\end{eqnarray}
acting on arbitrary ${\cal F}(\theta)$, with representation of the form
(2.27) by the rule
\begin{eqnarray}
{} & \bigl( {\rm deg}_{b(\theta)}, \min{\rm deg}_{b(\theta)},
{\rm deg}_{g(\theta)\partial_{\theta}{g}(\theta)},
\min{\rm deg}_{g(\theta)\partial_{\theta}{g}(\theta)}\bigr)
{\cal F}(\theta) = {} & \\ 
{} & \bigl(\max{p}, \min{p}, \max{(l+k)}, \min{(l+k)}\bigr),\
b(\theta) \in \{g(\theta), \partial_{\theta}{g}(\theta)\}\,, {} &\nonumber
\end{eqnarray}
where the symbols $\max\{p,(l+k)\}, \min\{p,(l+k)\}$ appear by the most and
the least values of degree order for $\vec{b}{}^{(j)_{p}}(\theta)$,
$\Bigl(\vec{g}{}^{(i)_{l}}\vec{\partial_{\theta}{g}}{}^{(j)_k}
\Bigr)(\theta)$, $i,j=1,\ldots,N$ respectively.
By means of the functions above the so-called  holonomic
$f^{j}_{1}\bigl(g(\theta),\theta\bigr)$ and linearized $f^{j}_{1{}{\rm lin}}(
\theta)$ constraints are singled out from $f^{j}_{1}(\theta)$ (A.12)
by the  relations respectively
\begin{eqnarray}
{\rm deg}_{\partial_{\theta}{g}(\theta)}
f^{j}_{1}(\theta) = 0,\ \ {\rm deg}_{g(\theta)\partial_{\theta}{g}(
\theta)}f^{j}_{1}(\theta) = 1\,. 
\end{eqnarray}

Not all from differential constraints (A.12) appear by (functionally)
independent. To investigate this problem let us assume the following
postulates fulfilment
\begin{eqnarray}
1)\hspace{0.2cm}\left(\tilde{g}{}^{j}(\theta),
\partial_{\theta}{\tilde{g}}{}^j(\theta) \right) = (0,0) \in T_{odd}\Phi ,\
\Phi \subset {\cal N}\,,  \hspace{8.1cm} 
\end{eqnarray}
with $\Phi$ being the set of solutions for Eqs.(A.12); \\
2) $ f_{1}^{j}(\theta) = 0$ define the 1st order
supersurface, which the conditions hold on
\begin{eqnarray}
f^{j}_{1}\bigl( g(\theta),
\partial_{\theta}{g}(\theta), \theta\bigr)_{\mid T_{odd}\Phi} =
0,\ \;{\rm rank} \left\|\frac{\delta_l f^{i}_{1}\bigl( g(\theta),
\partial_{\theta}{g}
(\theta),\theta\bigr)} {\delta
g^j(\theta_1)\phantom{xxxxxxxx}}\right\|_{\mid T_{odd} \Phi} \leq
N.
\end{eqnarray}
By definition under rank of supermatrix in (A.19) simultaneously
with allowance made for the connection of superfield variational
derivative of $f_{1}^{i}(\theta)$ w.r.t. $g^{j}({\theta}_{1})$
with partial superfield ones w.r.t. $g^{j}({\theta}_{1})$,
$\partial_{\theta_1}{g}^j({\theta}_{1})$ in correspondence with
(2.34) we mean
\begin{eqnarray}
{\rm rank} \left\|\left(\frac{\partial_l \phantom{xxx}}{\partial
g^j(\theta_1)} - (-1)^{\varepsilon(g^j)}
\partial_{\theta_1}\frac{\partial_l
\phantom{xxxx}}{{\partial(\partial_{\theta_1}{g}^j(\theta_1))}}\right)
\Bigl(f^{i}_1(\theta_1)\right\|_{\mid T_{odd} \Phi}
\hspace{-2.0em}\delta(\theta_1 - \theta)\Bigr)(-1)^{
\varepsilon(f^{i}_1)}. 
\end{eqnarray}
For holonomic constraints $f^{i}_1(\theta)$  the rank
(A.20) pass into almost standard form [39]
\begin{eqnarray}
{\rm rank} \left\|
\frac{\partial_l f^{i}_{1}\bigl( g(\theta_1), \theta_1\bigr)}{\partial
g^j(\theta_1)\phantom{xxxxx}} \right\|_{\mid \Phi}\delta(\theta_1 -
\theta)(-1)^{\varepsilon(f^{i}_1)}.
\end{eqnarray}
Hypothesis 2 permit to present $f_{1}^{j}({\theta})$ in the form
\begin{eqnarray}
{} \hspace{-1em} f^{j}_{1}(\theta) = f^{j}_{1{}{\rm lin}}(\theta) +
f^{j}_{1{} {\rm nl}}(\theta),\
\min{\rm deg}_{g(\theta) \partial_{\theta}{g}(\theta)}f^{j}_{1 {}{\rm
nl}}(\theta) \geq 2\,, 
\end{eqnarray}
so that $f^{j}_{1}(\theta)$ appear by perturbation of functionally dependent
linearized constraints by means of nonlinear components.

An effective analysis of the constraints (A.22), considered as the
1st order on $\theta$ $N$ ODE system not being reduced to NF of the form
(A.2), is based on the fundamental

\noindent
\underline{\bf Theorem:}\\
The 1st order on $\theta$  $N$ ODE system w.r.t. uknowns
$g^{j}(\theta)$ (A.12) subject to conditions (A.18), (A.19) is reduced
to equivalent system of independent equations in so-called
generalized normal form (GNF) under following parametrization for
$g^{j}(\theta) = \bigl(
{\alpha}^{\bar{j}}(\theta), {\beta}^{\underline{j}} (\theta),
{\gamma}^{\sigma}(\theta)\bigr) ,{}j=(\bar{j}, \underline{j}, \sigma)$
\begin{eqnarray}
\partial_{\theta}{\alpha}^{\bar{j}}(\theta)  =
\varphi^{\bar{j}}\bigl({\alpha}(\theta) , {\gamma}(\theta),
\partial_{\theta}{\gamma}(\theta), \theta\bigr),\ \;
{\beta}^{\underline{j}}(\theta) =
\kappa^{\underline{j}}\bigl({\alpha}(\theta) , {\gamma}(\theta),
\theta\bigr)\,, 
\end{eqnarray}
with arbitrary superfunctions ${\gamma}^{\sigma}(\theta)$. The number of
${\gamma}^{\sigma}$:
$[{\gamma}]$ coincides with one of differential identities among
Eqs.(A.12)
\begin{eqnarray}
\int d\theta f^{j}_{1}\bigl( g(\theta),
\partial_{\theta}{g}(\theta), \theta\bigr) \check{{\cal R}}_{j
\sigma}\left( g(\theta), \partial_{\theta}{g}(\theta), \theta;
{\theta}^{\prime}\right) = 0\,,
\end{eqnarray}
where quantities
$\check{{\cal R}}_{j \sigma}\left( g(\theta),
\partial_{\theta}{g}(\theta), \theta; {\theta}^{\prime}\right)$ are a)
local on $\theta$ and b) functionally independent operators
\begin{eqnarray}
{\rm a})\hspace{0.3cm} \check{{\cal R}}_{j \sigma}\left( g(\theta),
\partial_{\theta}{g}(\theta), \theta; {\theta}^{\prime}\right) \equiv
\check{{\cal R}}_{j \sigma}(\theta; {\theta}^{\prime}) =
\sum_{k=0}^{1}\left(\left(\partial_{\theta}\right)^{k}\delta(\theta -
\theta')\right) \check{{\cal R}}^{k}_{{}j \sigma}\bigl( g(\theta),
\partial_{\theta}{g}(\theta), \theta\bigr)\,, \hspace{0.9cm} 
\end{eqnarray}
b) functional equation
\begin{eqnarray}
\int d{\theta}'\check{{\cal R}}_{j \sigma}(\theta;
{\theta}^{\prime})u^{\sigma}\bigl( g(\theta'),
\partial_{\theta'}{g}(\theta'), {\theta}^{\prime}\bigr) = 0 
\end{eqnarray}
has unique solution $u^{\sigma}(\theta') = 0$.

In order to be solvable in superfield form the system (A.23) must satisfy to
the solvability conditions written additionally to these equations
\begin{eqnarray}
\partial_{\theta}{\varphi}^{\bar{j}}\bigl({\alpha}(\theta),
{\gamma}(\theta),\partial_{\theta}{\gamma}(\theta),\theta\bigr) = 0\,.
\end{eqnarray}
The detailed proof of  the Theorem and its related consequences will be
considered in another paper.

\noindent
\underline{\bf Corollary:}\\
For dependent  holonomic constraints $f_{1}^{j}(\theta)$ from the condition
(A.19) it follows the existence of equivalent
system of constraints under following parametrization for $g^{j}(\theta)$
\begin{eqnarray}
g^{j}(\theta) = \left( {\alpha}^{A}(\theta),
 {\gamma}^{\sigma}(\theta)\right){},\ j=(A,\sigma ),{}
\sigma=1,\ldots,[\gamma],{}A=1,\ldots,[\alpha]\,,
\end{eqnarray}
by means of relationships
\begin{eqnarray}
\Phi^{A}\bigl( {\alpha}(\theta),
{\gamma}(\theta), \theta\bigr) = 0\,.
\end{eqnarray}
The number $[\gamma]$
coincides with one of algebraic (in the sense of differentiation w.r.t.
$\theta$) identities among $f_{1}^{j}(\theta)$
\begin{eqnarray}
f^{j}_{1}\bigl( g(\theta), \theta\bigr) {{\cal R}}^{0}_{{}j
\sigma}\bigl( g(\theta), \theta\bigr) = 0\,,
\end{eqnarray}
being
obtained by integration on $\theta$ of Eqs.(A.24)  with allowance made for
the type (A.25) connection  of $\check{{\cal R}}_{j \sigma}(\theta;
{\theta}^{\prime})$ with  ${\cal R}^{0}_{j \sigma}(\theta)$
\begin{eqnarray}
\check{{\cal R}}_{j
\sigma}\bigl( g(\theta), \theta ; \theta'\bigr) = \delta(\theta - \theta')
{\cal R}^{0}_{{}j \sigma}\bigl( g(\theta),
\theta\bigr)(-1)^{\varepsilon( f_{1}^{j}(\theta))}\,.
\end{eqnarray}
A dependence on $\partial_{\theta}{g}^j(\theta)$ in (A.30), (A.31) may be
only parametric one.

\vspace{1ex}
\begin{center}
{\large{\bf References}}
\end{center}
\begin{enumerate}
\item C. Becchi, A. Rouet and R. Stora, Phys. Lett. B 25 (1974) 344; Comm.
Math. Phys. 42 (1975) 127;\\
I.V. Tyutin, P.N. Lebedev Physical Inst. of the
RF Academy of Sciences, preprint, No.39 (1975).
\item G. Curci and R.
Ferrari, Phys. Lett. B 63 (1976) 91;\\
I. Ojima, Prog. Theor. Phys. Suppl. 64
(1979) 625.
\item E.S. Fradkin and G.A. Vilkovisky, Phys. Lett. B 55 (1975)
224;\\
I.A. Batalin and G.A. Vilkovisky, Phys. Lett. B 69 (1977) 309;\\
E.S.Fradkin and T.E. Fradkina, Phys. Lett. B 72 (1978) 343;\\
I.A. Batalin and E.S. Fradkin, Phys. Lett. B 122 (1983) 157;\\
M. Henneaux, Phys. Rev. 126 (1985) 2.
\item I.A. Batalin, P.M. Lavrov and I.V. Tyutin, J. Math. Phys. 31
(1990) 6; 31 (1991) 2708;\\
I.A. Batalin, P.M. Lavrov and I.V. Tyutin, Int.
 J. Mod. Phys. A 6 (1991) 3599.
\item I.A. Batalin and G.A. Vilkovisky,
Phys. Lett. B 102 (1981) 27; Phys.  Rev. D 28 (1983) 2567.
\item I.A. Batalin, P.M. Lavrov and I.V. Tyutin, J. Math. Phys. 31 (1990)
1487; 32
 (1991) 532; 32 (1991) 2513.
\item E. Witten and B. Zwiebach, Nucl. Phys. B
377 (1992) 55;\\
B.H. Lian and G.J. Zuckerman, Comm. Math. Phys. 154 (1993)
613;\\
B. Zwiebach, Nucl. Phys. B 390 (1993) 33.
\item C. Chevalley and S. Eilenberg, Trans. Amer. Math. Soc. 63 (1948) 85.
\item J.M.L. Fisch and M.Henneaux, Comm. Math. Phys. 128 (1990) 627;\\
G. Barnich and M. Henneaux, Phys. Lett. B 311 (1993) 123;\\
M. Henneaux, Contemp. Math. 219 (1998) 93;\\
J.D. Stasheff, Deformation Theory and the Batalin-Vilkovisky Master Equation,
in: Deformation Theory and Symplectic Geometry, D. Sternheimer et al (Eds.),
Kluwer Acad. Publ., 1997 (q-alg/9702012).
\item F. Bayen, M. Flato, C.
Fronsdal, A. Lichnerowicz and D. Sternheimer, Ann. Phys. 111 (1978) 61;\\
M. Schlesinger and J.D. Stasheff, J. of Pure and Appl. Algebra 38
(1985) 313;\\
M. Kontsevich, Deformation Quantization of Poisson Manifolds, I,
q-alg/9709040.
\item B.F. Fedosov, J. Diff. Geom. 40 (1994) 213;\\
\vspace{-4ex}
\begin{sloppypar}
B. Fedosov, De\-for\-ma\-ti\-on Qu\-an\-ti\-za\-ti\-on and
In\-dex The\-ory (Ber\-lin, Ger\-ma\-ny:  Aka\-de\-mic-Ver\-lag, 1996).
\end{sloppypar}
\item M.A. Grigoriev and S.L. Lyakhovich, Comm. Math. Phys.
218 (2001) 437.
\item F.A. Berezin, Introduction to Algebra and Analysis
with Anticommuting Variables (Moscow State University Press, 1983);\\
F.A. Berezin, Introduction to Superanalysis, ed. by A.A. Kirillov (D. Reidel,
 Dordrecht 1987).
\item A.S. Schwarz, Comm. Math. Phys. 155 (1993) 249; 158 (1993)
373.
\item O.M. Khudaverdian and A.P. Nersessian, Mod. Phys. Lett.
A 8 (1993) 2377.
\item I.A. Batalin and I.V. Tyutin, Int. J. Mod.
Phys. A 8 (1993) 2333; Mod. Phys. Lett.  A 8 (1993) 3673; Int. J.
Mod. Phys. A 9 (1994) 1707.
\item I.A. Batalin and R. Marnelius,
Phys. Lett. B 350 (1995) 44.
\item I.A. Batalin, R. Marnelius and
A.M. Semikhatov, Nucl. Phys. B 446 (1995) 249.
\item B. Geyer,
D.M. Gitman and P.M. Lavrov, Mod. Phys. Lett. A 14 (1999) 661;
Theor. Math. Phys. 123 (2000) 813.
\item I.A. Batalin and R.
Marnelius, Nucl. Phys. B 465 (1996)
521;\\
M.A. Grigoriev and A.M. Semikhatov, Phys. Lett. B 417 (1998) 259;\\
B. Geyer, P. Lavrov and A. Nersessian, Phys. Lett. B 512 (2001)
211;
Int. J. Mod. Phys. A 17 (2002) 1183;\\
B. Geyer and P.M. Lavrov, Modified triplectic quantization in general
coordinates, hep-th/0304011.
\item V.N. Shander, Funct. analysis and its applications (on russian) 14
(1980) No.2 91; 17 (1983) No.1 89.
\item ${\rm \ddot{O}}$.F. Dayi, Mod.
Phys. Lett. A 4 (1989) 361; A 8 (1993) 811; ibid 2087; Int. J. Mod. Phys. A
11 (1996) 1.
\item L. Bonora and M. Tonin, Phys. Lett. B 98 (1981) 48;\\
L. Bonora, P. Pasti and M. Tonin, J. Math. Phys. 23 (1982) 839;\\
L. Baulieu, Phys. Rep. 129 (1985) 1.
\item C.M. Hull, B. Spence and J.L.Vazquez-Bello,
Nucl. Phys. B 348 (1991) 108.
\item P.M. Lavrov, P.Yu. Moshin and
A.A. Reshetnyak, Mod. Phys. Lett. A 10 (1995) 2687; JETP Lett. 62 (1995) 780.
\item P.M. Lavrov, Phys. Lett. B 366 (1996) 160.
\item B. Geyer, P.M. Lavrov
and P.Yu. Moshin, Phys. Lett. B 463 (1999) 188;\\
P.M. Lavrov and P.Yu.
Moshin, Superfield Covariant Quantization with BRST Symmetry, in: Proc. of
the Int. Conf. dedicated to Memory of Prof. E. Fradkin: Quantization, Gauge
Theories and Strings, A. Semikhatov, M. Vasiliev and V.Zaikin (Eds.),
Scientific World, 2 (2001) 205.  \item N.R.F. Braga and A. Das, Nucl. Phys. B
422 (1995) 655;\\
N.R.F. Braga and S.M. Souza, Phys. Rev. D 53 (1996) 916;\\
E.M.C. Abreu and N.R.F. Braga, Phys. Rev. D 54 (1996) 4080; Int. J. Mod.
Phys.  A 13 (1998) 493.
\item I.A. Batalin, K. Bering and P.H. Damgaard,
Nucl. Phys. B 515 (1998) 455; Phys. Lett. B 446 (1999) 175.
\item M.A. Grigoriev and P.H. Damgaard, Phys. Lett. B 474 (2000) 323.
\item I. Batalin and R. Marnelius, Phys. Lett. B 512 (2001) 225;
Superfield algorithm for topological field theories, in: Multiple Facets of
Quantization and Supersymmetry, M. Marinov Memorial Volume, ed.
M. Olshanetsky and A. Vainshtein, WSPC (2002) 233.
\item L. Edgren and N. Sandstr${\rm \ddot{o}}$m, JHEP 0209 (2002) 036;
Superfield algorithm for higher order gauge theories, hep-th/0306175.
\item M. Alexandrov, M. Kontsevich, A. Schwarz and O. Zaboronsky, Int. J.
Mod. Phys. A 12 (1997) 1405.
\item A.S. Cattaneo and G. Felder, Comm. Math. Phys. 212 (2000) 591;
Mod. Phys. Lett. A 16 (2001) 179.
\item M.A. Grigoriev, A.M. Semikhatov and I.Yu. Tipunin, J. Math. Phys.
40 (1999) 1792; 42 (2001) 3315;\\
M.A. Grigoriev, Zero locus reduction of the BRST differential, published in
Moscow 1999, High energy physics and quantum field theory, 603,
hep-th/9906209.
\item I. Batalin and R. Marnelius, Int. J. Mod. Phys. A 14 (1999) 5049.
\item B.S. De Witt, Dynamical Theory of Groups and Fields (Gordon and
Breach, New York, 1965).
\item I.L. Buchbinder and C.M. Kuzenko, Ideas and Methods of
Supersymmetry and Supergravity (Institute of Physics Publishing, Bristol and
Philadelphia, 1995).
\item D.M. Gitman and I.V. Tyutin, Izv. Vuzov SSSR,
Ser. Fiz. (on russian), No.5 (1983) 3;\\
D.M. Gitman and I.V. Tyutin,
Quantization of Fields with Constraints (Springer-Verlag, Berlin and
Heidelberg, 1990).
\item C.N. Yang and R. Mills, Phys. Rev. 96 (1954) 191.
\item  E. Witten, Phys. Lett. B 86 (1979) 283.
\end{enumerate}
\end{document}